\documentclass[twocolumn,nofootinbib,preprintnumbers,superscriptaddress,amsmath,amssymb,PRL]{revtex4-1}
\usepackage{graphicx}
\usepackage{dcolumn}
\usepackage{float}
\usepackage{mathptmx, courier, pifont}
\usepackage[scaled=0.92]{helvet}
\usepackage[T1]{fontenc}
\usepackage{textcomp}
\usepackage{color}
\usepackage[normalem]{ulem}

\usepackage[dvipsnames]{xcolor}
\usepackage[colorlinks=true,urlcolor=Blue,linkcolor=Blue]{hyperref}
\usepackage[all]{hypcap}
\usepackage{makecell}

\begin{document}


\title{$\beta^-$ decay $Q$-value measurement of $^{136}$Cs and its implications to neutrino studies}
\author{Z.~Ge}\thanks{Corresponding author: z.ge@gsi.de}
\affiliation{GSI Helmholtzzentrum f\"ur Schwerionenforschung GmbH, 64291 Darmstadt, Germany}
\affiliation{Department of Physics, University of Jyv\"askyl\"a, P.O. Box 35, FI-40014, Jyv\"askyl\"a, Finland}%
\author{T.~Eronen}\thanks{Corresponding author: tommi.eronen@jyu.fi}
\affiliation{Department of Physics, University of Jyv\"askyl\"a, P.O. Box 35, FI-40014, Jyv\"askyl\"a, Finland}%
\author{A.~de Roubin}\thanks{Present address: KU Leuven, Instituut voor Kern- en Stralingsfysica, B-3001 Leuven, Belgium}
\affiliation{Universit\'e de Bordeaux, CNRS/IN2P3, LP2I Bordeaux, UMR 5797, F-33170 Gradignan, France}%
\author{M.~Ramalho}
\affiliation{Department of Physics, University of Jyv\"askyl\"a, P.O. Box 35, FI-40014, Jyv\"askyl\"a, Finland}%
\author{J.~Kostensalo}
\affiliation{Natural Resources Institute Finland, Yliopistokatu 6B, FI-80100, Joensuu, Finland}%
\author{J.~Kotila}
\affiliation{Department of Physics, University of Jyv\"askyl\"a, P.O. Box 35, FI-40014, Jyv\"askyl\"a, Finland}%
\affiliation{Finnish Institute for Educational Research, University of Jyv\"askyl\"a, P.O. Box 35, FI-40014, Jyv\"askyl\"a, Finland}%
\affiliation{Center for Theoretical Physics, Sloane Physics Laboratory, Yale University, New Haven, Connecticut 06520-8120, USA}
\author{J.~Suhonen}\thanks{Corresponding author: jouni.t.suhonen@jyu.fi} 
\affiliation{Department of Physics, University of Jyv\"askyl\"a, P.O. Box 35, FI-40014, Jyv\"askyl\"a, Finland}%
\author{D.~A.~Nesterenko}
\affiliation{Department of Physics, University of Jyv\"askyl\"a, P.O. Box 35, FI-40014, Jyv\"askyl\"a, Finland}%
\author{A.~Kankainen}
\affiliation{Department of Physics, University of Jyv\"askyl\"a, P.O. Box 35, FI-40014, Jyv\"askyl\"a, Finland}%
\author{P.~Ascher}
\affiliation{Universit\'e de Bordeaux, CNRS/IN2P3, LP2I Bordeaux, UMR 5797, F-33170 Gradignan, France}%
\author{O.~Beliuskina}
\affiliation{Department of Physics, University of Jyv\"askyl\"a, P.O. Box 35, FI-40014, Jyv\"askyl\"a, Finland}%
\author{M.~Flayol}
\affiliation{Universit\'e de Bordeaux, CNRS/IN2P3, LP2I Bordeaux, UMR 5797, F-33170 Gradignan, France}%
%
\author{M.~Gerbaux}
\affiliation{Universit\'e de Bordeaux, CNRS/IN2P3, LP2I Bordeaux, UMR 5797, F-33170 Gradignan, France}%
\author{S.~Gr\'evy}
\affiliation{Universit\'e de Bordeaux, CNRS/IN2P3, LP2I Bordeaux, UMR 5797, F-33170 Gradignan, France}%
\author{M.~Hukkanen}
\affiliation{Department of Physics, University of Jyv\"askyl\"a, P.O. Box 35, FI-40014, Jyv\"askyl\"a, Finland}%
\affiliation{Universit\'e de Bordeaux, CNRS/IN2P3, LP2I Bordeaux, UMR 5797, F-33170 Gradignan, France}%
\author{A.~Husson} 
\affiliation{Universit\'e de Bordeaux, CNRS/IN2P3, LP2I Bordeaux, UMR 5797, F-33170 Gradignan, France}%
\author{A.~Jaries} 
\affiliation{Department of Physics, University of Jyv\"askyl\"a, P.O. Box 35, FI-40014, Jyv\"askyl\"a, Finland}%
\author{A.~Jokinen} 
\affiliation{Department of Physics, University of Jyv\"askyl\"a, P.O. Box 35, FI-40014, Jyv\"askyl\"a, Finland}%
\author{I.~D.~Moore}
\affiliation{Department of Physics, University of Jyv\"askyl\"a, P.O. Box 35, FI-40014, Jyv\"askyl\"a, Finland}%
\author{P.~Pirinen}
\affiliation{Department of Physics, University of Jyv\"askyl\"a, P.O. Box 35, FI-40014, Jyv\"askyl\"a, Finland}%
\author{J.~Romero}
\affiliation{Department of Physics, University of Jyv\"askyl\"a, P.O. Box 35, FI-40014, Jyv\"askyl\"a, Finland}%
\affiliation{Department of Physics, University of Liverpool, Liverpool, L69 7ZE,  United Kingdom}
\author{M.~Stryjczyk}
\affiliation{Department of Physics, University of Jyv\"askyl\"a, P.O. Box 35, FI-40014, Jyv\"askyl\"a, Finland}%
\author{V.~Virtanen}
\affiliation{Department of Physics, University of Jyv\"askyl\"a, P.O. Box 35, FI-40014, Jyv\"askyl\"a, Finland}%
\author{A.~Zadvornaya}\thanks{Present address: II. Physikalisches Institut, Justus-Liebig-Universit\"at Gie\ss en, 35392 Gie\ss en, Germany}
\affiliation{Department of Physics, University of Jyv\"askyl\"a, P.O. Box 35, FI-40014, Jyv\"askyl\"a, Finland}%
%
\date{\today}
\begin{abstract}
%
The $\beta^-$ decay $Q$-value of $^{136}$Cs ($J^\pi = 5^+$, $t_{1/2} \approx 13$~days) was measured with the JYFLTRAP Penning trap setup at the Ion Guide Isotope Separator On-Line (IGISOL) facility of the University of Jyväskylä, Finland. 
The mono-isotopic samples required in the measurements were prepared with a new scheme utilised for the cleaning, based on the coupling of dipolar excitation with Ramsey's method of time-separated oscillatory fields and the phase-imaging ion-cyclotron-resonance (PI-ICR) technique.
%
The $Q$ value is determined to be 2536.83(45) keV, which is $\sim$4 times more precise and 11.4(20) keV ($\sim$  6$\sigma$) smaller than the adopted value in the most recent Atomic Mass Evaluation AME2020. 
The daughter, $^{136}$Ba, has a 4$^+$ state at 2544.481(24) keV and a $3^-$ state at 2532.653(23) keV, both of which can potentially be ultralow $Q$-value end-states for the $^{136}$Cs decay. 
With our new ground-to-ground state $Q$ value, the decay energies to these two states become -7.65(45) keV and 4.18(45) keV, respectively. 
The former is confirmed to be  negative at the level of $\sim$ 17$\sigma$, which verifies that this transition is not a suitable candidate for neutrino mass determination. On the other hand, the slightly negative $Q$ value makes this transition an interesting candidate for the study of virtual $\beta$-$\gamma$ transitions. The decay to the  
3$^{-}$ state is validated to have a positive low $Q$ value which makes it a viable candidate for neutrino mass determination. For this transition, we obtained a shell-model-based half-life estimate of $2.1_{-0.8}^{+1.6}\times10^{12}$\,yr.
Furthermore, the newly determined low reaction threshold  of 79.08(54) keV for the charged-current $\nu_e$+$^{136}$Xe (0$^{+}$)$\rightarrow$ $^{136}$Cs$^*$+$e^-$ neutrino capture process is used to update the cross sections for a set of neutrino energies relevant to solar $^7$Be, pep, and CNO neutrinos. Based on our shell-model calculations, the new lower threshold shows event rates of 2--4 percent higher than the old threshold for several final states reached by the different species of solar neutrinos. 
\end{abstract}
\maketitle
\section{Introduction}
The standard model (SM) predicts that the neutrino is mass-less, and how neutrinos acquire their small masses, verified by the neutrino-oscillation experiments, is consequently a matter of great theoretical interest and may be evidence of new physics beyond the SM~\cite{Fukuda1998,SNOCollaboration2002,Gerbino2018a}. 
Assessing the neutrino mass scale has been an outstanding task for particle physics, as the absolute value of the neutrino mass would provide an important parameter to extend the SM of particle physics and to understand the origin of fermion masses beyond the Higgs mechanism. 
The neutrinoless double  $\beta^{}$ decay experiments aim to probe if neutrinos are of Dirac or Majorana nature and to measure the effective Majorana neutrino mass ~\cite{Suhonen1998,Avignone2008,Ejiri2019}.  This method is however nuclear-model dependent and strongly relies on the calculation of the involved nuclear matrix elements, sensitive to the details of the nuclear wave functions describing the initial, intermediate, and final nuclear states of the process~\cite{Ejiri2019}. Complementary ways to probe the involved wave functions have been devised, like the nuclear muon capture, charge-exchange and double charge-exchange reactions~\cite{Ejiri2019}. Nevertheless,  
$\beta^{-}$-decay  or electron-capture (EC) spectrum end-point study remains currently the only laboratory method to provide a model-independent measurement of the absolute scale of the (anti)neutrino mass. In these experiments the most sensitive upper limits on the mass of the electron neutrino m$_{\nu_e}$ have been achieved by investigating the end point of the $\beta^{-}$ electron spectrum. 
The most stringent upper limit of 0.8 eV/c$^2$ (90\% Confidence Level (C.L.)) for the electron-antineutrino mass  is obtained by studying the tritium decay  in the KATRIN (KArlsruhe TRitium Neutrino) experiment~\cite{Aker2022}, and an upper limit of 150 eV/c$^2$ (95\% C.L.) is obtained for the electron-neutrino mass, as achieved by studying the EC of $^{163}$Ho in the ECHo experiment~\cite{Velte2019}. 
In these decay experiments, as small as possible $Q$ value of the decay is essential to partially balance the limitation on the statistics when looking for the tiny (anti)neutrino-mass generated distortion close to the end-point energy. The preference for lower $Q$ values is based on the fact that the fraction of decays in a given energy interval $\Delta{E}$ below the end-point will be increased with a lower $Q$ value~\cite{Ferri2015}.  

Up to now, only ground-state-to-ground-state (gs-to-gs) decay cases of $^{3}$H, $^{187}$Re ($\beta^-$ decay)  and $^{163}$Ho (electron capture), having the lowest known gs-to-gs $Q$ values, have been used for direct neutrino-mass-determination experiments.
The $\beta^-$ decay of tritium, $^{3}$H(1/2$^{+}$)$\rightarrow$ $^{3}$He(1/2$^{+}$), which is of the allowed type (a Fermi and/or Gamow–Teller transition) with a $Q$ value ($Q_{\beta^-}^0$) of $\sim$18.6 keV~\cite{Myers2015}, is utilised to measure the electron antineutrino mass. 
In an EC transition, like $^{163}_{\ 67}$Ho + $e^-$ $\rightarrow$ $^{163}_{\ 66}$Dy$^*$ + $\nu_e$, one can determine the electron neutrino mass from the upper end of the decay spectrum of the excited Dy, which is given by the $Q$ value ($\sim$2.8 keV) minus the neutrino mass. 

The possibility to utilize transitions to excited final states has recently attracted a lot of attention, as reviewed in \cite{Redshaw2023}. Intensive search for isotopes featuring  $\beta^-$/EC transitions from ground-state-to-excited-states (gs-to-es) with a positive low $Q$ value, preferably ultra-low (< 1 keV), has recently been carried out~\cite{Haaranen2013,Suhonen2014,Sandler2019,Karthein2019a,DeRoubin2020,ge2021,ge2021b,Ge2022a,ERONEN2022,Ge2022a,Ge2022b,Ramalho2022,Keblbeck2023}. In addition to the slightly positive $Q$ values,  the slightly negative $Q$ values can also be of interest in seeking for a new type of transition process, like the virtual radiative "detour" transitions (RDT). A
recent study of this type of transition in $^{59}$Ni was carried out in Ref.~\cite{Pfutzner2015}, where a virtual transition via a state 26 keV higher than allowed by the $Q$ value of the transition was found to contribute about 4\% to the experimental gamma spectrum. This result highlights that a slightly energetically forbidden transition will open a door to the possibility to study RDTs. Since the probability of such a detour transition is proportional to $(E^*-E_{\gamma})^{-2}$ \cite{Longmire49}, where  $E_{\gamma}$ is the energy of the emitted gamma ray, a transition with an ultra-low negative $Q$ value would make the RDT a relatively strong channel and thus easier to detect. 

Special attention is given to
possible alterations in neutrino-capture cross sections of low-energy neutrinos, for example those from the Sun, by the more precise $Q$-value measurements. Of interest are the charged-current $\nu_e$+$^{136}$Xe (0$^{+}$)$\rightarrow$ $^{136}$Cs$^*$+$e^-$ neutrino-capture cross sections for the solar $^7$Be, pep, and CNO neutrinos where our improved threshold value could alter the cross sections and thus the detection potential of these neutrinos in xenon-based solar-neutrino observatories \cite{Haselschwardt20}.

In summary, a precise and accurate determination of the transition $Q$ value is extremely important to validate the possible further usage of  low $Q$-value-decay candidate transitions in the context of searches for the absolute (anti)neutrino mass scale or for radiative ``detour'' transitions. Also implications for the low-energy solar-neutrino detection could potentially be of relevance.
The allowed transition $^{136}$Cs (5$^+$, $t_{1/2} \sim $ 13 days)  $\rightarrow$ $^{136}$Ba$^*$ (4$^+$, 2544.481(24) keV~\cite{NNDC}),  is of paramount interest for the antineutrino-mass studies because of its small gs-to-es $Q$ value $Q_{\beta^-}^*$ ($=Q_{\beta^-}^0 - E^*$) of 3.7(19) keV~\cite{Wang2021}. This transition is proposed to be one of the most promising candidates for neutrino mass determination~\cite{Keblbeck2023}. The $Q_{\beta^-}^*$ value for this transition can be deduced from the sub-keV-precision energy-level E$^*$ data in ~\cite{NNDC} and the gs-to-gs $Q$ value of 2548.2(19) keV from AME2020~\cite{Wang2021}.
The gs-to-gs $Q$ value of $^{136}$Cs in AME2020 is evaluated 
%
primarily using data from two $^{136}$Cs(${\beta^-}$)$^{136}$Ba-decay experiments performed more than 60 years ago ~\cite{136Csa,136Csb}.
Previous studies have already demonstrated that $Q$ values derived in indirect methods, such as decay spectroscopy, show large discrepancies with those from direct mass measurements and can be inaccurate over a wide range of mass numbers ~\cite{Fink2012,Nesterenko2019,ge2021}. 
The AME2020 $Q$ value with its large uncertainty of 1.9 keV, and its possible inaccuracy, requires verification to
unambiguously identify energetically allowed or forbidden low-$Q$ transitions. 
To confirm whether there are $\beta^-$-decay transitions from $^{136}$Cs that can serve as potential candidates for future antineutrino-mass determination experiments or be eligible for studies of RDTs, the gs-to-gs $Q$ value needs to be measured directly with a sub-keV uncertainty. 

Penning trap mass spectrometry (PTMS) is the leading technique for accurate and precise mass and $Q$-value determination. It relies on the determination of the cyclotron frequency ratio of parent and daughter ions, from which the mass difference can be extracted. 
In this article, we report on the first-time direct determination of the gs-to-gs $\beta^{-}$-decay $Q$ value of  $^{136}$Cs  with the JYFLTRAP PTMS. A method based on utilisation of a dipolar RF-excitation of ion motion with time-separated oscillatory fields in the precision trap coupled with the phase-imaging ion-cyclotron-resonance (PI-ICR) technique, is used to prepare mono-isotopic ions to ensure a contaminant-free high-precision $Q$-value determination. The new scheme allows for an efficient isobaric ion separation of $^{136}$Cs from the small mass-difference (90 keV/$c^2$) contaminant of  $^{136}$Xe, and  isomeric ion separation of $^{136}$Cs from its co-produced low-lying isomeric state at 518~keV.

\section{Experimental method}
\begin{figure}[!htb]
\centering
\includegraphics[width=1.0\columnwidth]{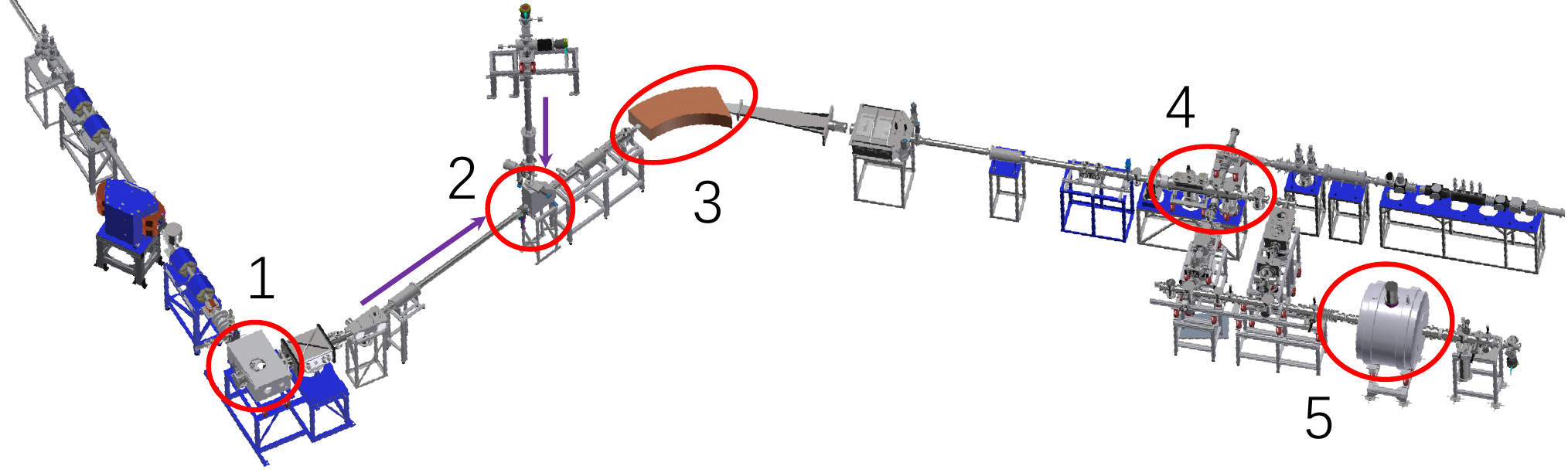}
\caption{(Color online).  Schematic view of the IGISOL facility. The  $^{136}$Cs$^{+}$  and $^{136}$Ba$^{+}$ ions were produced with proton-induced  fission reactions on a natural uranium target within the IGISOL target chamber (1). The online beam  was selected with an electrostatic kicker (2) and the dipole magnet (3) was used to transport only ions with $A/q = 136$. The ion cooling and bunching was carried out in the RFQ cooler-buncher (4) and the final $Q$-value and mass measurement was performed with the JYFLTRAP Penning trap setup (5). }
\label{fig:igisol}
\end{figure}

The measurement was performed at the Ion Guide Isotope Separator On-Line facility (IGISOL) ~\cite{Moore2013} with the JYFLTRAP double Penning trap mass spectrometer~\cite{Eronen2012,Kolhinen2013} at the University of Jyv\"askyl\"a, Finland.  A schematic view of the experimental setup is shown in Fig.~\ref{fig:igisol}.
The two ion species of the decay pair, $^{136}$Cs and $^{136}$Ba, were produced by irradiating a natural uranium target foil with a few $\mu$A proton beam at 30 MeV from the the K-130 cyclotron. 
The produced ions were stopped and thermalized in a helium-filled gas cell, and extracted out with the gas flow and electric fields via a sextupole ion guide~\cite{Karvonen2008}. 
The extracted ions were accelerated to 30 keV of energy 
and transported further to the 55$^\circ$ dipole magnet having a mass resolving power of $M/\Delta{M}$ $\sim$ 500. This allows isobaric separation to select only ions with $A/q=136$, including $^{136}$Cs, $^{136m}$Cs, $^{136}$Xe, $^{136}$Ba,  $^{136}$Te and  $^{136}$I that are all produced in the fission reaction. The ions are then delivered to a radiofrequency quadrupole cooler-buncher~\cite{Nieminen2001}, where they are accumulated, cooled and bunched prior to sending the bunches to the JYFLTRAP double Penning trap mass spectrometer for further purification and the final mass-difference measurements. 


\begin{figure}[!htb]
   \includegraphics[width=0.75\columnwidth]{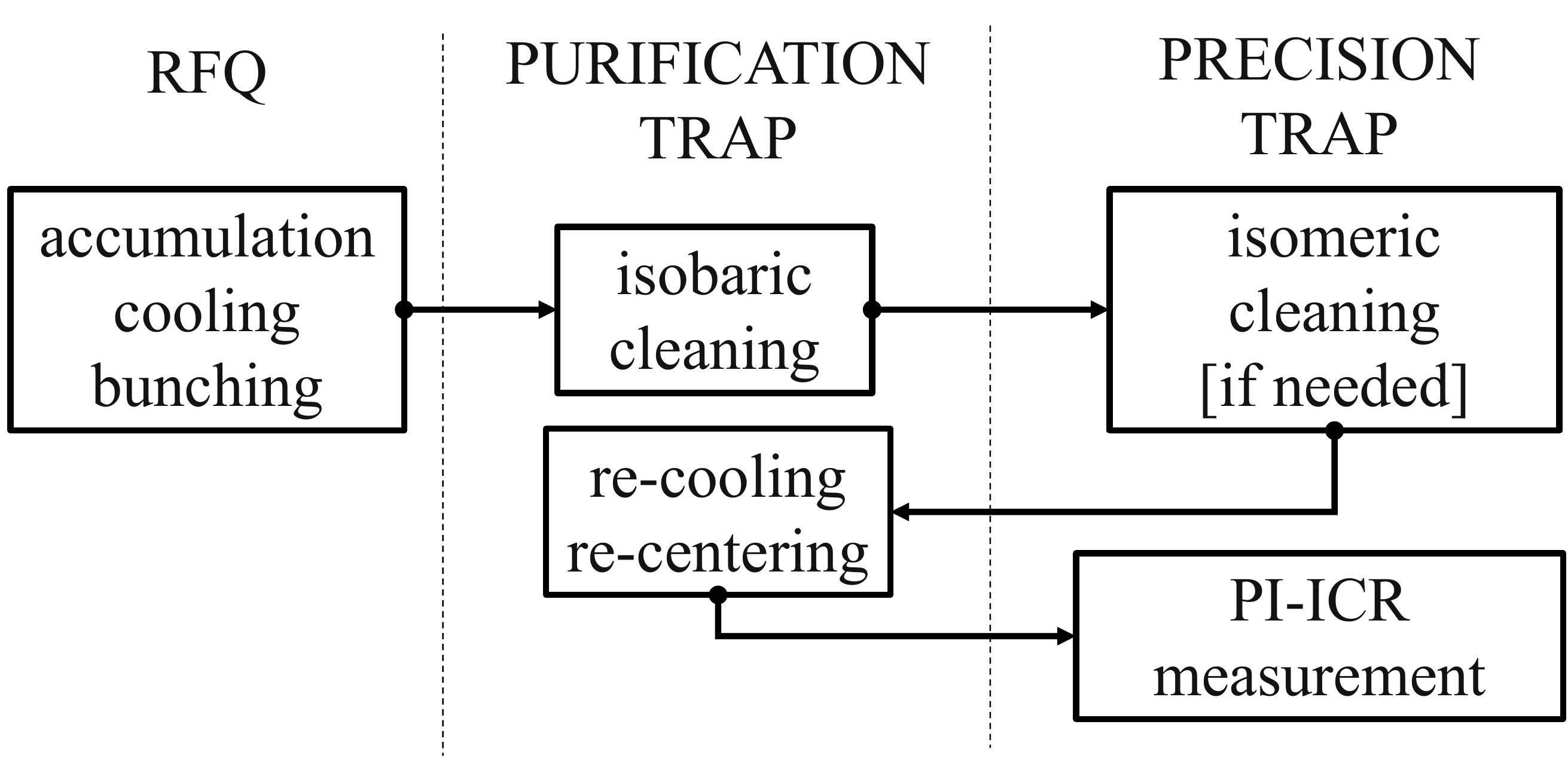}
   \caption{(Color online). Schematic of the measurement cycle at JYFLTRAP~\cite{Eronen2008a,Eronen2012,Nesterenko2018,nesterenko2021}. 
   The purification trap is used for isobaric cleaning, and it is often sufficient to provide contaminant-free samples in most of the cases studied. The precision trap is used for further isomeric cleaning when higher resolving power is needed and final high-precision mass or $Q$-value measurements.}  
   \label{fig:cleaning-scheme}
\end{figure}

\begin{figure}[!htb]
   \includegraphics[width=0.95\columnwidth]{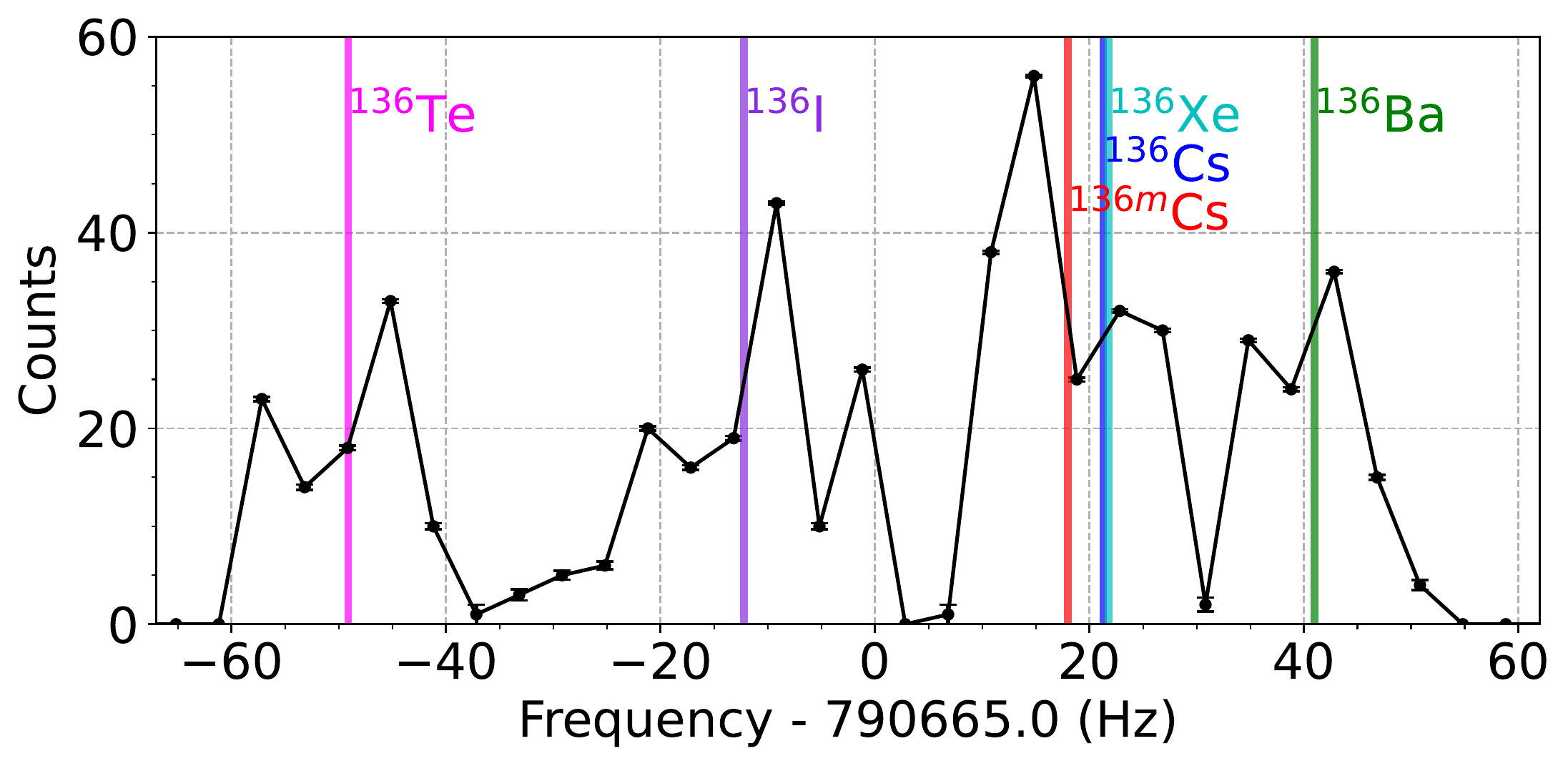}
   \caption{(Color online).  Detected number of ions downstream from the PTMS as a function of quadrupole excitation frequency in the purification trap. The vertical lines in various colors indicate the excitation frequency to be applied for the selection of the corresponding ion species of singly-charged ions of mass $A = 136$. 
}  
   \label{fig:RF2}
\end{figure}

JYFLTRAP consists of two cylindrical Penning traps in  a 7~T magnetic field. The first trap (purification trap) is filled with helium buffer gas and is used for isobaric purification via the buffer-gas cooling technique~\cite{Savard1991}. This technique can provide a mass purification with a resolving power of around $10^{5}$. For higher mass resolving power, the Ramsey cleaning method~\cite{Eronen2008a} can be employed.  Fig.~\ref{fig:cleaning-scheme} shows the schematic diagram of the steps employed 
prior to the actual mass and $Q$-value measurements in the second trap (precision trap).

In this experiment, a purified sample of decay-daughter ions $^{136}$Ba$^{+}$ was prepared with the buffer-gas cooling technique, which was enough to remove all other ion species. This is shown in Fig.~\ref{fig:RF2}, where the mass-sensitive quadrupole excitation frequency was scanned over the resonance frequencies of the $A/q = 136$ ion species.

For the preparation of clean samples of $^{136}$Cs$^{+}$ decay-parent ions, higher resolving power is needed. As indicated in Fig.~\ref{fig:RF2}, the selection frequencies to center ions of $^{136}$Cs, $^{136m}$Cs, $^{136}$Xe are too close to completely separate them from each other. In this case, the Ramsey cleaning technique~\cite{Eronen2008a} is employed right after the sideband buffer-gas cooling.
Due to the closeness in mass of $^{136}$Cs$^{+}$ to both  $^{136m}$Cs$^{+}$ and $^{136}$Xe$^{+}$, it is still challenging by the use of the conventional Ramsey cleaning technique~\cite{Eronen2008a} to completely purify the ion sample of  $^{136}$Cs$^{+}$. Here we introduce a new cleaning scheme, which relies on scanning the dipolar excitation (so-called cleaning excitation) frequency over the $\nu_{+}$ frequency of the ion species present in the precision trap while applying 
the phase-imaging ion-cyclotron-resonance (PI-ICR) technique~\cite{Nesterenko2018,nesterenko2021} to identify which ions are ultimately transmitted.
%
%

The dipolar excitation was applied as two 22-ms fringes interrupted for 762 ms. Depending on the applied frequency, the ions are left with different cyclotron motion amplitude. If this amplitude is high enough, the ions will hit the electrode of the diaphram between the two traps in the subsequent transfer back to the first trap for re-cooling and centering. To assess the composition of the remaining ion bunch, the ions are transferred again to the precision trap where the PI-ICR method is utilized. 

\begin{figure}[!htb]
   \includegraphics[width=0.99\columnwidth]{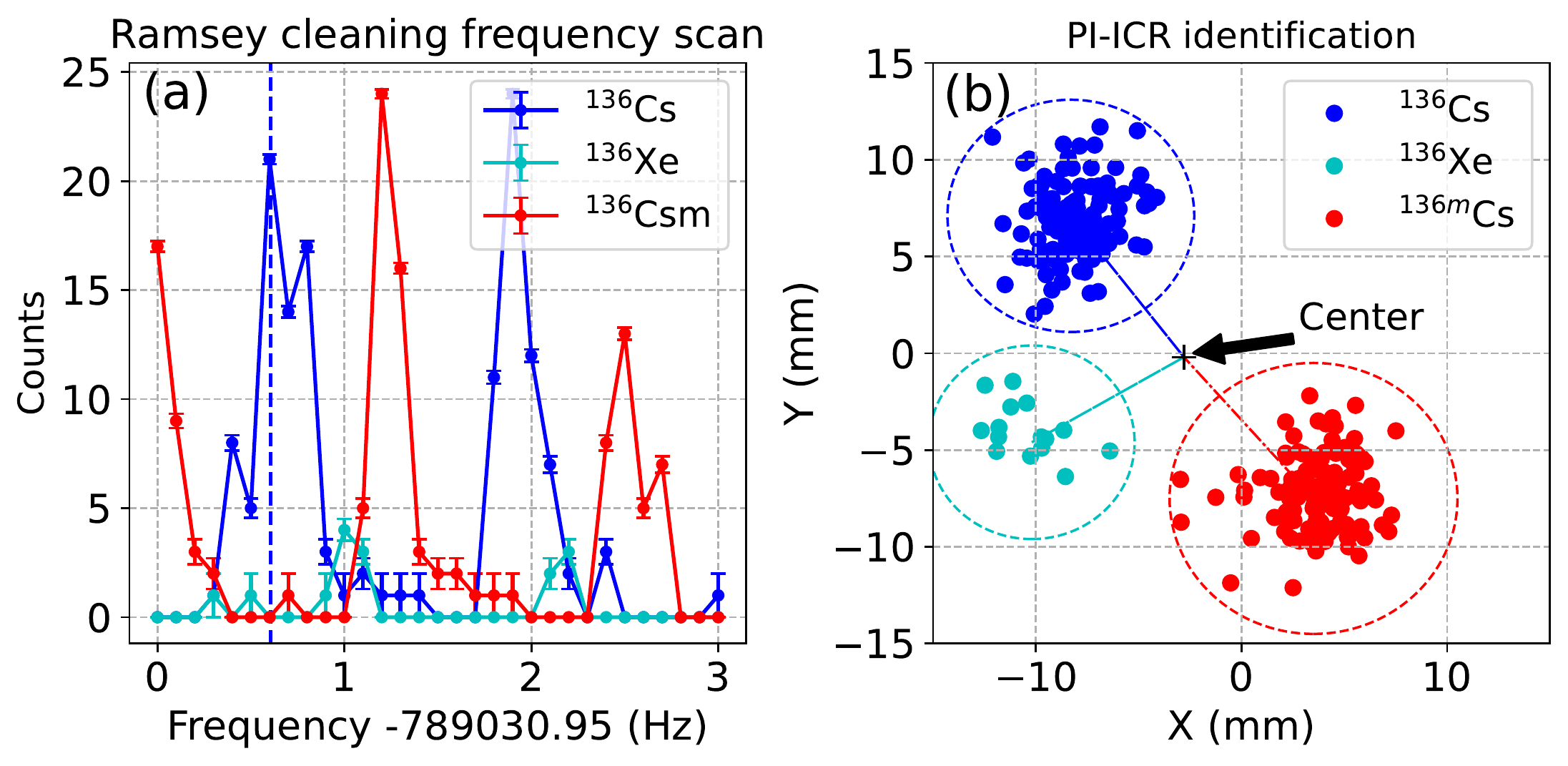}
   \caption{(Color online). A Ramsey-type dipole excitation frequency scan (a)  with a  22 ms (On) - 762 ms (Off) - 22 ms (On) excitation pattern in the second trap gated by the PI-ICR identification (458 ms phase accumulation time) plot (b). The used angular gates are highlighted. 
  The vertical dashed line shows the optimal frequency to transmit $^{136}$Cs ions while suppressing the others.
}  
   \label{fig:Ramesy}
\end{figure}

The phase accumulation time in the PI-ICR identification was chosen to be 458 ms. This allowed sufficient angular separation to unambiguously observe all three ion species. 
Fig.~\ref{fig:Ramesy} shows the dipolar excitation scan while gating on the well-resolved spots of different species. Setting the excitation frequency to maximally transmit $^{136}$Cs$^+$ ions, the other two are, if not completely, at least heavily suppressed.
%
%
After the verification, the cleaning settings are locked and the final mass measurement with the PI-ICR technique commenced. The actual PI-ICR mass measurement was performed with phase accumulation times chosen such that the spots of different ions did not overlap and thus interfere with spot position fitting.


The PI-ICR technique used in this work for the $Q$ value measurement is the state-of-the-art Penning trap mass measurement technique for short-lived ions 
\cite{Nesterenko2018,Eliseev2014,Eliseev2013}.
This technique allows extraction of the free-space ion-cyclotron frequency
\begin{equation}
    \nu_c = \frac{1}{2\pi}\frac{q}{m}B,
\end{equation}
where $q$ is the charge of the ion, $m$ the mass and $B$ the magnetic field of the trap, through observation of the final motional phase of the ions. The measurement begins by initial excitation of cyclotron motion of the ions with a short ($\sim 1$ms) dipolar pulse at the $\nu_+$ frequency. This is followed by a cyclotron-to-magnetron motion quadrupolar conversion pulse at frequency $\nu_c$. Finally, ions are extracted from the trap to be detected with the position-sensitive MCP detector.

The quadrupolar conversion pulse needs to be applied with two different delay times while keeping the overall cycle identical. One short delay is used to record the so-called magnetron phase and the other, longer, for the cyclotron phase. The delay difference of these settings define the phase-accumulation time $t_{acc}$. The cycle is
described in detail in~\cite{Nesterenko2018,nesterenko2021}. The phase angle detected between the two cycles with respect to the center spot is $\alpha_c$ = $\alpha_+$-$\alpha_-$, where $\alpha_+$ and $\alpha_-$ are the polar angles of the cyclotron and magnetron motion phases. The cyclotron frequency $\nu_{c}$ is derived from: 
\begin{equation}
\label{eq:nuc2}
\nu_{c}=\frac{\alpha_{c}+2\pi n_{c}}{2\pi{t_{acc}}},
\end{equation}
%
where $n_{c}$ is the number of complete revolutions of the measured ions during the phase accumulation time $t_{acc}$. Two different accumulation times, 458 ms and 428 ms,
were used  in this measurement. These times were chosen to ensure contaminant ions (especially $^{136m}$Cs and $^{136}$Xe for $^{136}$Cs frequency determination) do not appear on the same angle with the ion of interest in case of leakage from the trap.

\begin{figure}[!htb]
   \includegraphics[width=0.9\columnwidth]{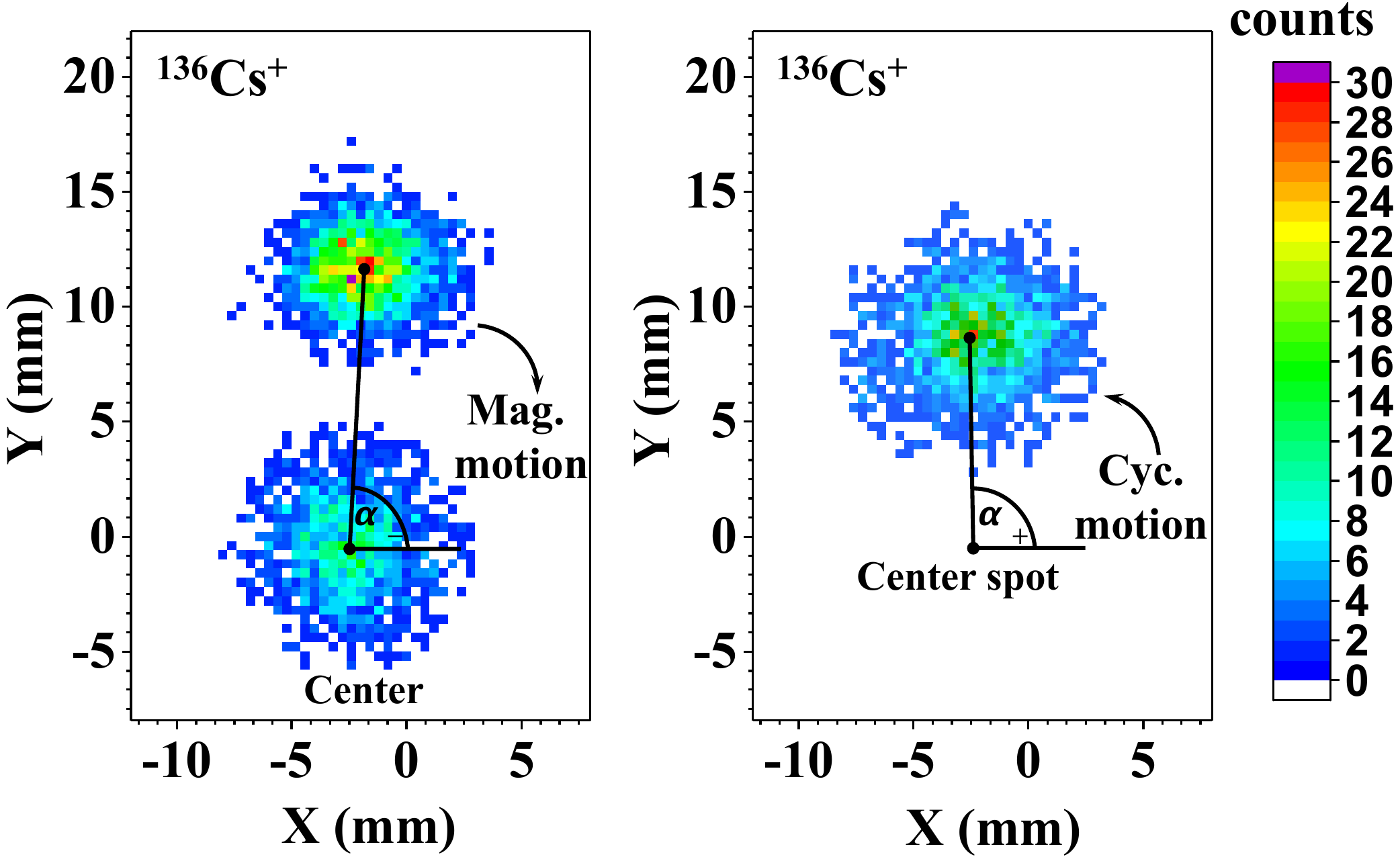}
   \caption{(Color online). $^{136}$Cs$^{+}$ ion spots of center, cyclotron phase and magnetron phase on the 2-dimensional position-sensitive MCP detector after a PI-ICR excitation pattern with an accumulation time of 458 ms. The magnetron phase spot along with a center spot is illustrated on the left and the cyclotron phase spot on the right. The cyclotron frequency $\nu_c$ is deduced from  angle difference between the two spots relative to the center spot. The color bar indicates  the number of detected ions for each pixel.
   }
   \label{fig:2-phases}
\end{figure}

The excitation time was fine-tuned to be multiple integers of $\nu_c$ period such that the angle $\alpha_c$  did not exceed a few degrees. This reduces the shift in the $\nu_{c}$ measurement due to the conversion of the cyclotron motion to magnetron motion and the possible distortion of the ion-motion projection onto the detector to a level well below 10$^{-10}$~\cite{Eliseev2014}. 
Additionally, the start time of the initial cyclotron motion excitation was scanned over one magnetron period and the extraction delay was varied over one cyclotron period to account for any residual magnetron and cyclotron motion that could shift the different spots. 
An example of  phase spots collected is shown in Fig.~\ref{fig:2-phases}. In total, $\sim 13$~h of data was collected in interleaved $\nu_{c}$ measurements of 
$^{136}$Cs$^{+}$ and $^{136}$Ba$^{+}$ ions. 

The $Q_{\beta^-}$ value can be derived using the cyclotron frequency ratio of the measured ion pair:
\begin{equation}
\label{eq:Qec}
Q_{\beta^-}=(M_p - M_d)c^2 = (R-1)(M_d - qm_e)c^2+(R \cdot B_{d} - B_{p}),
\end{equation}
where $M_p$ and  $M_d$ are the masses of the parent  ($^{136}$Cs$^{+}$) and daughter  ($^{136}$Ba$^{+}$) atoms, respectively, and $R$ their cyclotron frequency ratio ($\frac{\nu_{c,d}}{\nu_{c,m}}$) for singly-charged ions ($q=1$). $m_{e}$ is the mass of an electron.  $B_{p}$ and $B_{d}$ are the electron binding energies of the parent and daughter atoms, which is neglected as it is on the order of a few eV~\cite{NIST_ASD} and $R$ is off from unity by less than $10^{-4}$. Since both the parent and daughter have the same $A/q$,  mass-dependent shifts effectively become inferior compared to the statistical uncertainty achieved in the measurements. Moreover, due to the very small relative mass difference of the parent and daughter ($\Delta M/M$ < $10^{-4}$), the contribution of the uncertainty to the $Q$ value from the mass uncertainty of the reference (daughter), 0.24 keV/c$^2$, can be neglected.

\begin{table}[!htb]
\caption{
Mean cyclotron frequency ratio $\overline R$ between the daughter  $^{136}$Ba (0$^{+}$) and parent $^{136}$Cs (5$^{+}$) nuclei, $Q_{\beta^-}$  values (in keV) and the mass-excess (in keV/c$^2$) of parent nuclei determined in this work in comparison with the AME2020 values~\cite{Wang2021}.}
\begin{ruledtabular}
   \begin{tabular*}{\textwidth}{cccc}
& $\overline{R}$ & $Q_{\beta^-}$ & \makecell[c]{mass-excess \\( $^{136}$Cs (5$^{+}_{\rm g.s.}$))}\\
\hline\noalign{\smallskip}
AME2020 &  &  2548.2(19) & -86338.9	(19) \\
This Work&  1.000 020 039 1(35) &  2536.83(45) & -86350.09(54) \\
   \end{tabular*}
   \label{table:Q-value}
\end{ruledtabular}
\end{table}

\begin{table}[!htb]
   \caption{Potential candidate transitions of initial state of parent nucleus $^{136}$Cs (5$^{+}$, ground state), to the excited states of daughter $^{136}$Ba with ultra-low $Q$ values. The first column gives  the spin and parity of excited final state of $^{136}$Ba for the low $Q$-value transition. The second column gives the decay type.  The third column gives the derived decay $Q^*_{\beta^-}$ value  in units of keV  from literature (Lit.)~\cite{Wang2021} and the fourth  column from this work (new). The fifth column gives the experimental excitation energy with the experimental uncertainty~\cite{NNDC} in units of keV. The last column shows the confidence ($\sigma$) of the $Q$ value being non-zero. A negative value indicates a negative $Q$ value. 1st FU represents 1st forbidden unique.}
  \begin{ruledtabular}
   \begin{tabular*}{\textwidth}{lccccc}
 \makecell[c]{Final state\\ of $^{136}$Ba} &Decay type & \makecell[c]{$Q^*_{\beta^-}$ \\(Lit.)} &\makecell[c]{$Q^*_{\beta^-}$ \\ (new)}& $E^{*}$ & \makecell[c]{$Q/\delta Q$ \\(new)} \\
\hline\noalign{\smallskip}
   4$^{+}$&  allowed &   3.7(19)&-7.65(45)    &2544.481(24) & -17\\
   3$^{-}$&  1st FU &   15.5(19)& 4.18(45)    &2532.653(23)& 9 \\
   \end{tabular*}
   \label{table:low-Q}
   \end{ruledtabular}
\end{table}

\section{Results and discussion}
\begin{figure}[!htb]
   \includegraphics[width=0.95\columnwidth]{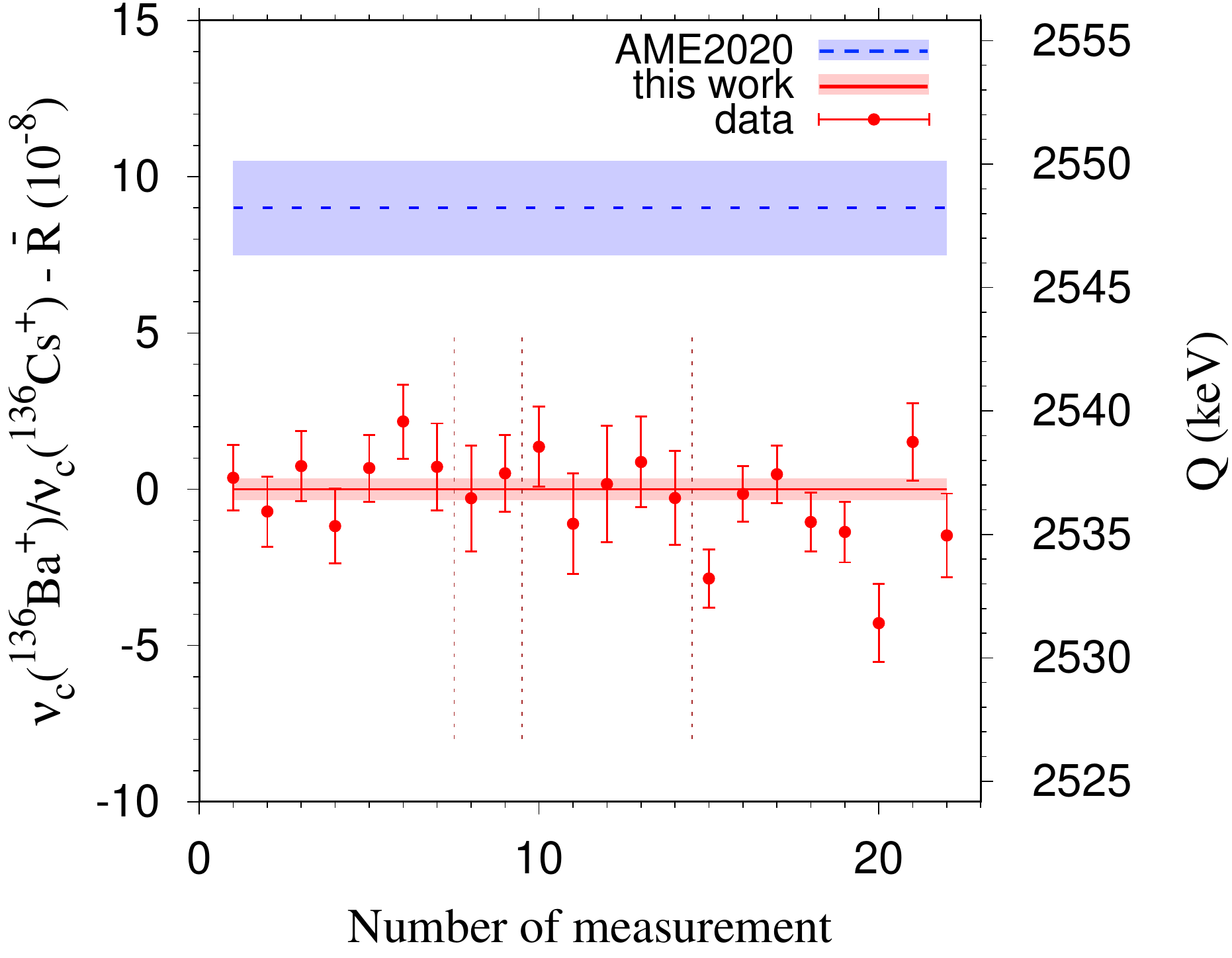}
   \caption{(Color online). The deviation  (left axis) of the individually measured cyclotron frequency ratios  $R$ ($\nu_c$($^{136}$Ba$^{+}$)/$\nu_c$($^{136}$Cs$^{+}$)) from the measured value $\overline{R}$ and (right axis)) $Q$ values in this work compared to value adopted from AME2020~\cite{Huang2021,Wang2021}. The red points with uncertainties are measured individual data collected in 4 different time slots, which are separated with vertical brown dashed lines. The weighted average value from this work  $\overline{R}$ = 1.000 020 039 1(35) is illustrated by the horizontal solid red line with its 1$\sigma$ uncertainty band. The dashed blue line is the value in AME2020 with its 1$\sigma$  uncertainty area shaded in blue.}
   \label{fig:ratio}
\end{figure}

In total, 13.5 hours of PI-ICR measurement data with two different accumulation times were recorded. The full sequence, consisting of measurement of magnetron phase, cyclotron phase and center spots required about 3 minutes to complete. This was sequentially repeated for both ion species $^{136}$Cs$^{+}$  and $^{136}$Ba$^{+}$. In the analysis, the position of each spot was fitted with the maximum-likelihood method. A few rounds were summed to have a reasonable number of detected ions for fitting. The phase angles were calculated accordingly based on the determined positions of the phases to deduce the $\nu_c$ frequency of each ion species. 
The $\nu_{c}$ of the daughter $^{136}$Ba$^{+}$ as a reference was linearly interpolated to the time of the measurement of the parent $^{136}$Cs$^{+}$ to deduce the cyclotron frequency ratio $R$. Ion bunches containing no more than five detected ions were considered in the data analysis in order to reduce a possible cyclotron frequency shift due to ion-ion interactions~\cite{Kellerbauer2003,Roux2013}. The count-rate related frequency shifts were not observed in the analysis. 
The temporal fluctuation of the magnetic field has been measured to be $\delta_B(\nu_{c})/\nu_{c}=  \Delta t \times 2.01(25)  \times  10^{-12}$/min~\cite{nesterenko2021}, where $\Delta t$ is the time interval between two consecutive reference measurements.
Contribution of  temporal fluctuations of the magnetic field to the final frequency ratio uncertainty was less than 10$^{-10}$. 
The frequency shifts in the PI-ICR measurement due to ion image distortions, which were well below the statistical uncertainty, were ignored in the calculation of the final uncertainty.  
The weighted mean ratio $\overline{R}$ of all single ratios was calculated along with the inner and outer errors to deduce the Birge rato~\cite{Birge1932}.  The maximum of the inner and outer errors was taken as the weight to calculate $\overline{R}$. The determination of $Q_{\beta^-}$ from $\overline{R}$ depends on the measured cyclotron  frequency $\nu_{c}$ via Eq.~\ref{eq:Qec}. 
In Fig.~\ref{fig:ratio}, results of the analysis including all data with comparison to literature values are demonstrated. The final frequency ratio $\overline{R}$ with its  uncertainty as well as the corresponding $Q$ value are $\overline{R}$ =  1.000 020 039 1(35) and Q$_{\beta^-}$ = 2536.83(45) keV, respectively.

\begin{figure}[!htb]
   \includegraphics[width=0.85\columnwidth]{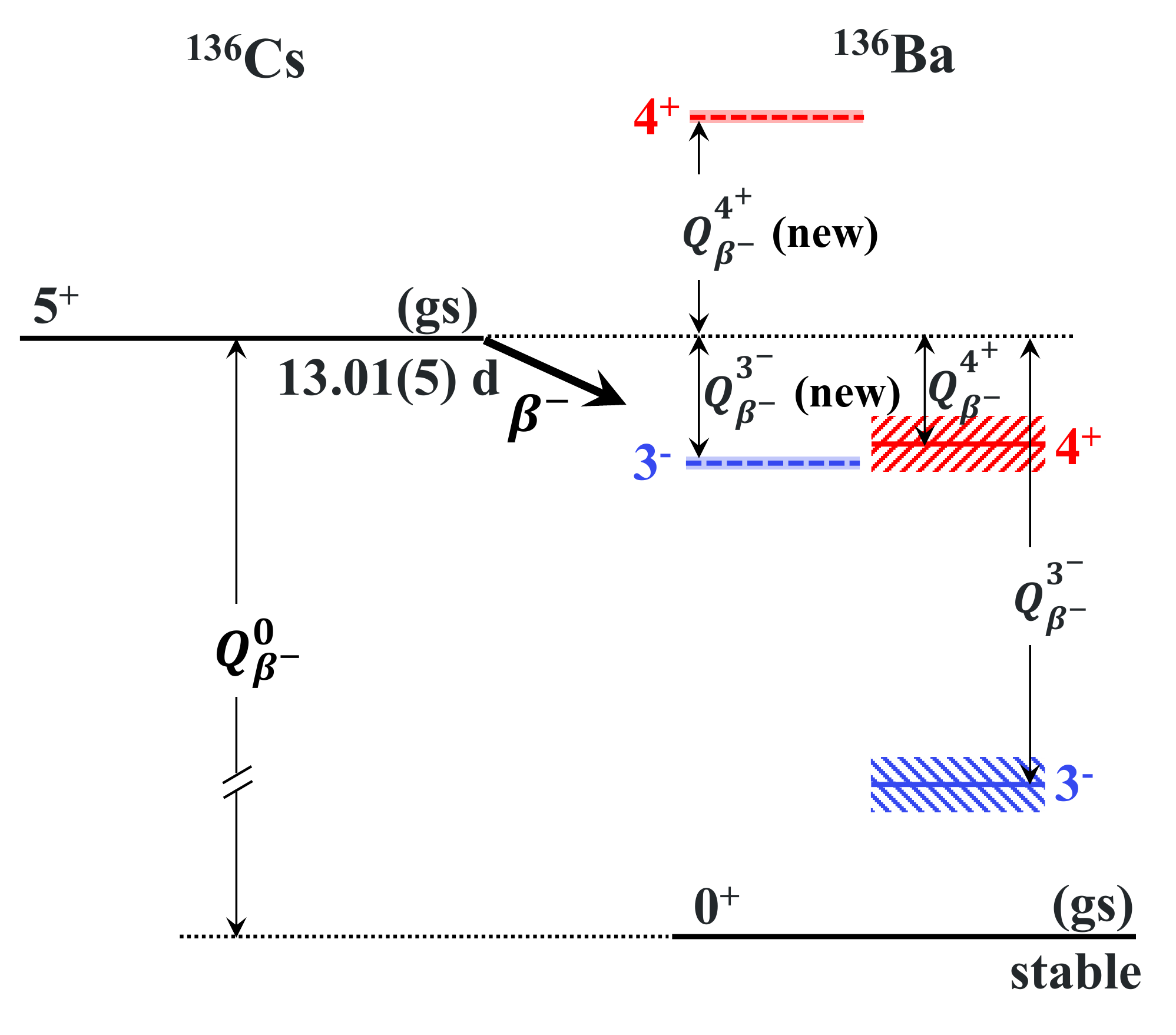}
   \caption{(Color online). Partial decay diagram for the $^{136}$Cs ground state to ground state and possible ultra-low $Q$-value excited states of $4^+$ and $3^-$ in $^{136}$Ba using $Q$ values from AME2020~\cite{Huang2021,Wang2021} in comparison to this work. 
   The levels drawn with solid lines show the the excited states with the $Q$ values from AME2020 and dashed lines from the refined  $Q$ values in this work (new). 
   The hatched areas (in blue for the $3^-$ and in red for the $4^+$ state) illustrate the corresponding 1$\sigma$ uncertainty in the $Q$ values. 
Table~\ref{table:low-Q} lists the $Q$ values in detail.
   }
   \label{fig:level-scheme}
\end{figure}

A comparison of our results with the literature values is tabulated in  Table~\ref{table:Q-value}. The mass-excess of the parent nucleus  $^{136}$Cs (5$^{+}$) was deduced to be  -86350.09(54) keV.
The gs-to-gs $Q$ value ($Q^0_{\beta^-}$), determined to be 2536.83(45) keV from this work, is  $\sim$4 times more precise than that derived from the evaluated masses in AME2020~\cite{Huang2021,Wang2021}. The new  $Q^0_{\beta^-}$ value has a deviation of -11.4(20) keV from the AME2020 value and is $\sim$6$\sigma$ smaller. 
The high-precision $\beta^-$ decay energy from this work, together with the nuclear energy level data from~\cite{NNDC} of the excited states of $^{136}$Ba as tabulated in Table~\ref{table:low-Q}, was used to determine gs-to-es $Q$ value ($Q^*_{\beta^-}$)  of these two states, see Fig~\ref {fig:level-scheme}. The calculated $Q$ values of potential candidate transitions of the  ground state of parent  nuclei $^{136}$Cs  to the excited states of daughter  $^{136}$Ba are tabulated in Table~\ref{table:low-Q}. Our results confirm that the decay of the ground-state of $^{136}$Cs to the 4$^{+}$ excited state in $^{136}$Ba with an excitation energy of 2544.481(24) keV is energetically forbidden. The $Q_{\beta^-}$ value is  negative with $\sim 17\sigma$ confidence. The decay channel to the 3$^{-}$ excited state at 2532.653(23) keV, having a refined $Q$ value of 4.18(45) keV, is energetically allowed and serves as a possible low $Q$-value transition to be used for neutrino-mass determination. 
The unexpectedly large deviation of the $Q^0_{\beta^-}$, which lowers the gs-to-es $Q$ value of 15.5(19) keV by more than 10 keV for the excited state of 2532.653(23) keV, makes the decay to this state of considerable interest.

\begin{figure}[!htb]
   \includegraphics[width=0.9\columnwidth]{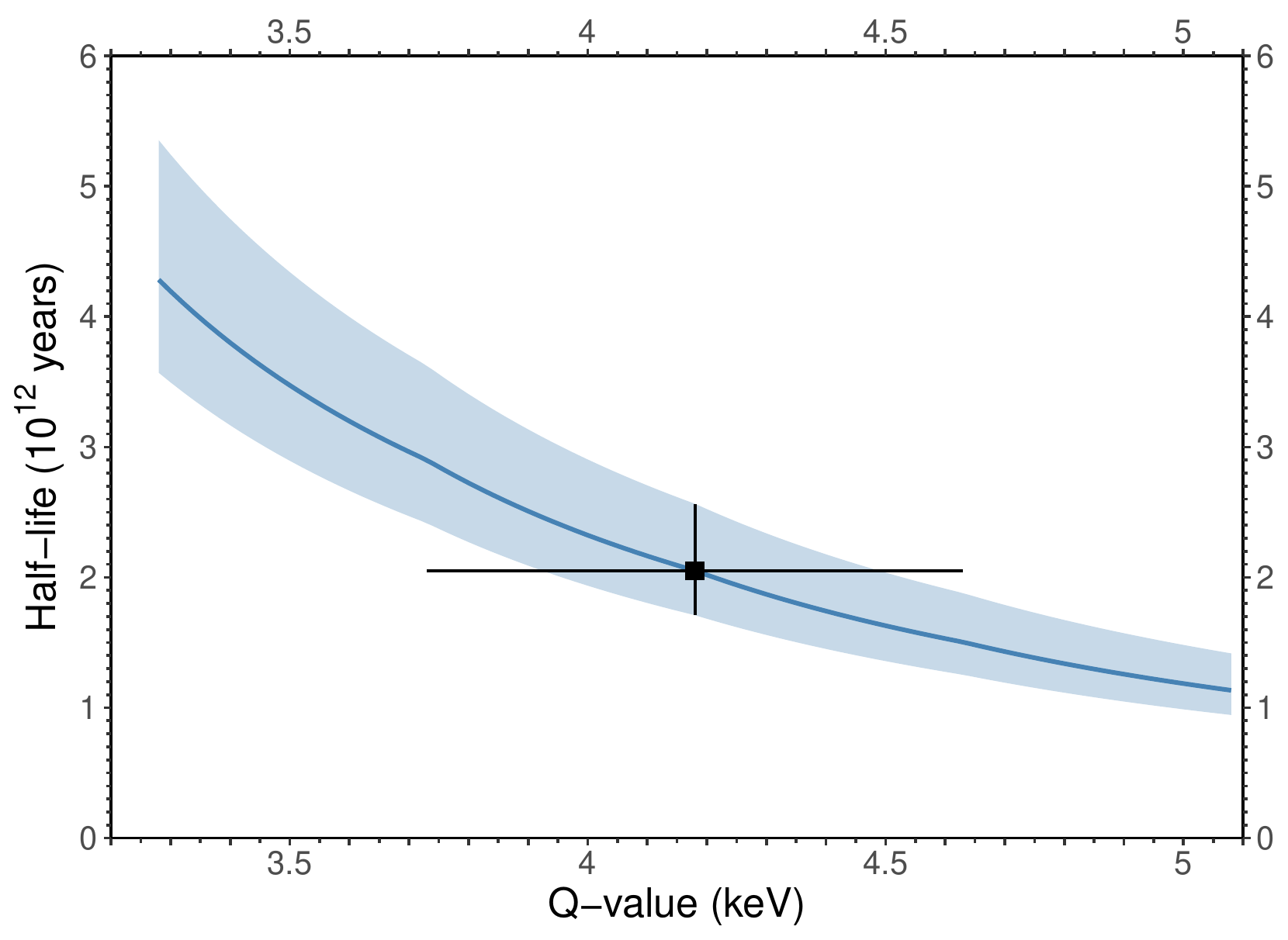}
   \caption{(Color online). Theoretical estimate for the partial half-life of the first-forbidden unique transition $^{136}{\rm Cs}(5^+_{\rm g.s.})\rightarrow\,^{136}{\rm Ba}(3^-)$ with a $Q$-value of 4.18(45) keV. The shaded area represents a 1$\sigma$ uncertainty for a given $Q$ value, while the horizontal error bar represents the 1$\sigma$ uncertainty of the $Q$ value, and the vertical error bar the 1$\sigma$ nuclear structure uncertainty for the best estimate of the $Q$ value.  
   }
   \label{fig:theor_hl}
\end{figure}

The partial half-life of the transition, which is of first-forbidden unique type,
can be estimated with a microscopic nuclear model. It depends on the $Q$ value through a phase-space factor and on nuclear structure through the involved nuclear matrix element (NME). The relevant NME was calculated using the nuclear shell model in the full $0g_{9/2}-1d-2s-0h_{11/2}$ model space using the effective interaction SN100PN \cite{SN100PN}. The calculation was carried out using the shell-model code NUSHELLX@MSU \cite{Brown2014}. To account for the well-known problem of the shell model, underestimation of the half-lives of beta-decay transitions \cite{Suhonen2017}, we adopt an effective value of the axial-vector coupling constant $g_{\rm A}^{\rm eff}$=1, while the 1$\sigma$ uncertainties related to the shell-model calculation are estimated by varying $g_{\rm A}^{\rm eff}$ between 0.8 and 1.2 (see, e.g., \cite{Suhonen2017}). The phase-space factor was calculated using exact Dirac electron wave functions with finite nuclear size and electron screening as was previously done for double $\beta$ decays \cite{Kotila2012} and allowed $\beta$ decay \cite{ERONEN2022}. The used formalism for calculating phase-space factors for first-forbidden unique transitions was adopted from \cite{GOVE1971205}. The resulting theoretical half-life estimate is $2.1_{-0.8}^{+1.6}\times10^{12}$\,yr. The half-life as a function of $Q$ value is presented in Fig.~\ref{fig:theor_hl}. The best estimate corresponds to a branching ratio of about $1.7\times 10^{-12}$\,\%.

As an isotope which undergoes double $\beta$ decay, $^{136}$Xe is particularly well-suited as a target for study of the charged-current (CC) neutrino capture process $\nu_e$+$^{136}$Xe(0$^{+}$)$\rightarrow$ $^{136}$Cs$^*$+$e^-$~\cite{Davis68,Raghavan97}. It features a low reaction threshold of $Q$ = 90.3(19) keV (mass difference from AME2020~\cite{Huang2021,Wang2021}) and a relatively large cross section due to the sizable Gamow-Teller transition strengths connecting the 0$^+$ $^{136}$Xe ground state and the lowest-lying 1$^+$ excited states of $^{136}$Cs.
The signal generated in the detector is the combination of the outgoing electron and any $\gamma$ rays or conversion electrons emitted as the Cs nucleus relaxes to its ground state. 
%
Recently, many new low-lying states in $^{136}$Cs have been identified, several of which are isomeric and potentially can be used in filtering events~\cite{Haselschwardt2023}.
As the reaction threshold $Q$ of $^{136}$Xe is low enough (lowest among all naturally occurring isotope of xenon), this reaction can be used to search for neutrinos from the solar carbon-nitrogen-oxygen (CNO) cycle~\cite{Newstead19,Appel22}, and can also provide a unique measurement of $^7$Be neutrinos, which may enable novel measurements of temperature of the solar core~\cite{Bahcall94}. With the mass excess of $^{136}$Cs from our measurements combined with the precise mass value of $^{136}$Xe measured at FSU Penning trap~\cite{Matthew07,Huang2021,Wang2021}, we refined the $Q$ value to be 79.1(5)~keV. This value is 11.2(19)~keV lower than the evaluated value from AME2020, which will increase the solar neutrino capture rates in the CC neutrino capture process. The same final state of $^{136}$Cs with a lower $Q$ value will indicate a higher sensitivity to search for CC absorption of MeV-scale fermionic dark matter on nuclei as well~\cite{Newstead19,Dror2020}.

The $\nu_e$+$^{136}$Xe (0$^{+}$)$\rightarrow$ $^{136}$Cs$^*$+$e^-$ neutrino capture process to the two lowest-lying $1^+$ states of $^{136}$Cs has been studied earlier in Ref. \cite{Haselschwardt20}. The wave functions of the initial and final states were computed in the nuclear shell model. Here we update the cross sections with the new $Q$ value for a set of neutrino energies relevant to solar $^7$Be, pep, and CNO neutrinos. The results are shown in Table \ref{table:136cccs}. The new lower threshold will result in event rates roughly two to four percent higher than the old threshold for the given final states and listed species of solar neutrinos. 

\begin{table}[!htb]
\caption{Total cross sections of the $\nu_e$+$^{136}$Xe (0$^{+}$)$\rightarrow$ $^{136}$Cs$^*$+$e^-$ neutrino capture process for the two lowest-lying $1^+$ final states of $^{136}$Cs (column 1) with discrete neutrino energies (column 2) in a range relevant to solar $^7$Be, pep, and CNO neutrinos. Results are shown for the new threshold 79.08 keV (column 3) and the old threshold 90 keV as used in Ref. \cite{Haselschwardt20} (column 4). The calculations were made in the nuclear shell model as described in Ref. \cite{Haselschwardt20}.} 
\label{table:136cccs}
\begin{ruledtabular}
    \begin{tabular}{cccc} 
          Final  & $E_\nu$  & NEW $\sigma_\mathrm{tot}$ (cm$^2$)  &  OLD $\sigma_\mathrm{tot}$  (cm$^2$)  \\
	state  & (MeV)  &   ($Q = 79.08$ keV)  & ($Q = 90 $ keV)  	\\
	\hline\noalign{\smallskip}
     $1^+_1$ & 0.7 & $ 1.16 \times 10^{-44} $  & $1.12 \times 10^{-44}$  \\
     (590 keV)&0.8 & $1.57 \times 10^{-44}  $  & $1.52\times 10^{-44}$ \\
	  &0.9 & $ 2.06 \times 10^{-44} $  & $ 2.00 \times 10^{-44} $ \\
		&1.0 & $ 2.60 \times 10^{-44} $  & $ 2.54 \times 10^{-44} $ \\
		&1.1 & $3.21 \times 10^{-44}$    &  $3.14 \times 10^{-44} $ \\
		&1.2 & $3.88 \times 10^{-44}$    &  $3.80 \times 10^{-44}$ \\ 
		&1.3 & $4.60 \times 10^{-44}$    &  $4.52\times 10^{-44} $ \\
		&1.4 & $5.38 \times 10^{-44}$    &  $5.29 \times 10^{-44}$\\
\\
    $1^+_2$ &1.0 & $7.57\times 10^{-45}$ & $7.32\times 10^{-45}$  \\
   (890 keV)&1.1 & $1.01\times 10^{-44}$ & $9.82\times 10^{-45}$  \\
			&1.2 & $1.30\times 10^{-44}$ & $1.27\times 10^{-44}$  \\
			&1.3 & $1.63\times 10^{-44}$  &$ 1.59\times 10^{-44}$ \\
			&1.4 & $1.99\times 10^{-44}$ & $1.95\times 10^{-44}$  \\
			&1.5 & $2.39\times 10^{-44}$ & $2.34\times 10^{-44}$ \\
			&1.6 & $2.81\times 10^{-44}$ & $2.76\times 10^{-44}$ \\
			&1.7 & $3.27\times 10^{-44}$ & $3.22\times 10^{-44}$  \\
    \end{tabular}
\end{ruledtabular}
\end{table}

\section{Conclusion}
A new scheme of preparing mono-isotopic samples of $^{136}$Cs and $^{136}$Ba, based on the  coupling of the Ramsey cleaning method and the PI-ICR technique to enhance the separation capability of JYFLTRAP, has been employed. 
A direct high-precision ground-state to ground-state  $\beta^-$ decay $Q$-value measurement of $^{136}$Cs  (5$^{+}$)$\rightarrow $$^{136}$Ba (0$^{+}$) was performed using the PI-ICR technique at the JYFLTRAP double Penning trap mass spectrometer. A $Q$ value of 2536.83(45)~keV  was obtained and its  precision  is improved by a factor of four.
A discrepancy of around 6 standard deviations is found compared to the adopted value in the AME2020.
We confirm that one of the two potential ultra-low $Q$-value $\beta^-$-decay transitions, $^{136}$Cs  (5$^{+}$)$\rightarrow $$^{136}$Ba$^{*}$ (4$^{+}$, 2544.481(24) keV), is energetically forbidden at the 17$\sigma$ level. 
This finding underlines the need to measure the $Q$ values to high precision before attempts to detect such possible low $Q$-value decay branches is made with the goal to realize these decays for neutrino mass determination. While the negative $Q$ values exclude the use of this transition to study neutrino mass, the small negative $Q$ values could make it a candidate for the study of $\beta$-$\gamma$ detour transitions proceeding via virtual states. Moreover, we verify that another transition,  $^{136}$Cs  (5$^{+}$)$\rightarrow $$^{136}$Ba$^{*}$ (3$^{-}$, 2532.653(23) keV), as a first-forbidden unique transition with a simple universal spectral shape, is positively allowed at a level of 9$\sigma$ with a small low $Q$ value and thus is a possible candidate for future neutrino mass determination experiment. The refined mass difference of ground states of $^{136}$Xe and $^{136}$Cs indicates a higher sensitivity of $^{136}$Xe as a target for study of charged-current (CC) neutrino capture processes.

\acknowledgments 

We acknowledge the staff of the Accelerator Laboratory of University of Jyv\"askyl\"a (JYFL-ACCLAB) for providing stable online beam. We thank the support by the Academy of Finland under the Finnish Centre of Excellence Programme 2012-2017 (Nuclear and Accelerator Based Physics Research at JYFL) and projects No. 306980, 312544, 275389, 284516, 295207, 314733, 315179, 327629, 320062 and 345869. The support by the EU Horizon 2020 research and innovation program under grant No. 771036 (ERC CoG MAIDEN) is acknowledged.  This project has received funding from the European Union’s Horizon 2020 research and innovation programme under grant agreement No. 861198–LISA–H2020-MSCA-ITN-2019.

\bibliography{my-final-bib-from-jabref}

\begin{thebibliography}{61}%
\makeatletter
\providecommand \@ifxundefined [1]{%
 \@ifx{#1\undefined}
}%
\providecommand \@ifnum [1]{%
 \ifnum #1\expandafter \@firstoftwo
 \else \expandafter \@secondoftwo
 \fi
}%
\providecommand \@ifx [1]{%
 \ifx #1\expandafter \@firstoftwo
 \else \expandafter \@secondoftwo
 \fi
}%
\providecommand \natexlab [1]{#1}%
\providecommand \enquote  [1]{``#1''}%
\providecommand \bibnamefont  [1]{#1}%
\providecommand \bibfnamefont [1]{#1}%
\providecommand \citenamefont [1]{#1}%
\providecommand \href@noop [0]{\@secondoftwo}%
\providecommand \href [0]{\begingroup \@sanitize@url \@href}%
\providecommand \@href[1]{\@@startlink{#1}\@@href}%
\providecommand \@@href[1]{\endgroup#1\@@endlink}%
\providecommand \@sanitize@url [0]{\catcode `\\12\catcode `\$12\catcode
  `\&12\catcode `\#12\catcode `\^12\catcode `\_12\catcode `\%12\relax}%
\providecommand \@@startlink[1]{}%
\providecommand \@@endlink[0]{}%
\providecommand \url  [0]{\begingroup\@sanitize@url \@url }%
\providecommand \@url [1]{\endgroup\@href {#1}{\urlprefix }}%
\providecommand \urlprefix  [0]{URL }%
\providecommand \Eprint [0]{\href }%
\providecommand \doibase [0]{http://dx.doi.org/}%
\providecommand \selectlanguage [0]{\@gobble}%
\providecommand \bibinfo  [0]{\@secondoftwo}%
\providecommand \bibfield  [0]{\@secondoftwo}%
\providecommand \translation [1]{[#1]}%
\providecommand \BibitemOpen [0]{}%
\providecommand \bibitemStop [0]{}%
\providecommand \bibitemNoStop [0]{.\EOS\space}%
\providecommand \EOS [0]{\spacefactor3000\relax}%
\providecommand \BibitemShut  [1]{\csname bibitem#1\endcsname}%
\let\auto@bib@innerbib\@empty
\bibitem [{\citenamefont {Fukuda}\ \emph {et~al.}(1998)\citenamefont {Fukuda},
  \citenamefont {Hayakawa}, \citenamefont {Ichihara}, \citenamefont {Inoue},
  \citenamefont {Ishihara}, \citenamefont {Ishino}, \citenamefont {Itow},
  \citenamefont {Kajita}, \citenamefont {Kameda}, \citenamefont {Kasuga},
  \citenamefont {Kobayashi}, \citenamefont {Kobayashi}, \citenamefont {Koshio},
  \citenamefont {Miura}, \citenamefont {Nakahata}, \citenamefont {Nakayama},
  \citenamefont {Okada}, \citenamefont {Okumura}, \citenamefont {Sakurai},
  \citenamefont {Shiozawa}, \citenamefont {Suzuki}, \citenamefont {Takeuchi},
  \citenamefont {Totsuka}, \citenamefont {Yamada}, \citenamefont {Earl},
  \citenamefont {Habig}, \citenamefont {Kearns}, \citenamefont {Messier},
  \citenamefont {Scholberg}, \citenamefont {Stone}, \citenamefont {Sulak},
  \citenamefont {Walter}, \citenamefont {Goldhaber}, \citenamefont
  {Barszczxak}, \citenamefont {Casper}, \citenamefont {Gajewski}, \citenamefont
  {Halverson}, \citenamefont {Hsu}, \citenamefont {Kropp}, \citenamefont
  {Price}, \citenamefont {Reines}, \citenamefont {Smy}, \citenamefont {Sobel},
  \citenamefont {Vagins}, \citenamefont {Ganezer}, \citenamefont {Keig},
  \citenamefont {Ellsworth}, \citenamefont {Tasaka}, \citenamefont {Flanagan},
  \citenamefont {Kibayashi}, \citenamefont {Learned}, \citenamefont {Matsuno},
  \citenamefont {Stenger}, \citenamefont {Takemori}, \citenamefont {Ishii},
  \citenamefont {Kanzaki}, \citenamefont {Kobayashi}, \citenamefont {Mine},
  \citenamefont {Nakamura}, \citenamefont {Nishikawa}, \citenamefont {Oyama},
  \citenamefont {Sakai}, \citenamefont {Sakuda}, \citenamefont {Sasaki},
  \citenamefont {Echigo}, \citenamefont {Kohama}, \citenamefont {Suzuki},
  \citenamefont {Haines}, \citenamefont {Blaufuss}, \citenamefont {Kim},
  \citenamefont {Sanford}, \citenamefont {Svoboda}, \citenamefont {Chen},
  \citenamefont {Conner}, \citenamefont {Goodman}, \citenamefont {Sullivan},
  \citenamefont {Hill}, \citenamefont {Jung}, \citenamefont {Martens},
  \citenamefont {Mauger}, \citenamefont {{Mc Grew}}, \citenamefont {Sharkey},
  \citenamefont {Viren}, \citenamefont {Yanagisawa}, \citenamefont {Doki},
  \citenamefont {Miyano}, \citenamefont {Okazawa}, \citenamefont {Saji},
  \citenamefont {Takahata}, \citenamefont {Nagashima}, \citenamefont {Takita},
  \citenamefont {Yamaguchi}, \citenamefont {Yoshida}, \citenamefont {Kim},
  \citenamefont {Etoh}, \citenamefont {Fujita}, \citenamefont {Hasegawa},
  \citenamefont {Hasegawa}, \citenamefont {Hatakeyama}, \citenamefont
  {Iwamoto}, \citenamefont {Koga}, \citenamefont {Maruyama}, \citenamefont
  {Ogawa}, \citenamefont {Shirai}, \citenamefont {Suzuki}, \citenamefont
  {Tsushima}, \citenamefont {Koshiba}, \citenamefont {Nemoto}, \citenamefont
  {Nishijima}, \citenamefont {Futagami}, \citenamefont {Hayato}, \citenamefont
  {Kanaya}, \citenamefont {Kaneyuki}, \citenamefont {Watanabe}, \citenamefont
  {Kielczewska}, \citenamefont {Doyle}, \citenamefont {George}, \citenamefont
  {Stachyra}, \citenamefont {Wai}, \citenamefont {Wilkes},\ and\ \citenamefont
  {Young}}]{Fukuda1998}%
  \BibitemOpen
  \bibfield  {author} {\bibinfo {author} {\bibfnamefont {Y.}~\bibnamefont
  {Fukuda}}, \bibinfo {author} {\bibfnamefont {T.}~\bibnamefont {Hayakawa}},
  \bibinfo {author} {\bibfnamefont {E.}~\bibnamefont {Ichihara}}, \bibinfo
  {author} {\bibfnamefont {K.}~\bibnamefont {Inoue}}, \bibinfo {author}
  {\bibfnamefont {K.}~\bibnamefont {Ishihara}}, \bibinfo {author}
  {\bibfnamefont {H.}~\bibnamefont {Ishino}}, \bibinfo {author} {\bibfnamefont
  {Y.}~\bibnamefont {Itow}}, \bibinfo {author} {\bibfnamefont {T.}~\bibnamefont
  {Kajita}}, \bibinfo {author} {\bibfnamefont {J.}~\bibnamefont {Kameda}},
  \bibinfo {author} {\bibfnamefont {S.}~\bibnamefont {Kasuga}}, \bibinfo
  {author} {\bibfnamefont {K.}~\bibnamefont {Kobayashi}}, \bibinfo {author}
  {\bibfnamefont {Y.}~\bibnamefont {Kobayashi}}, \bibinfo {author}
  {\bibfnamefont {Y.}~\bibnamefont {Koshio}}, \bibinfo {author} {\bibfnamefont
  {M.}~\bibnamefont {Miura}}, \bibinfo {author} {\bibfnamefont
  {M.}~\bibnamefont {Nakahata}}, \bibinfo {author} {\bibfnamefont
  {S.}~\bibnamefont {Nakayama}}, \bibinfo {author} {\bibfnamefont
  {A.}~\bibnamefont {Okada}}, \bibinfo {author} {\bibfnamefont
  {K.}~\bibnamefont {Okumura}}, \bibinfo {author} {\bibfnamefont
  {N.}~\bibnamefont {Sakurai}}, \bibinfo {author} {\bibfnamefont
  {M.}~\bibnamefont {Shiozawa}}, \bibinfo {author} {\bibfnamefont
  {Y.}~\bibnamefont {Suzuki}}, \bibinfo {author} {\bibfnamefont
  {Y.}~\bibnamefont {Takeuchi}}, \bibinfo {author} {\bibfnamefont
  {Y.}~\bibnamefont {Totsuka}}, \bibinfo {author} {\bibfnamefont
  {S.}~\bibnamefont {Yamada}}, \bibinfo {author} {\bibfnamefont
  {M.}~\bibnamefont {Earl}}, \bibinfo {author} {\bibfnamefont {A.}~\bibnamefont
  {Habig}}, \bibinfo {author} {\bibfnamefont {E.}~\bibnamefont {Kearns}},
  \bibinfo {author} {\bibfnamefont {M.~D.}\ \bibnamefont {Messier}}, \bibinfo
  {author} {\bibfnamefont {K.}~\bibnamefont {Scholberg}}, \bibinfo {author}
  {\bibfnamefont {J.~L.}\ \bibnamefont {Stone}}, \bibinfo {author}
  {\bibfnamefont {L.~R.}\ \bibnamefont {Sulak}}, \bibinfo {author}
  {\bibfnamefont {C.~W.}\ \bibnamefont {Walter}}, \bibinfo {author}
  {\bibfnamefont {M.}~\bibnamefont {Goldhaber}}, \bibinfo {author}
  {\bibfnamefont {T.}~\bibnamefont {Barszczxak}}, \bibinfo {author}
  {\bibfnamefont {D.}~\bibnamefont {Casper}}, \bibinfo {author} {\bibfnamefont
  {W.}~\bibnamefont {Gajewski}}, \bibinfo {author} {\bibfnamefont {P.~G.}\
  \bibnamefont {Halverson}}, \bibinfo {author} {\bibfnamefont {J.}~\bibnamefont
  {Hsu}}, \bibinfo {author} {\bibfnamefont {W.~R.}\ \bibnamefont {Kropp}},
  \bibinfo {author} {\bibfnamefont {L.~R.}\ \bibnamefont {Price}}, \bibinfo
  {author} {\bibfnamefont {F.}~\bibnamefont {Reines}}, \bibinfo {author}
  {\bibfnamefont {M.}~\bibnamefont {Smy}}, \bibinfo {author} {\bibfnamefont
  {H.~W.}\ \bibnamefont {Sobel}}, \bibinfo {author} {\bibfnamefont {M.~R.}\
  \bibnamefont {Vagins}}, \bibinfo {author} {\bibfnamefont {K.~S.}\
  \bibnamefont {Ganezer}}, \bibinfo {author} {\bibfnamefont {W.~E.}\
  \bibnamefont {Keig}}, \bibinfo {author} {\bibfnamefont {R.~W.}\ \bibnamefont
  {Ellsworth}}, \bibinfo {author} {\bibfnamefont {S.}~\bibnamefont {Tasaka}},
  \bibinfo {author} {\bibfnamefont {J.~W.}\ \bibnamefont {Flanagan}}, \bibinfo
  {author} {\bibfnamefont {A.}~\bibnamefont {Kibayashi}}, \bibinfo {author}
  {\bibfnamefont {J.~G.}\ \bibnamefont {Learned}}, \bibinfo {author}
  {\bibfnamefont {S.}~\bibnamefont {Matsuno}}, \bibinfo {author} {\bibfnamefont
  {V.~J.}\ \bibnamefont {Stenger}}, \bibinfo {author} {\bibfnamefont
  {D.}~\bibnamefont {Takemori}}, \bibinfo {author} {\bibfnamefont
  {T.}~\bibnamefont {Ishii}}, \bibinfo {author} {\bibfnamefont
  {J.}~\bibnamefont {Kanzaki}}, \bibinfo {author} {\bibfnamefont
  {T.}~\bibnamefont {Kobayashi}}, \bibinfo {author} {\bibfnamefont
  {S.}~\bibnamefont {Mine}}, \bibinfo {author} {\bibfnamefont {K.}~\bibnamefont
  {Nakamura}}, \bibinfo {author} {\bibfnamefont {K.}~\bibnamefont {Nishikawa}},
  \bibinfo {author} {\bibfnamefont {Y.}~\bibnamefont {Oyama}}, \bibinfo
  {author} {\bibfnamefont {A.}~\bibnamefont {Sakai}}, \bibinfo {author}
  {\bibfnamefont {M.}~\bibnamefont {Sakuda}}, \bibinfo {author} {\bibfnamefont
  {O.}~\bibnamefont {Sasaki}}, \bibinfo {author} {\bibfnamefont
  {S.}~\bibnamefont {Echigo}}, \bibinfo {author} {\bibfnamefont
  {M.}~\bibnamefont {Kohama}}, \bibinfo {author} {\bibfnamefont {A.~T.}\
  \bibnamefont {Suzuki}}, \bibinfo {author} {\bibfnamefont {T.~J.}\
  \bibnamefont {Haines}}, \bibinfo {author} {\bibfnamefont {E.}~\bibnamefont
  {Blaufuss}}, \bibinfo {author} {\bibfnamefont {B.~K.}\ \bibnamefont {Kim}},
  \bibinfo {author} {\bibfnamefont {R.}~\bibnamefont {Sanford}}, \bibinfo
  {author} {\bibfnamefont {R.}~\bibnamefont {Svoboda}}, \bibinfo {author}
  {\bibfnamefont {M.~L.}\ \bibnamefont {Chen}}, \bibinfo {author}
  {\bibfnamefont {Z.}~\bibnamefont {Conner}}, \bibinfo {author} {\bibfnamefont
  {J.~A.}\ \bibnamefont {Goodman}}, \bibinfo {author} {\bibfnamefont {G.~W.}\
  \bibnamefont {Sullivan}}, \bibinfo {author} {\bibfnamefont {J.}~\bibnamefont
  {Hill}}, \bibinfo {author} {\bibfnamefont {C.~K.}\ \bibnamefont {Jung}},
  \bibinfo {author} {\bibfnamefont {K.}~\bibnamefont {Martens}}, \bibinfo
  {author} {\bibfnamefont {C.}~\bibnamefont {Mauger}}, \bibinfo {author}
  {\bibfnamefont {C.}~\bibnamefont {{Mc Grew}}}, \bibinfo {author}
  {\bibfnamefont {E.}~\bibnamefont {Sharkey}}, \bibinfo {author} {\bibfnamefont
  {B.}~\bibnamefont {Viren}}, \bibinfo {author} {\bibfnamefont
  {C.}~\bibnamefont {Yanagisawa}}, \bibinfo {author} {\bibfnamefont
  {W.}~\bibnamefont {Doki}}, \bibinfo {author} {\bibfnamefont {K.}~\bibnamefont
  {Miyano}}, \bibinfo {author} {\bibfnamefont {H.}~\bibnamefont {Okazawa}},
  \bibinfo {author} {\bibfnamefont {C.}~\bibnamefont {Saji}}, \bibinfo {author}
  {\bibfnamefont {M.}~\bibnamefont {Takahata}}, \bibinfo {author}
  {\bibfnamefont {Y.}~\bibnamefont {Nagashima}}, \bibinfo {author}
  {\bibfnamefont {M.}~\bibnamefont {Takita}}, \bibinfo {author} {\bibfnamefont
  {T.}~\bibnamefont {Yamaguchi}}, \bibinfo {author} {\bibfnamefont
  {M.}~\bibnamefont {Yoshida}}, \bibinfo {author} {\bibfnamefont {S.~B.}\
  \bibnamefont {Kim}}, \bibinfo {author} {\bibfnamefont {M.}~\bibnamefont
  {Etoh}}, \bibinfo {author} {\bibfnamefont {K.}~\bibnamefont {Fujita}},
  \bibinfo {author} {\bibfnamefont {A.}~\bibnamefont {Hasegawa}}, \bibinfo
  {author} {\bibfnamefont {T.}~\bibnamefont {Hasegawa}}, \bibinfo {author}
  {\bibfnamefont {S.}~\bibnamefont {Hatakeyama}}, \bibinfo {author}
  {\bibfnamefont {T.}~\bibnamefont {Iwamoto}}, \bibinfo {author} {\bibfnamefont
  {M.}~\bibnamefont {Koga}}, \bibinfo {author} {\bibfnamefont {T.}~\bibnamefont
  {Maruyama}}, \bibinfo {author} {\bibfnamefont {H.}~\bibnamefont {Ogawa}},
  \bibinfo {author} {\bibfnamefont {J.}~\bibnamefont {Shirai}}, \bibinfo
  {author} {\bibfnamefont {A.}~\bibnamefont {Suzuki}}, \bibinfo {author}
  {\bibfnamefont {F.}~\bibnamefont {Tsushima}}, \bibinfo {author}
  {\bibfnamefont {M.}~\bibnamefont {Koshiba}}, \bibinfo {author} {\bibfnamefont
  {M.}~\bibnamefont {Nemoto}}, \bibinfo {author} {\bibfnamefont
  {K.}~\bibnamefont {Nishijima}}, \bibinfo {author} {\bibfnamefont
  {T.}~\bibnamefont {Futagami}}, \bibinfo {author} {\bibfnamefont
  {Y.}~\bibnamefont {Hayato}}, \bibinfo {author} {\bibfnamefont
  {Y.}~\bibnamefont {Kanaya}}, \bibinfo {author} {\bibfnamefont
  {K.}~\bibnamefont {Kaneyuki}}, \bibinfo {author} {\bibfnamefont
  {Y.}~\bibnamefont {Watanabe}}, \bibinfo {author} {\bibfnamefont
  {D.}~\bibnamefont {Kielczewska}}, \bibinfo {author} {\bibfnamefont {R.~A.}\
  \bibnamefont {Doyle}}, \bibinfo {author} {\bibfnamefont {J.~S.}\ \bibnamefont
  {George}}, \bibinfo {author} {\bibfnamefont {A.~L.}\ \bibnamefont
  {Stachyra}}, \bibinfo {author} {\bibfnamefont {L.~L.}\ \bibnamefont {Wai}},
  \bibinfo {author} {\bibfnamefont {R.~J.}\ \bibnamefont {Wilkes}}, \ and\
  \bibinfo {author} {\bibfnamefont {K.~K.}\ \bibnamefont {Young}},\ }\href
  {\doibase 10.1103/PhysRevLett.81.1562} {\bibfield  {journal} {\bibinfo
  {journal} {Physical Review Letters}\ }\textbf {\bibinfo {volume} {81}},\
  \bibinfo {pages} {1562} (\bibinfo {year} {1998})},\ \Eprint
  {http://arxiv.org/abs/9807003} {arXiv:9807003 [hep-ex]} \BibitemShut
  {NoStop}%
\bibitem [{\citenamefont {{SNO Collaboration}}(2002)}]{SNOCollaboration2002}%
  \BibitemOpen
  \bibfield  {author} {\bibinfo {author} {\bibnamefont {{SNO Collaboration}}},\
  }\href {\doibase 10.1103/PhysRevLett.89.011301} {\bibfield  {journal}
  {\bibinfo  {journal} {Physical Review Letters}\ }\textbf {\bibinfo {volume}
  {89}},\ \bibinfo {pages} {1} (\bibinfo {year} {2002})},\ \Eprint
  {http://arxiv.org/abs/0204008} {arXiv:0204008 [nucl-ex]} \BibitemShut
  {NoStop}%
\bibitem [{\citenamefont {Gerbino}\ and\ \citenamefont
  {Lattanzi}(2018)}]{Gerbino2018a}%
  \BibitemOpen
  \bibfield  {author} {\bibinfo {author} {\bibfnamefont {M.}~\bibnamefont
  {Gerbino}}\ and\ \bibinfo {author} {\bibfnamefont {M.}~\bibnamefont
  {Lattanzi}},\ }\href {\doibase 10.3389/fphy.2017.00070} {\bibfield  {journal}
  {\bibinfo  {journal} {Frontiers in Physics}\ }\textbf {\bibinfo {volume} {5}}
  (\bibinfo {year} {2018}),\ 10.3389/fphy.2017.00070}\BibitemShut {NoStop}%
\bibitem [{\citenamefont {Suhonen}\ and\ \citenamefont
  {Civitarese}(1998)}]{Suhonen1998}%
  \BibitemOpen
  \bibfield  {author} {\bibinfo {author} {\bibfnamefont {J.}~\bibnamefont
  {Suhonen}}\ and\ \bibinfo {author} {\bibfnamefont {O.}~\bibnamefont
  {Civitarese}},\ }\href {\doibase 10.1016/S0370-1573(97)00087-2} {\bibfield
  {journal} {\bibinfo  {journal} {Physics Reports}\ }\textbf {\bibinfo {volume}
  {300}},\ \bibinfo {pages} {123} (\bibinfo {year} {1998})}\BibitemShut
  {NoStop}%
\bibitem [{\citenamefont {Avignone}\ \emph {et~al.}(2008)\citenamefont
  {Avignone}, \citenamefont {Elliott},\ and\ \citenamefont
  {Engel}}]{Avignone2008}%
  \BibitemOpen
  \bibfield  {author} {\bibinfo {author} {\bibfnamefont {F.~T.}\ \bibnamefont
  {Avignone}}, \bibinfo {author} {\bibfnamefont {S.~R.}\ \bibnamefont
  {Elliott}}, \ and\ \bibinfo {author} {\bibfnamefont {J.}~\bibnamefont
  {Engel}},\ }\href {\doibase 10.1103/RevModPhys.80.481} {\bibfield  {journal}
  {\bibinfo  {journal} {Reviews of Modern Physics}\ }\textbf {\bibinfo {volume}
  {80}},\ \bibinfo {pages} {481} (\bibinfo {year} {2008})},\ \Eprint
  {http://arxiv.org/abs/0708.1033} {arXiv:0708.1033} \BibitemShut {NoStop}%
\bibitem [{\citenamefont {Ejiri}\ \emph {et~al.}(2019)\citenamefont {Ejiri},
  \citenamefont {Suhonen},\ and\ \citenamefont {Zuber}}]{Ejiri2019}%
  \BibitemOpen
  \bibfield  {author} {\bibinfo {author} {\bibfnamefont {H.}~\bibnamefont
  {Ejiri}}, \bibinfo {author} {\bibfnamefont {J.}~\bibnamefont {Suhonen}}, \
  and\ \bibinfo {author} {\bibfnamefont {K.}~\bibnamefont {Zuber}},\ }\href
  {\doibase 10.1016/j.physrep.2018.12.001} {\bibfield  {journal} {\bibinfo
  {journal} {Physics Reports}\ }\textbf {\bibinfo {volume} {797}},\ \bibinfo
  {pages} {1} (\bibinfo {year} {2019})}\BibitemShut {NoStop}%
\bibitem [{\citenamefont {Aker}\ \emph {et~al.}(2022)\citenamefont {Aker},
  \citenamefont {Beglarian}, \citenamefont {Behrens}, \citenamefont {Berlev},
  \citenamefont {Besserer}, \citenamefont {Bieringer}, \citenamefont {Block},
  \citenamefont {Bornschein}, \citenamefont {Bornschein}, \citenamefont
  {Böttcher}, \citenamefont {Brunst}, \citenamefont {Caldwell}, \citenamefont
  {Carney}, \citenamefont {Cascio}, \citenamefont {Chilingaryan}, \citenamefont
  {Choi}, \citenamefont {Debowski}, \citenamefont {Deffert}, \citenamefont
  {Descher}, \citenamefont {Barrero}, \citenamefont {Doe}, \citenamefont
  {Dragoun}, \citenamefont {Drexlin}, \citenamefont {Eitel}, \citenamefont
  {Ellinger}, \citenamefont {Engel}, \citenamefont {Enomoto}, \citenamefont
  {Felden}, \citenamefont {Formaggio}, \citenamefont {Fränkle}, \citenamefont
  {Franklin}, \citenamefont {Friedel}, \citenamefont {Fulst}, \citenamefont
  {Gauda}, \citenamefont {Gil}, \citenamefont {Glück}, \citenamefont
  {Grössle}, \citenamefont {Gumbsheimer}, \citenamefont {Gupta}, \citenamefont
  {Höhn}, \citenamefont {Hannen}, \citenamefont {Haußmann}, \citenamefont
  {Helbing}, \citenamefont {Hickford}, \citenamefont {Hiller}, \citenamefont
  {Hillesheimer}, \citenamefont {Hinz}, \citenamefont {Houdy}, \citenamefont
  {Huber}, \citenamefont {Jansen}, \citenamefont {Karl}, \citenamefont
  {Kellerer}, \citenamefont {Kellerer}, \citenamefont {Klein}, \citenamefont
  {Köhler}, \citenamefont {Köllenberger}, \citenamefont {Kopmann},
  \citenamefont {Korzeczek}, \citenamefont {Koval\'ik}, \citenamefont {Krasch},
  \citenamefont {Krause}, \citenamefont {Kunka}, \citenamefont {Lasserre},
  \citenamefont {Le}, \citenamefont {Lebeda}, \citenamefont {Lehnert},
  \citenamefont {Lokhov}, \citenamefont {Machatschek}, \citenamefont
  {Malcherek}, \citenamefont {Mark}, \citenamefont {Marsteller}, \citenamefont
  {Martin}, \citenamefont {Melzer}, \citenamefont {Menshikov}, \citenamefont
  {Mertens}, \citenamefont {Mostafa}, \citenamefont {Müller}, \citenamefont
  {Niemes}, \citenamefont {Oelpmann}, \citenamefont {Parno}, \citenamefont
  {Poon}, \citenamefont {Poyato}, \citenamefont {Priester}, \citenamefont
  {Röllig}, \citenamefont {Röttele}, \citenamefont {Robertson}, \citenamefont
  {Rodejohann}, \citenamefont {Rodenbeck}, \citenamefont {Ryšavý},
  \citenamefont {Sack}, \citenamefont {Saenz}, \citenamefont {Schäfer},
  \citenamefont {Schaller}, \citenamefont {Schimpf}, \citenamefont
  {Schlösser}, \citenamefont {Schlösser}, \citenamefont {Schlüter},
  \citenamefont {Schneidewind}, \citenamefont {Schrank}, \citenamefont
  {Schulz}, \citenamefont {Schwemmer}, \citenamefont {Šef\^c\'ik},
  \citenamefont {Sibille}, \citenamefont {Siegmann}, \citenamefont {Slezák},
  \citenamefont {Steidl}, \citenamefont {Sturm}, \citenamefont {Sun},
  \citenamefont {Tcherniakhovski}, \citenamefont {Telle}, \citenamefont
  {Thorne}, \citenamefont {Thümmler}, \citenamefont {Titov}, \citenamefont
  {Tkachev}, \citenamefont {Urban}, \citenamefont {Valerius}, \citenamefont
  {Vénos}, \citenamefont {Hernández}, \citenamefont {Weinheimer},
  \citenamefont {Welte}, \citenamefont {Wendel}, \citenamefont {Wilkerson},
  \citenamefont {Wolf}, \citenamefont {Wüstling}, \citenamefont {Xu},
  \citenamefont {Yen}, \citenamefont {Zadoroghny},\ and\ \citenamefont
  {Zeller}}]{Aker2022}%
  \BibitemOpen
  \bibfield  {author} {\bibinfo {author} {\bibfnamefont {M.}~\bibnamefont
  {Aker}}, \bibinfo {author} {\bibfnamefont {A.}~\bibnamefont {Beglarian}},
  \bibinfo {author} {\bibfnamefont {J.}~\bibnamefont {Behrens}}, \bibinfo
  {author} {\bibfnamefont {A.}~\bibnamefont {Berlev}}, \bibinfo {author}
  {\bibfnamefont {U.}~\bibnamefont {Besserer}}, \bibinfo {author}
  {\bibfnamefont {B.}~\bibnamefont {Bieringer}}, \bibinfo {author}
  {\bibfnamefont {F.}~\bibnamefont {Block}}, \bibinfo {author} {\bibfnamefont
  {B.}~\bibnamefont {Bornschein}}, \bibinfo {author} {\bibfnamefont
  {L.}~\bibnamefont {Bornschein}}, \bibinfo {author} {\bibfnamefont
  {M.}~\bibnamefont {Böttcher}}, \bibinfo {author} {\bibfnamefont
  {T.}~\bibnamefont {Brunst}}, \bibinfo {author} {\bibfnamefont {T.~S.}\
  \bibnamefont {Caldwell}}, \bibinfo {author} {\bibfnamefont {R.~M.~D.}\
  \bibnamefont {Carney}}, \bibinfo {author} {\bibfnamefont {L.~L.}\
  \bibnamefont {Cascio}}, \bibinfo {author} {\bibfnamefont {S.}~\bibnamefont
  {Chilingaryan}}, \bibinfo {author} {\bibfnamefont {W.}~\bibnamefont {Choi}},
  \bibinfo {author} {\bibfnamefont {K.}~\bibnamefont {Debowski}}, \bibinfo
  {author} {\bibfnamefont {M.}~\bibnamefont {Deffert}}, \bibinfo {author}
  {\bibfnamefont {M.}~\bibnamefont {Descher}}, \bibinfo {author} {\bibfnamefont
  {D.~D.}\ \bibnamefont {Barrero}}, \bibinfo {author} {\bibfnamefont {P.~J.}\
  \bibnamefont {Doe}}, \bibinfo {author} {\bibfnamefont {O.}~\bibnamefont
  {Dragoun}}, \bibinfo {author} {\bibfnamefont {G.}~\bibnamefont {Drexlin}},
  \bibinfo {author} {\bibfnamefont {K.}~\bibnamefont {Eitel}}, \bibinfo
  {author} {\bibfnamefont {E.}~\bibnamefont {Ellinger}}, \bibinfo {author}
  {\bibfnamefont {R.}~\bibnamefont {Engel}}, \bibinfo {author} {\bibfnamefont
  {S.}~\bibnamefont {Enomoto}}, \bibinfo {author} {\bibfnamefont
  {A.}~\bibnamefont {Felden}}, \bibinfo {author} {\bibfnamefont {J.~A.}\
  \bibnamefont {Formaggio}}, \bibinfo {author} {\bibfnamefont {F.~M.}\
  \bibnamefont {Fränkle}}, \bibinfo {author} {\bibfnamefont {G.~B.}\
  \bibnamefont {Franklin}}, \bibinfo {author} {\bibfnamefont {F.}~\bibnamefont
  {Friedel}}, \bibinfo {author} {\bibfnamefont {A.}~\bibnamefont {Fulst}},
  \bibinfo {author} {\bibfnamefont {K.}~\bibnamefont {Gauda}}, \bibinfo
  {author} {\bibfnamefont {W.}~\bibnamefont {Gil}}, \bibinfo {author}
  {\bibfnamefont {F.}~\bibnamefont {Glück}}, \bibinfo {author} {\bibfnamefont
  {R.}~\bibnamefont {Grössle}}, \bibinfo {author} {\bibfnamefont
  {R.}~\bibnamefont {Gumbsheimer}}, \bibinfo {author} {\bibfnamefont
  {V.}~\bibnamefont {Gupta}}, \bibinfo {author} {\bibfnamefont
  {T.}~\bibnamefont {Höhn}}, \bibinfo {author} {\bibfnamefont
  {V.}~\bibnamefont {Hannen}}, \bibinfo {author} {\bibfnamefont
  {N.}~\bibnamefont {Haußmann}}, \bibinfo {author} {\bibfnamefont
  {K.}~\bibnamefont {Helbing}}, \bibinfo {author} {\bibfnamefont
  {S.}~\bibnamefont {Hickford}}, \bibinfo {author} {\bibfnamefont
  {R.}~\bibnamefont {Hiller}}, \bibinfo {author} {\bibfnamefont
  {D.}~\bibnamefont {Hillesheimer}}, \bibinfo {author} {\bibfnamefont
  {D.}~\bibnamefont {Hinz}}, \bibinfo {author} {\bibfnamefont {T.}~\bibnamefont
  {Houdy}}, \bibinfo {author} {\bibfnamefont {A.}~\bibnamefont {Huber}},
  \bibinfo {author} {\bibfnamefont {A.}~\bibnamefont {Jansen}}, \bibinfo
  {author} {\bibfnamefont {C.}~\bibnamefont {Karl}}, \bibinfo {author}
  {\bibfnamefont {F.}~\bibnamefont {Kellerer}}, \bibinfo {author}
  {\bibfnamefont {J.}~\bibnamefont {Kellerer}}, \bibinfo {author}
  {\bibfnamefont {M.}~\bibnamefont {Klein}}, \bibinfo {author} {\bibfnamefont
  {C.}~\bibnamefont {Köhler}}, \bibinfo {author} {\bibfnamefont
  {L.}~\bibnamefont {Köllenberger}}, \bibinfo {author} {\bibfnamefont
  {A.}~\bibnamefont {Kopmann}}, \bibinfo {author} {\bibfnamefont
  {M.}~\bibnamefont {Korzeczek}}, \bibinfo {author} {\bibfnamefont
  {A.}~\bibnamefont {Koval\'ik}}, \bibinfo {author} {\bibfnamefont
  {B.}~\bibnamefont {Krasch}}, \bibinfo {author} {\bibfnamefont
  {H.}~\bibnamefont {Krause}}, \bibinfo {author} {\bibfnamefont
  {N.}~\bibnamefont {Kunka}}, \bibinfo {author} {\bibfnamefont
  {T.}~\bibnamefont {Lasserre}}, \bibinfo {author} {\bibfnamefont {T.~L.}\
  \bibnamefont {Le}}, \bibinfo {author} {\bibfnamefont {O.}~\bibnamefont
  {Lebeda}}, \bibinfo {author} {\bibfnamefont {B.}~\bibnamefont {Lehnert}},
  \bibinfo {author} {\bibfnamefont {A.}~\bibnamefont {Lokhov}}, \bibinfo
  {author} {\bibfnamefont {M.}~\bibnamefont {Machatschek}}, \bibinfo {author}
  {\bibfnamefont {E.}~\bibnamefont {Malcherek}}, \bibinfo {author}
  {\bibfnamefont {M.}~\bibnamefont {Mark}}, \bibinfo {author} {\bibfnamefont
  {A.}~\bibnamefont {Marsteller}}, \bibinfo {author} {\bibfnamefont {E.~L.}\
  \bibnamefont {Martin}}, \bibinfo {author} {\bibfnamefont {C.}~\bibnamefont
  {Melzer}}, \bibinfo {author} {\bibfnamefont {A.}~\bibnamefont {Menshikov}},
  \bibinfo {author} {\bibfnamefont {S.}~\bibnamefont {Mertens}}, \bibinfo
  {author} {\bibfnamefont {J.}~\bibnamefont {Mostafa}}, \bibinfo {author}
  {\bibfnamefont {K.}~\bibnamefont {Müller}}, \bibinfo {author} {\bibfnamefont
  {S.}~\bibnamefont {Niemes}}, \bibinfo {author} {\bibfnamefont
  {P.}~\bibnamefont {Oelpmann}}, \bibinfo {author} {\bibfnamefont {D.~S.}\
  \bibnamefont {Parno}}, \bibinfo {author} {\bibfnamefont {A.~W.~P.}\
  \bibnamefont {Poon}}, \bibinfo {author} {\bibfnamefont {J.~M.~L.}\
  \bibnamefont {Poyato}}, \bibinfo {author} {\bibfnamefont {F.}~\bibnamefont
  {Priester}}, \bibinfo {author} {\bibfnamefont {M.}~\bibnamefont {Röllig}},
  \bibinfo {author} {\bibfnamefont {C.}~\bibnamefont {Röttele}}, \bibinfo
  {author} {\bibfnamefont {R.~G.~H.}\ \bibnamefont {Robertson}}, \bibinfo
  {author} {\bibfnamefont {W.}~\bibnamefont {Rodejohann}}, \bibinfo {author}
  {\bibfnamefont {C.}~\bibnamefont {Rodenbeck}}, \bibinfo {author}
  {\bibfnamefont {M.}~\bibnamefont {Ryšavý}}, \bibinfo {author}
  {\bibfnamefont {R.}~\bibnamefont {Sack}}, \bibinfo {author} {\bibfnamefont
  {A.}~\bibnamefont {Saenz}}, \bibinfo {author} {\bibfnamefont
  {P.}~\bibnamefont {Schäfer}}, \bibinfo {author} {\bibfnamefont
  {A.}~\bibnamefont {Schaller}}, \bibinfo {author} {\bibfnamefont
  {L.}~\bibnamefont {Schimpf}}, \bibinfo {author} {\bibfnamefont
  {K.}~\bibnamefont {Schlösser}}, \bibinfo {author} {\bibfnamefont
  {M.}~\bibnamefont {Schlösser}}, \bibinfo {author} {\bibfnamefont
  {L.}~\bibnamefont {Schlüter}}, \bibinfo {author} {\bibfnamefont
  {S.}~\bibnamefont {Schneidewind}}, \bibinfo {author} {\bibfnamefont
  {M.}~\bibnamefont {Schrank}}, \bibinfo {author} {\bibfnamefont
  {B.}~\bibnamefont {Schulz}}, \bibinfo {author} {\bibfnamefont
  {A.}~\bibnamefont {Schwemmer}}, \bibinfo {author} {\bibfnamefont
  {M.}~\bibnamefont {Šef\^c\'ik}}, \bibinfo {author} {\bibfnamefont
  {V.}~\bibnamefont {Sibille}}, \bibinfo {author} {\bibfnamefont
  {D.}~\bibnamefont {Siegmann}}, \bibinfo {author} {\bibfnamefont
  {M.}~\bibnamefont {Slezák}}, \bibinfo {author} {\bibfnamefont
  {M.}~\bibnamefont {Steidl}}, \bibinfo {author} {\bibfnamefont
  {M.}~\bibnamefont {Sturm}}, \bibinfo {author} {\bibfnamefont
  {M.}~\bibnamefont {Sun}}, \bibinfo {author} {\bibfnamefont {D.}~\bibnamefont
  {Tcherniakhovski}}, \bibinfo {author} {\bibfnamefont {H.~H.}\ \bibnamefont
  {Telle}}, \bibinfo {author} {\bibfnamefont {L.~A.}\ \bibnamefont {Thorne}},
  \bibinfo {author} {\bibfnamefont {T.}~\bibnamefont {Thümmler}}, \bibinfo
  {author} {\bibfnamefont {N.}~\bibnamefont {Titov}}, \bibinfo {author}
  {\bibfnamefont {I.}~\bibnamefont {Tkachev}}, \bibinfo {author} {\bibfnamefont
  {K.}~\bibnamefont {Urban}}, \bibinfo {author} {\bibfnamefont
  {K.}~\bibnamefont {Valerius}}, \bibinfo {author} {\bibfnamefont
  {D.}~\bibnamefont {Vénos}}, \bibinfo {author} {\bibfnamefont {A.~P.~V.}\
  \bibnamefont {Hernández}}, \bibinfo {author} {\bibfnamefont
  {C.}~\bibnamefont {Weinheimer}}, \bibinfo {author} {\bibfnamefont
  {S.}~\bibnamefont {Welte}}, \bibinfo {author} {\bibfnamefont
  {J.}~\bibnamefont {Wendel}}, \bibinfo {author} {\bibfnamefont {J.~F.}\
  \bibnamefont {Wilkerson}}, \bibinfo {author} {\bibfnamefont {J.}~\bibnamefont
  {Wolf}}, \bibinfo {author} {\bibfnamefont {S.}~\bibnamefont {Wüstling}},
  \bibinfo {author} {\bibfnamefont {W.}~\bibnamefont {Xu}}, \bibinfo {author}
  {\bibfnamefont {Y.~R.}\ \bibnamefont {Yen}}, \bibinfo {author} {\bibfnamefont
  {S.}~\bibnamefont {Zadoroghny}}, \ and\ \bibinfo {author} {\bibfnamefont
  {G.}~\bibnamefont {Zeller}},\ }\href {\doibase
  https://10.1038/s41567-021-01463-1} {\bibfield  {journal} {\bibinfo
  {journal} {Nature Physics}\ }\textbf {\bibinfo {volume} {18}},\ \bibinfo
  {pages} {160} (\bibinfo {year} {2022})}\BibitemShut {NoStop}%
\bibitem [{\citenamefont {Velte}\ \emph {et~al.}(2019)\citenamefont {Velte},
  \citenamefont {Ahrens}, \citenamefont {Barth}, \citenamefont {Blaum},
  \citenamefont {Bra{\ss}}, \citenamefont {Door}, \citenamefont {Dorrer},
  \citenamefont {Düllmann}, \citenamefont {Eliseev}, \citenamefont {Enss},
  \citenamefont {Filianin}, \citenamefont {Fleischmann}, \citenamefont
  {Gastaldo}, \citenamefont {Goeggelmann}, \citenamefont {Goodacre},
  \citenamefont {Haverkort}, \citenamefont {Hengstler}, \citenamefont {Jochum},
  \citenamefont {Johnston}, \citenamefont {Keller}, \citenamefont {Kempf},
  \citenamefont {Kieck}, \citenamefont {König}, \citenamefont {Köster},
  \citenamefont {Kromer}, \citenamefont {Mantegazzini}, \citenamefont {Marsh},
  \citenamefont {Novikov}, \citenamefont {Piquemal}, \citenamefont {Riccio},
  \citenamefont {Richter}, \citenamefont {Rischka}, \citenamefont {Rothe},
  \citenamefont {Schüssler}, \citenamefont {Schweiger}, \citenamefont {Stora},
  \citenamefont {Wegner}, \citenamefont {Wendt}, \citenamefont {Zampaolo},\
  and\ \citenamefont {Zuber}}]{Velte2019}%
  \BibitemOpen
  \bibfield  {author} {\bibinfo {author} {\bibfnamefont {C.}~\bibnamefont
  {Velte}}, \bibinfo {author} {\bibfnamefont {F.}~\bibnamefont {Ahrens}},
  \bibinfo {author} {\bibfnamefont {A.}~\bibnamefont {Barth}}, \bibinfo
  {author} {\bibfnamefont {K.}~\bibnamefont {Blaum}}, \bibinfo {author}
  {\bibfnamefont {M.}~\bibnamefont {Bra{\ss}}}, \bibinfo {author}
  {\bibfnamefont {M.}~\bibnamefont {Door}}, \bibinfo {author} {\bibfnamefont
  {H.}~\bibnamefont {Dorrer}}, \bibinfo {author} {\bibfnamefont {C.~E.}\
  \bibnamefont {Düllmann}}, \bibinfo {author} {\bibfnamefont {S.}~\bibnamefont
  {Eliseev}}, \bibinfo {author} {\bibfnamefont {C.}~\bibnamefont {Enss}},
  \bibinfo {author} {\bibfnamefont {P.}~\bibnamefont {Filianin}}, \bibinfo
  {author} {\bibfnamefont {A.}~\bibnamefont {Fleischmann}}, \bibinfo {author}
  {\bibfnamefont {L.}~\bibnamefont {Gastaldo}}, \bibinfo {author}
  {\bibfnamefont {A.}~\bibnamefont {Goeggelmann}}, \bibinfo {author}
  {\bibfnamefont {T.~D.}\ \bibnamefont {Goodacre}}, \bibinfo {author}
  {\bibfnamefont {M.~W.}\ \bibnamefont {Haverkort}}, \bibinfo {author}
  {\bibfnamefont {D.}~\bibnamefont {Hengstler}}, \bibinfo {author}
  {\bibfnamefont {J.}~\bibnamefont {Jochum}}, \bibinfo {author} {\bibfnamefont
  {K.}~\bibnamefont {Johnston}}, \bibinfo {author} {\bibfnamefont
  {M.}~\bibnamefont {Keller}}, \bibinfo {author} {\bibfnamefont
  {S.}~\bibnamefont {Kempf}}, \bibinfo {author} {\bibfnamefont
  {T.}~\bibnamefont {Kieck}}, \bibinfo {author} {\bibfnamefont {C.~M.}\
  \bibnamefont {König}}, \bibinfo {author} {\bibfnamefont {U.}~\bibnamefont
  {Köster}}, \bibinfo {author} {\bibfnamefont {K.}~\bibnamefont {Kromer}},
  \bibinfo {author} {\bibfnamefont {F.}~\bibnamefont {Mantegazzini}}, \bibinfo
  {author} {\bibfnamefont {B.}~\bibnamefont {Marsh}}, \bibinfo {author}
  {\bibfnamefont {Y.~N.}\ \bibnamefont {Novikov}}, \bibinfo {author}
  {\bibfnamefont {F.}~\bibnamefont {Piquemal}}, \bibinfo {author}
  {\bibfnamefont {C.}~\bibnamefont {Riccio}}, \bibinfo {author} {\bibfnamefont
  {D.}~\bibnamefont {Richter}}, \bibinfo {author} {\bibfnamefont
  {A.}~\bibnamefont {Rischka}}, \bibinfo {author} {\bibfnamefont
  {S.}~\bibnamefont {Rothe}}, \bibinfo {author} {\bibfnamefont {R.~X.}\
  \bibnamefont {Schüssler}}, \bibinfo {author} {\bibfnamefont
  {C.}~\bibnamefont {Schweiger}}, \bibinfo {author} {\bibfnamefont
  {T.}~\bibnamefont {Stora}}, \bibinfo {author} {\bibfnamefont
  {M.}~\bibnamefont {Wegner}}, \bibinfo {author} {\bibfnamefont
  {K.}~\bibnamefont {Wendt}}, \bibinfo {author} {\bibfnamefont
  {M.}~\bibnamefont {Zampaolo}}, \ and\ \bibinfo {author} {\bibfnamefont
  {K.}~\bibnamefont {Zuber}},\ }\href {\doibase 10.1140/epjc/s10052-019-7513-x}
  {\bibfield  {journal} {\bibinfo  {journal} {The European Physical Journal C}\
  }\textbf {\bibinfo {volume} {79}} (\bibinfo {year} {2019}),\
  10.1140/epjc/s10052-019-7513-x}\BibitemShut {NoStop}%
\bibitem [{\citenamefont {Ferri}\ \emph {et~al.}(2015)\citenamefont {Ferri},
  \citenamefont {Bagliani}, \citenamefont {Biasotti}, \citenamefont {Ceruti},
  \citenamefont {Corsini}, \citenamefont {Faverzani}, \citenamefont {Gatti},
  \citenamefont {Giachero}, \citenamefont {Gotti}, \citenamefont {Kilbourne},
  \citenamefont {Kling}, \citenamefont {Maino}, \citenamefont {Manfrinetti},
  \citenamefont {Nucciotti}, \citenamefont {Pessina}, \citenamefont
  {Pizzigoni}, \citenamefont {{Ribeiro Gomes}},\ and\ \citenamefont
  {Sisti}}]{Ferri2015}%
  \BibitemOpen
  \bibfield  {author} {\bibinfo {author} {\bibfnamefont {E.}~\bibnamefont
  {Ferri}}, \bibinfo {author} {\bibfnamefont {D.}~\bibnamefont {Bagliani}},
  \bibinfo {author} {\bibfnamefont {M.}~\bibnamefont {Biasotti}}, \bibinfo
  {author} {\bibfnamefont {G.}~\bibnamefont {Ceruti}}, \bibinfo {author}
  {\bibfnamefont {D.}~\bibnamefont {Corsini}}, \bibinfo {author} {\bibfnamefont
  {M.}~\bibnamefont {Faverzani}}, \bibinfo {author} {\bibfnamefont
  {F.}~\bibnamefont {Gatti}}, \bibinfo {author} {\bibfnamefont
  {A.}~\bibnamefont {Giachero}}, \bibinfo {author} {\bibfnamefont
  {C.}~\bibnamefont {Gotti}}, \bibinfo {author} {\bibfnamefont
  {C.}~\bibnamefont {Kilbourne}}, \bibinfo {author} {\bibfnamefont
  {A.}~\bibnamefont {Kling}}, \bibinfo {author} {\bibfnamefont
  {M.}~\bibnamefont {Maino}}, \bibinfo {author} {\bibfnamefont
  {P.}~\bibnamefont {Manfrinetti}}, \bibinfo {author} {\bibfnamefont
  {A.}~\bibnamefont {Nucciotti}}, \bibinfo {author} {\bibfnamefont
  {G.}~\bibnamefont {Pessina}}, \bibinfo {author} {\bibfnamefont
  {G.}~\bibnamefont {Pizzigoni}}, \bibinfo {author} {\bibfnamefont
  {M.}~\bibnamefont {{Ribeiro Gomes}}}, \ and\ \bibinfo {author} {\bibfnamefont
  {M.}~\bibnamefont {Sisti}},\ }\href {\doibase 10.1016/j.phpro.2014.12.037}
  {\bibfield  {journal} {\bibinfo  {journal} {Physics Procedia}\ }\textbf
  {\bibinfo {volume} {61}},\ \bibinfo {pages} {227} (\bibinfo {year}
  {2015})}\BibitemShut {NoStop}%
\bibitem [{\citenamefont {Myers}\ \emph {et~al.}(2015)\citenamefont {Myers},
  \citenamefont {Wagner}, \citenamefont {Kracke},\ and\ \citenamefont
  {Wesson}}]{Myers2015}%
  \BibitemOpen
  \bibfield  {author} {\bibinfo {author} {\bibfnamefont {E.~G.}\ \bibnamefont
  {Myers}}, \bibinfo {author} {\bibfnamefont {A.}~\bibnamefont {Wagner}},
  \bibinfo {author} {\bibfnamefont {H.}~\bibnamefont {Kracke}}, \ and\ \bibinfo
  {author} {\bibfnamefont {B.~A.}\ \bibnamefont {Wesson}},\ }\href {\doibase
  10.1103/PhysRevLett.114.013003} {\bibfield  {journal} {\bibinfo  {journal}
  {Physical Review Letters}\ }\textbf {\bibinfo {volume} {114}} (\bibinfo
  {year} {2015}),\ 10.1103/PhysRevLett.114.013003}\BibitemShut {NoStop}%
\bibitem [{\citenamefont {Redshaw}(2023)}]{Redshaw2023}%
  \BibitemOpen
  \bibfield  {author} {\bibinfo {author} {\bibfnamefont {M.}~\bibnamefont
  {Redshaw}},\ }\href {\doibase 10.1140/epja/s10050-023-00925-9} {\bibfield
  {journal} {\bibinfo  {journal} {The European Physical Journal A}\ }\textbf
  {\bibinfo {volume} {59}},\ \bibinfo {pages} {18} (\bibinfo {year}
  {2023})}\BibitemShut {NoStop}%
\bibitem [{\citenamefont {Haaranen}\ and\ \citenamefont
  {Suhonen}(2013)}]{Haaranen2013}%
  \BibitemOpen
  \bibfield  {author} {\bibinfo {author} {\bibfnamefont {M.}~\bibnamefont
  {Haaranen}}\ and\ \bibinfo {author} {\bibfnamefont {J.}~\bibnamefont
  {Suhonen}},\ }\href {\doibase 10.1140/epja/i2013-13093-8} {\bibfield
  {journal} {\bibinfo  {journal} {The European Physical Journal A}\ }\textbf
  {\bibinfo {volume} {49}},\ \bibinfo {pages} {1} (\bibinfo {year}
  {2013})}\BibitemShut {NoStop}%
\bibitem [{\citenamefont {Suhonen}(2014)}]{Suhonen2014}%
  \BibitemOpen
  \bibfield  {author} {\bibinfo {author} {\bibfnamefont {J.}~\bibnamefont
  {Suhonen}},\ }\href {\doibase 10.1088/0031-8949/89/5/054032} {\bibfield
  {journal} {\bibinfo  {journal} {Physica Scripta}\ }\textbf {\bibinfo {volume}
  {89}},\ \bibinfo {pages} {54032} (\bibinfo {year} {2014})}\BibitemShut
  {NoStop}%
\bibitem [{\citenamefont {Sandler}\ \emph {et~al.}(2019)\citenamefont
  {Sandler}, \citenamefont {Bollen}, \citenamefont {Gamage}, \citenamefont
  {Hamaker}, \citenamefont {Izzo}, \citenamefont {Puentes}, \citenamefont
  {Redshaw}, \citenamefont {Ringle},\ and\ \citenamefont
  {Yandow}}]{Sandler2019}%
  \BibitemOpen
  \bibfield  {author} {\bibinfo {author} {\bibfnamefont {R.}~\bibnamefont
  {Sandler}}, \bibinfo {author} {\bibfnamefont {G.}~\bibnamefont {Bollen}},
  \bibinfo {author} {\bibfnamefont {N.~D.}\ \bibnamefont {Gamage}}, \bibinfo
  {author} {\bibfnamefont {A.}~\bibnamefont {Hamaker}}, \bibinfo {author}
  {\bibfnamefont {C.}~\bibnamefont {Izzo}}, \bibinfo {author} {\bibfnamefont
  {D.}~\bibnamefont {Puentes}}, \bibinfo {author} {\bibfnamefont
  {M.}~\bibnamefont {Redshaw}}, \bibinfo {author} {\bibfnamefont
  {R.}~\bibnamefont {Ringle}}, \ and\ \bibinfo {author} {\bibfnamefont
  {I.}~\bibnamefont {Yandow}},\ }\href {\doibase 10.1103/PhysRevC.100.024309}
  {\bibfield  {journal} {\bibinfo  {journal} {Physical Review C}\ }\textbf
  {\bibinfo {volume} {100}},\ \bibinfo {pages} {1} (\bibinfo {year} {2019})},\
  \Eprint {http://arxiv.org/abs/1906.03335} {arXiv:1906.03335} \BibitemShut
  {NoStop}%
\bibitem [{\citenamefont {Karthein}\ \emph {et~al.}(2019)\citenamefont
  {Karthein}, \citenamefont {Atanasov}, \citenamefont {Blaum}, \citenamefont
  {Eliseev}, \citenamefont {Filianin}, \citenamefont {Lunney}, \citenamefont
  {Manea}, \citenamefont {Mougeot}, \citenamefont {Neidherr}, \citenamefont
  {Novikov}, \citenamefont {Schweikhard}, \citenamefont {Welker}, \citenamefont
  {Wienholtz},\ and\ \citenamefont {Zuber}}]{Karthein2019a}%
  \BibitemOpen
  \bibfield  {author} {\bibinfo {author} {\bibfnamefont {J.}~\bibnamefont
  {Karthein}}, \bibinfo {author} {\bibfnamefont {D.}~\bibnamefont {Atanasov}},
  \bibinfo {author} {\bibfnamefont {K.}~\bibnamefont {Blaum}}, \bibinfo
  {author} {\bibfnamefont {S.}~\bibnamefont {Eliseev}}, \bibinfo {author}
  {\bibfnamefont {P.}~\bibnamefont {Filianin}}, \bibinfo {author}
  {\bibfnamefont {D.}~\bibnamefont {Lunney}}, \bibinfo {author} {\bibfnamefont
  {V.}~\bibnamefont {Manea}}, \bibinfo {author} {\bibfnamefont
  {M.}~\bibnamefont {Mougeot}}, \bibinfo {author} {\bibfnamefont
  {D.}~\bibnamefont {Neidherr}}, \bibinfo {author} {\bibfnamefont
  {Y.}~\bibnamefont {Novikov}}, \bibinfo {author} {\bibfnamefont
  {L.}~\bibnamefont {Schweikhard}}, \bibinfo {author} {\bibfnamefont
  {A.}~\bibnamefont {Welker}}, \bibinfo {author} {\bibfnamefont
  {F.}~\bibnamefont {Wienholtz}}, \ and\ \bibinfo {author} {\bibfnamefont
  {K.}~\bibnamefont {Zuber}},\ }\href {\doibase 10.1007/s10751-019-1601-z}
  {\bibfield  {journal} {\bibinfo  {journal} {Hyperfine Interactions}\ }\textbf
  {\bibinfo {volume} {240}},\ \bibinfo {pages} {1} (\bibinfo {year} {2019})},\
  \Eprint {http://arxiv.org/abs/1905.05510} {arXiv:1905.05510} \BibitemShut
  {NoStop}%
\bibitem [{\citenamefont {{De Roubin}}\ \emph {et~al.}(2020)\citenamefont {{De
  Roubin}}, \citenamefont {Kostensalo}, \citenamefont {Eronen}, \citenamefont
  {Canete}, \citenamefont {{De Groote}}, \citenamefont {Jokinen}, \citenamefont
  {Kankainen}, \citenamefont {Nesterenko}, \citenamefont {Moore}, \citenamefont
  {Rinta-Antila}, \citenamefont {Suhonen},\ and\ \citenamefont
  {Vil{\'{e}}n}}]{DeRoubin2020}%
  \BibitemOpen
  \bibfield  {author} {\bibinfo {author} {\bibfnamefont {A.}~\bibnamefont {{De
  Roubin}}}, \bibinfo {author} {\bibfnamefont {J.}~\bibnamefont {Kostensalo}},
  \bibinfo {author} {\bibfnamefont {T.}~\bibnamefont {Eronen}}, \bibinfo
  {author} {\bibfnamefont {L.}~\bibnamefont {Canete}}, \bibinfo {author}
  {\bibfnamefont {R.~P.}\ \bibnamefont {{De Groote}}}, \bibinfo {author}
  {\bibfnamefont {A.}~\bibnamefont {Jokinen}}, \bibinfo {author} {\bibfnamefont
  {A.}~\bibnamefont {Kankainen}}, \bibinfo {author} {\bibfnamefont {D.~A.}\
  \bibnamefont {Nesterenko}}, \bibinfo {author} {\bibfnamefont {I.~D.}\
  \bibnamefont {Moore}}, \bibinfo {author} {\bibfnamefont {S.}~\bibnamefont
  {Rinta-Antila}}, \bibinfo {author} {\bibfnamefont {J.}~\bibnamefont
  {Suhonen}}, \ and\ \bibinfo {author} {\bibfnamefont {M.}~\bibnamefont
  {Vil{\'{e}}n}},\ }\href {\doibase 10.1103/PhysRevLett.124.222503} {\bibfield
  {journal} {\bibinfo  {journal} {Physical Review Letters}\ }\textbf {\bibinfo
  {volume} {124}},\ \bibinfo {pages} {1} (\bibinfo {year} {2020})},\ \Eprint
  {http://arxiv.org/abs/2002.08282} {arXiv:2002.08282} \BibitemShut {NoStop}%
\bibitem [{\citenamefont {Ge}\ \emph {et~al.}(2021{\natexlab{a}})\citenamefont
  {Ge}, \citenamefont {Eronen}, \citenamefont {de~Roubin}, \citenamefont
  {Nesterenko}, \citenamefont {Hukkanen}, \citenamefont {Beliuskina},
  \citenamefont {de~Groote}, \citenamefont {Geldhof}, \citenamefont {Gins},
  \citenamefont {Kankainen}, \citenamefont {Koszor\'us}, \citenamefont
  {Kotila}, \citenamefont {Kostensalo}, \citenamefont {Moore}, \citenamefont
  {Raggio}, \citenamefont {Rinta-Antila}, \citenamefont {Suhonen},
  \citenamefont {Virtanen}, \citenamefont {Weaver}, \citenamefont
  {Zadvornaya},\ and\ \citenamefont {Jokinen}}]{ge2021}%
  \BibitemOpen
  \bibfield  {author} {\bibinfo {author} {\bibfnamefont {Z.}~\bibnamefont
  {Ge}}, \bibinfo {author} {\bibfnamefont {T.}~\bibnamefont {Eronen}}, \bibinfo
  {author} {\bibfnamefont {A.}~\bibnamefont {de~Roubin}}, \bibinfo {author}
  {\bibfnamefont {D.~A.}\ \bibnamefont {Nesterenko}}, \bibinfo {author}
  {\bibfnamefont {M.}~\bibnamefont {Hukkanen}}, \bibinfo {author}
  {\bibfnamefont {O.}~\bibnamefont {Beliuskina}}, \bibinfo {author}
  {\bibfnamefont {R.}~\bibnamefont {de~Groote}}, \bibinfo {author}
  {\bibfnamefont {S.}~\bibnamefont {Geldhof}}, \bibinfo {author} {\bibfnamefont
  {W.}~\bibnamefont {Gins}}, \bibinfo {author} {\bibfnamefont {A.}~\bibnamefont
  {Kankainen}}, \bibinfo {author} {\bibfnamefont {A.}~\bibnamefont
  {Koszor\'us}}, \bibinfo {author} {\bibfnamefont {J.}~\bibnamefont {Kotila}},
  \bibinfo {author} {\bibfnamefont {J.}~\bibnamefont {Kostensalo}}, \bibinfo
  {author} {\bibfnamefont {I.~D.}\ \bibnamefont {Moore}}, \bibinfo {author}
  {\bibfnamefont {A.}~\bibnamefont {Raggio}}, \bibinfo {author} {\bibfnamefont
  {S.}~\bibnamefont {Rinta-Antila}}, \bibinfo {author} {\bibfnamefont
  {J.}~\bibnamefont {Suhonen}}, \bibinfo {author} {\bibfnamefont
  {V.}~\bibnamefont {Virtanen}}, \bibinfo {author} {\bibfnamefont {A.~P.}\
  \bibnamefont {Weaver}}, \bibinfo {author} {\bibfnamefont {A.}~\bibnamefont
  {Zadvornaya}}, \ and\ \bibinfo {author} {\bibfnamefont {A.}~\bibnamefont
  {Jokinen}},\ }\href {\doibase 10.1103/PhysRevC.103.065502} {\bibfield
  {journal} {\bibinfo  {journal} {Physical Review C}\ }\textbf {\bibinfo
  {volume} {103}},\ \bibinfo {pages} {065502} (\bibinfo {year}
  {2021}{\natexlab{a}})}\BibitemShut {NoStop}%
\bibitem [{\citenamefont {Ge}\ \emph {et~al.}(2021{\natexlab{b}})\citenamefont
  {Ge}, \citenamefont {Eronen}, \citenamefont {Tyrin}, \citenamefont {Kotila},
  \citenamefont {Kostensalo}, \citenamefont {Nesterenko}, \citenamefont
  {Beliuskina}, \citenamefont {de~Groote}, \citenamefont {de~Roubin},
  \citenamefont {Geldhof}, \citenamefont {Gins}, \citenamefont {Hukkanen},
  \citenamefont {Jokinen}, \citenamefont {Kankainen}, \citenamefont
  {Koszor\'us}, \citenamefont {Krivoruchenko}, \citenamefont {Kujanp\"a\"a},
  \citenamefont {Moore}, \citenamefont {Raggio}, \citenamefont {Rinta-Antila},
  \citenamefont {Suhonen}, \citenamefont {Virtanen}, \citenamefont {Weaver},\
  and\ \citenamefont {Zadvornaya}}]{ge2021b}%
  \BibitemOpen
  \bibfield  {author} {\bibinfo {author} {\bibfnamefont {Z.}~\bibnamefont
  {Ge}}, \bibinfo {author} {\bibfnamefont {T.}~\bibnamefont {Eronen}}, \bibinfo
  {author} {\bibfnamefont {K.~S.}\ \bibnamefont {Tyrin}}, \bibinfo {author}
  {\bibfnamefont {J.}~\bibnamefont {Kotila}}, \bibinfo {author} {\bibfnamefont
  {J.}~\bibnamefont {Kostensalo}}, \bibinfo {author} {\bibfnamefont {D.~A.}\
  \bibnamefont {Nesterenko}}, \bibinfo {author} {\bibfnamefont
  {O.}~\bibnamefont {Beliuskina}}, \bibinfo {author} {\bibfnamefont
  {R.}~\bibnamefont {de~Groote}}, \bibinfo {author} {\bibfnamefont
  {A.}~\bibnamefont {de~Roubin}}, \bibinfo {author} {\bibfnamefont
  {S.}~\bibnamefont {Geldhof}}, \bibinfo {author} {\bibfnamefont
  {W.}~\bibnamefont {Gins}}, \bibinfo {author} {\bibfnamefont {M.}~\bibnamefont
  {Hukkanen}}, \bibinfo {author} {\bibfnamefont {A.}~\bibnamefont {Jokinen}},
  \bibinfo {author} {\bibfnamefont {A.}~\bibnamefont {Kankainen}}, \bibinfo
  {author} {\bibfnamefont {A.}~\bibnamefont {Koszor\'us}}, \bibinfo {author}
  {\bibfnamefont {M.~I.}\ \bibnamefont {Krivoruchenko}}, \bibinfo {author}
  {\bibfnamefont {S.}~\bibnamefont {Kujanp\"a\"a}}, \bibinfo {author}
  {\bibfnamefont {I.~D.}\ \bibnamefont {Moore}}, \bibinfo {author}
  {\bibfnamefont {A.}~\bibnamefont {Raggio}}, \bibinfo {author} {\bibfnamefont
  {S.}~\bibnamefont {Rinta-Antila}}, \bibinfo {author} {\bibfnamefont
  {J.}~\bibnamefont {Suhonen}}, \bibinfo {author} {\bibfnamefont
  {V.}~\bibnamefont {Virtanen}}, \bibinfo {author} {\bibfnamefont {A.~P.}\
  \bibnamefont {Weaver}}, \ and\ \bibinfo {author} {\bibfnamefont
  {A.}~\bibnamefont {Zadvornaya}},\ }\href {\doibase
  10.1103/PhysRevLett.127.272301} {\bibfield  {journal} {\bibinfo  {journal}
  {Phys. Rev. Lett.}\ }\textbf {\bibinfo {volume} {127}},\ \bibinfo {pages}
  {272301} (\bibinfo {year} {2021}{\natexlab{b}})}\BibitemShut {NoStop}%
\bibitem [{\citenamefont {Ge}\ \emph {et~al.}(2022{\natexlab{a}})\citenamefont
  {Ge}, \citenamefont {Eronen}, \citenamefont {{de Roubin}}, \citenamefont
  {Tyrin}, \citenamefont {Canete}, \citenamefont {Geldhof}, \citenamefont
  {Jokinen}, \citenamefont {Kankainen}, \citenamefont {Kostensalo},
  \citenamefont {Kotila}, \citenamefont {Krivoruchenko}, \citenamefont {Moore},
  \citenamefont {Nesterenko}, \citenamefont {Suhonen},\ and\ \citenamefont
  {Vilén}}]{Ge2022a}%
  \BibitemOpen
  \bibfield  {author} {\bibinfo {author} {\bibfnamefont {Z.}~\bibnamefont
  {Ge}}, \bibinfo {author} {\bibfnamefont {T.}~\bibnamefont {Eronen}}, \bibinfo
  {author} {\bibfnamefont {A.}~\bibnamefont {{de Roubin}}}, \bibinfo {author}
  {\bibfnamefont {K.}~\bibnamefont {Tyrin}}, \bibinfo {author} {\bibfnamefont
  {L.}~\bibnamefont {Canete}}, \bibinfo {author} {\bibfnamefont
  {S.}~\bibnamefont {Geldhof}}, \bibinfo {author} {\bibfnamefont
  {A.}~\bibnamefont {Jokinen}}, \bibinfo {author} {\bibfnamefont
  {A.}~\bibnamefont {Kankainen}}, \bibinfo {author} {\bibfnamefont
  {J.}~\bibnamefont {Kostensalo}}, \bibinfo {author} {\bibfnamefont
  {J.}~\bibnamefont {Kotila}}, \bibinfo {author} {\bibfnamefont
  {M.}~\bibnamefont {Krivoruchenko}}, \bibinfo {author} {\bibfnamefont
  {I.}~\bibnamefont {Moore}}, \bibinfo {author} {\bibfnamefont
  {D.}~\bibnamefont {Nesterenko}}, \bibinfo {author} {\bibfnamefont
  {J.}~\bibnamefont {Suhonen}}, \ and\ \bibinfo {author} {\bibfnamefont
  {M.}~\bibnamefont {Vilén}},\ }\href {\doibase
  https://doi.org/10.1016/j.physletb.2022.137226} {\bibfield  {journal}
  {\bibinfo  {journal} {Physics Letters B}\ }\textbf {\bibinfo {volume}
  {832}},\ \bibinfo {pages} {137226} (\bibinfo {year}
  {2022}{\natexlab{a}})}\BibitemShut {NoStop}%
\bibitem [{\citenamefont {Eronen}\ \emph {et~al.}(2022)\citenamefont {Eronen},
  \citenamefont {Ge}, \citenamefont {{de Roubin}}, \citenamefont {Ramalho},
  \citenamefont {Kostensalo}, \citenamefont {Kotila}, \citenamefont
  {Beliushkina}, \citenamefont {Delafosse}, \citenamefont {Geldhof},
  \citenamefont {Gins}, \citenamefont {Hukkanen}, \citenamefont {Jokinen},
  \citenamefont {Kankainen}, \citenamefont {Moore}, \citenamefont {Nesterenko},
  \citenamefont {Stryjczyk},\ and\ \citenamefont {Suhonen}}]{ERONEN2022}%
  \BibitemOpen
  \bibfield  {author} {\bibinfo {author} {\bibfnamefont {T.}~\bibnamefont
  {Eronen}}, \bibinfo {author} {\bibfnamefont {Z.}~\bibnamefont {Ge}}, \bibinfo
  {author} {\bibfnamefont {A.}~\bibnamefont {{de Roubin}}}, \bibinfo {author}
  {\bibfnamefont {M.}~\bibnamefont {Ramalho}}, \bibinfo {author} {\bibfnamefont
  {J.}~\bibnamefont {Kostensalo}}, \bibinfo {author} {\bibfnamefont
  {J.}~\bibnamefont {Kotila}}, \bibinfo {author} {\bibfnamefont
  {O.}~\bibnamefont {Beliushkina}}, \bibinfo {author} {\bibfnamefont
  {C.}~\bibnamefont {Delafosse}}, \bibinfo {author} {\bibfnamefont
  {S.}~\bibnamefont {Geldhof}}, \bibinfo {author} {\bibfnamefont
  {W.}~\bibnamefont {Gins}}, \bibinfo {author} {\bibfnamefont {M.}~\bibnamefont
  {Hukkanen}}, \bibinfo {author} {\bibfnamefont {A.}~\bibnamefont {Jokinen}},
  \bibinfo {author} {\bibfnamefont {A.}~\bibnamefont {Kankainen}}, \bibinfo
  {author} {\bibfnamefont {I.}~\bibnamefont {Moore}}, \bibinfo {author}
  {\bibfnamefont {D.}~\bibnamefont {Nesterenko}}, \bibinfo {author}
  {\bibfnamefont {M.}~\bibnamefont {Stryjczyk}}, \ and\ \bibinfo {author}
  {\bibfnamefont {J.}~\bibnamefont {Suhonen}},\ }\href {\doibase
  https://doi.org/10.1016/j.physletb.2022.137135} {\bibfield  {journal}
  {\bibinfo  {journal} {Physics Letters B}\ }\textbf {\bibinfo {volume}
  {830}},\ \bibinfo {pages} {137135} (\bibinfo {year} {2022})}\BibitemShut
  {NoStop}%
\bibitem [{\citenamefont {Ge}\ \emph {et~al.}(2022{\natexlab{b}})\citenamefont
  {Ge}, \citenamefont {Eronen}, \citenamefont {de~Roubin}, \citenamefont
  {Kostensalo}, \citenamefont {Suhonen}, \citenamefont {Nesterenko},
  \citenamefont {Beliuskina}, \citenamefont {de~Groote}, \citenamefont
  {Delafosse}, \citenamefont {Geldhof}, \citenamefont {Gins}, \citenamefont
  {Hukkanen}, \citenamefont {Jokinen}, \citenamefont {Kankainen}, \citenamefont
  {Kotila}, \citenamefont {Koszor\'us}, \citenamefont {Moore}, \citenamefont
  {Raggio}, \citenamefont {Rinta-Antila}, \citenamefont {Virtanen},
  \citenamefont {Weaver},\ and\ \citenamefont {Zadvornaya}}]{Ge2022b}%
  \BibitemOpen
  \bibfield  {author} {\bibinfo {author} {\bibfnamefont {Z.}~\bibnamefont
  {Ge}}, \bibinfo {author} {\bibfnamefont {T.}~\bibnamefont {Eronen}}, \bibinfo
  {author} {\bibfnamefont {A.}~\bibnamefont {de~Roubin}}, \bibinfo {author}
  {\bibfnamefont {J.}~\bibnamefont {Kostensalo}}, \bibinfo {author}
  {\bibfnamefont {J.}~\bibnamefont {Suhonen}}, \bibinfo {author} {\bibfnamefont
  {D.~A.}\ \bibnamefont {Nesterenko}}, \bibinfo {author} {\bibfnamefont
  {O.}~\bibnamefont {Beliuskina}}, \bibinfo {author} {\bibfnamefont
  {R.}~\bibnamefont {de~Groote}}, \bibinfo {author} {\bibfnamefont
  {C.}~\bibnamefont {Delafosse}}, \bibinfo {author} {\bibfnamefont
  {S.}~\bibnamefont {Geldhof}}, \bibinfo {author} {\bibfnamefont
  {W.}~\bibnamefont {Gins}}, \bibinfo {author} {\bibfnamefont {M.}~\bibnamefont
  {Hukkanen}}, \bibinfo {author} {\bibfnamefont {A.}~\bibnamefont {Jokinen}},
  \bibinfo {author} {\bibfnamefont {A.}~\bibnamefont {Kankainen}}, \bibinfo
  {author} {\bibfnamefont {J.}~\bibnamefont {Kotila}}, \bibinfo {author}
  {\bibfnamefont {A.}~\bibnamefont {Koszor\'us}}, \bibinfo {author}
  {\bibfnamefont {I.~D.}\ \bibnamefont {Moore}}, \bibinfo {author}
  {\bibfnamefont {A.}~\bibnamefont {Raggio}}, \bibinfo {author} {\bibfnamefont
  {S.}~\bibnamefont {Rinta-Antila}}, \bibinfo {author} {\bibfnamefont
  {V.}~\bibnamefont {Virtanen}}, \bibinfo {author} {\bibfnamefont {A.~P.}\
  \bibnamefont {Weaver}}, \ and\ \bibinfo {author} {\bibfnamefont
  {A.}~\bibnamefont {Zadvornaya}},\ }\href {\doibase
  10.1103/PhysRevC.106.015502} {\bibfield  {journal} {\bibinfo  {journal}
  {Phys. Rev. C}\ }\textbf {\bibinfo {volume} {106}},\ \bibinfo {pages}
  {015502} (\bibinfo {year} {2022}{\natexlab{b}})}\BibitemShut {NoStop}%
\bibitem [{\citenamefont {Ramalho}\ \emph {et~al.}(2022)\citenamefont
  {Ramalho}, \citenamefont {Ge}, \citenamefont {Eronen}, \citenamefont
  {Nesterenko}, \citenamefont {Jaatinen}, \citenamefont {Jokinen},
  \citenamefont {Kankainen}, \citenamefont {Kostensalo}, \citenamefont
  {Kotila}, \citenamefont {Krivoruchenko}, \citenamefont {Suhonen},
  \citenamefont {Tyrin},\ and\ \citenamefont {Virtanen}}]{Ramalho2022}%
  \BibitemOpen
  \bibfield  {author} {\bibinfo {author} {\bibfnamefont {M.}~\bibnamefont
  {Ramalho}}, \bibinfo {author} {\bibfnamefont {Z.}~\bibnamefont {Ge}},
  \bibinfo {author} {\bibfnamefont {T.}~\bibnamefont {Eronen}}, \bibinfo
  {author} {\bibfnamefont {D.~A.}\ \bibnamefont {Nesterenko}}, \bibinfo
  {author} {\bibfnamefont {J.}~\bibnamefont {Jaatinen}}, \bibinfo {author}
  {\bibfnamefont {A.}~\bibnamefont {Jokinen}}, \bibinfo {author} {\bibfnamefont
  {A.}~\bibnamefont {Kankainen}}, \bibinfo {author} {\bibfnamefont
  {J.}~\bibnamefont {Kostensalo}}, \bibinfo {author} {\bibfnamefont
  {J.}~\bibnamefont {Kotila}}, \bibinfo {author} {\bibfnamefont {M.~I.}\
  \bibnamefont {Krivoruchenko}}, \bibinfo {author} {\bibfnamefont
  {J.}~\bibnamefont {Suhonen}}, \bibinfo {author} {\bibfnamefont {K.~S.}\
  \bibnamefont {Tyrin}}, \ and\ \bibinfo {author} {\bibfnamefont
  {V.}~\bibnamefont {Virtanen}},\ }\href {\doibase 10.1103/PhysRevC.106.015501}
  {\bibfield  {journal} {\bibinfo  {journal} {Phys. Rev. C}\ }\textbf {\bibinfo
  {volume} {106}},\ \bibinfo {pages} {015501} (\bibinfo {year}
  {2022})}\BibitemShut {NoStop}%
\bibitem [{\citenamefont {Keblbeck}\ \emph {et~al.}(2023)\citenamefont
  {Keblbeck}, \citenamefont {Bhandari}, \citenamefont {Gamage}, \citenamefont
  {Horana~Gamage}, \citenamefont {Leach}, \citenamefont {Mougeot},\ and\
  \citenamefont {Redshaw}}]{Keblbeck2023}%
  \BibitemOpen
  \bibfield  {author} {\bibinfo {author} {\bibfnamefont {D.~K.}\ \bibnamefont
  {Keblbeck}}, \bibinfo {author} {\bibfnamefont {R.}~\bibnamefont {Bhandari}},
  \bibinfo {author} {\bibfnamefont {N.~D.}\ \bibnamefont {Gamage}}, \bibinfo
  {author} {\bibfnamefont {M.}~\bibnamefont {Horana~Gamage}}, \bibinfo {author}
  {\bibfnamefont {K.~G.}\ \bibnamefont {Leach}}, \bibinfo {author}
  {\bibfnamefont {X.}~\bibnamefont {Mougeot}}, \ and\ \bibinfo {author}
  {\bibfnamefont {M.}~\bibnamefont {Redshaw}},\ }\href {\doibase
  10.1103/PhysRevC.107.015504} {\bibfield  {journal} {\bibinfo  {journal}
  {Phys. Rev. C}\ }\textbf {\bibinfo {volume} {107}},\ \bibinfo {pages}
  {015504} (\bibinfo {year} {2023})}\BibitemShut {NoStop}%
\bibitem [{\citenamefont {Pf\"utzner}\ \emph {et~al.}(2015)\citenamefont
  {Pf\"utzner}, \citenamefont {Pachucki},\ and\ \citenamefont
  {\ifmmode~\dot{Z}\else \.{Z}\fi{}ylicz}}]{Pfutzner2015}%
  \BibitemOpen
  \bibfield  {author} {\bibinfo {author} {\bibfnamefont {M.}~\bibnamefont
  {Pf\"utzner}}, \bibinfo {author} {\bibfnamefont {K.}~\bibnamefont
  {Pachucki}}, \ and\ \bibinfo {author} {\bibfnamefont {J.}~\bibnamefont
  {\ifmmode~\dot{Z}\else \.{Z}\fi{}ylicz}},\ }\href {\doibase
  10.1103/PhysRevC.92.044305} {\bibfield  {journal} {\bibinfo  {journal}
  {Physical Review C}\ }\textbf {\bibinfo {volume} {92}},\ \bibinfo {pages}
  {044305} (\bibinfo {year} {2015})}\BibitemShut {NoStop}%
\bibitem [{\citenamefont {Longmire}(1949)}]{Longmire49}%
  \BibitemOpen
  \bibfield  {author} {\bibinfo {author} {\bibfnamefont {C.~L.}\ \bibnamefont
  {Longmire}},\ }\href {\doibase 10.1103/PhysRev.75.15} {\bibfield  {journal}
  {\bibinfo  {journal} {Phys. Rev.}\ }\textbf {\bibinfo {volume} {75}},\
  \bibinfo {pages} {15} (\bibinfo {year} {1949})}\BibitemShut {NoStop}%
\bibitem [{\citenamefont {Haselschwardt}\ \emph {et~al.}(2020)\citenamefont
  {Haselschwardt}, \citenamefont {Lenardo}, \citenamefont {Pirinen},\ and\
  \citenamefont {Suhonen}}]{Haselschwardt20}%
  \BibitemOpen
  \bibfield  {author} {\bibinfo {author} {\bibfnamefont {S.}~\bibnamefont
  {Haselschwardt}}, \bibinfo {author} {\bibfnamefont {B.}~\bibnamefont
  {Lenardo}}, \bibinfo {author} {\bibfnamefont {P.}~\bibnamefont {Pirinen}}, \
  and\ \bibinfo {author} {\bibfnamefont {J.}~\bibnamefont {Suhonen}},\ }\href
  {\doibase 10.1103/PhysRevD.102.072009} {\bibfield  {journal} {\bibinfo
  {journal} {Phys. Rev. D}\ }\textbf {\bibinfo {volume} {102}},\ \bibinfo
  {pages} {072009} (\bibinfo {year} {2020})}\BibitemShut {NoStop}%
\bibitem [{NND(2021)}]{NNDC}%
  \BibitemOpen
  \href {https://www.nndc.bnl.gov/} {\enquote {\bibinfo {title} {National
  nuclear data center},}\ }\bibinfo {howpublished} {Available at
  \url{https://www.nndc.bnl.gov/} (2020/4/7)} (\bibinfo {year}
  {2021})\BibitemShut {NoStop}%
\bibitem [{\citenamefont {Wang}\ \emph {et~al.}(2021)\citenamefont {Wang},
  \citenamefont {Huang}, \citenamefont {Kondev}, \citenamefont {Audi},\ and\
  \citenamefont {Naimi}}]{Wang2021}%
  \BibitemOpen
  \bibfield  {author} {\bibinfo {author} {\bibfnamefont {M.}~\bibnamefont
  {Wang}}, \bibinfo {author} {\bibfnamefont {W.}~\bibnamefont {Huang}},
  \bibinfo {author} {\bibfnamefont {F.}~\bibnamefont {Kondev}}, \bibinfo
  {author} {\bibfnamefont {G.}~\bibnamefont {Audi}}, \ and\ \bibinfo {author}
  {\bibfnamefont {S.}~\bibnamefont {Naimi}},\ }\href {\doibase
  10.1088/1674-1137/abddaf} {\bibfield  {journal} {\bibinfo  {journal} {Chinese
  Physics C}\ }\textbf {\bibinfo {volume} {45}},\ \bibinfo {pages} {030003}
  (\bibinfo {year} {2021})}\BibitemShut {NoStop}%
\bibitem [{\citenamefont {Olsen}\ and\ \citenamefont
  {O'Kelley}(1954)}]{136Csa}%
  \BibitemOpen
  \bibfield  {author} {\bibinfo {author} {\bibfnamefont {J.~L.}\ \bibnamefont
  {Olsen}}\ and\ \bibinfo {author} {\bibfnamefont {G.~D.}\ \bibnamefont
  {O'Kelley}},\ }\href {\doibase 10.1103/PhysRev.95.1539} {\bibfield  {journal}
  {\bibinfo  {journal} {Phys. Rev.}\ }\textbf {\bibinfo {volume} {95}},\
  \bibinfo {pages} {1539} (\bibinfo {year} {1954})}\BibitemShut {NoStop}%
\bibitem [{\citenamefont {Reising}\ and\ \citenamefont {Pate}(1965)}]{136Csb}%
  \BibitemOpen
  \bibfield  {author} {\bibinfo {author} {\bibfnamefont {R.}~\bibnamefont
  {Reising}}\ and\ \bibinfo {author} {\bibfnamefont {B.}~\bibnamefont {Pate}},\
  }\href {\doibase https://doi.org/10.1016/0029-5582(65)90328-7} {\bibfield
  {journal} {\bibinfo  {journal} {Nuclear Physics}\ }\textbf {\bibinfo {volume}
  {65}},\ \bibinfo {pages} {609} (\bibinfo {year} {1965})}\BibitemShut
  {NoStop}%
\bibitem [{\citenamefont {Fink}\ \emph {et~al.}(2012)\citenamefont {Fink},
  \citenamefont {Barea}, \citenamefont {Beck}, \citenamefont {Blaum},
  \citenamefont {B{\"{o}}hm}, \citenamefont {Borgmann}, \citenamefont
  {Breitenfeldt}, \citenamefont {Herfurth}, \citenamefont {Herlert},
  \citenamefont {Kotila}, \citenamefont {Kowalska}, \citenamefont {Kreim},
  \citenamefont {Lunney}, \citenamefont {Naimi}, \citenamefont {Rosenbusch},
  \citenamefont {Schwarz}, \citenamefont {Schweikhard}, \citenamefont
  {{\v{S}}imkovic}, \citenamefont {Stanja},\ and\ \citenamefont
  {Zuber}}]{Fink2012}%
  \BibitemOpen
  \bibfield  {author} {\bibinfo {author} {\bibfnamefont {D.}~\bibnamefont
  {Fink}}, \bibinfo {author} {\bibfnamefont {J.}~\bibnamefont {Barea}},
  \bibinfo {author} {\bibfnamefont {D.}~\bibnamefont {Beck}}, \bibinfo {author}
  {\bibfnamefont {K.}~\bibnamefont {Blaum}}, \bibinfo {author} {\bibfnamefont
  {C.}~\bibnamefont {B{\"{o}}hm}}, \bibinfo {author} {\bibfnamefont
  {C.}~\bibnamefont {Borgmann}}, \bibinfo {author} {\bibfnamefont
  {M.}~\bibnamefont {Breitenfeldt}}, \bibinfo {author} {\bibfnamefont
  {F.}~\bibnamefont {Herfurth}}, \bibinfo {author} {\bibfnamefont
  {A.}~\bibnamefont {Herlert}}, \bibinfo {author} {\bibfnamefont
  {J.}~\bibnamefont {Kotila}}, \bibinfo {author} {\bibfnamefont
  {M.}~\bibnamefont {Kowalska}}, \bibinfo {author} {\bibfnamefont
  {S.}~\bibnamefont {Kreim}}, \bibinfo {author} {\bibfnamefont
  {D.}~\bibnamefont {Lunney}}, \bibinfo {author} {\bibfnamefont
  {S.}~\bibnamefont {Naimi}}, \bibinfo {author} {\bibfnamefont
  {M.}~\bibnamefont {Rosenbusch}}, \bibinfo {author} {\bibfnamefont
  {S.}~\bibnamefont {Schwarz}}, \bibinfo {author} {\bibfnamefont
  {L.}~\bibnamefont {Schweikhard}}, \bibinfo {author} {\bibfnamefont
  {F.}~\bibnamefont {{\v{S}}imkovic}}, \bibinfo {author} {\bibfnamefont
  {J.}~\bibnamefont {Stanja}}, \ and\ \bibinfo {author} {\bibfnamefont
  {K.}~\bibnamefont {Zuber}},\ }\href {\doibase 10.1103/PhysRevLett.108.062502}
  {\bibfield  {journal} {\bibinfo  {journal} {Physical Review Letters}\
  }\textbf {\bibinfo {volume} {108}},\ \bibinfo {pages} {1} (\bibinfo {year}
  {2012})}\BibitemShut {NoStop}%
\bibitem [{\citenamefont {Nesterenko}\ \emph {et~al.}(2019)\citenamefont
  {Nesterenko}, \citenamefont {Canete}, \citenamefont {Eronen}, \citenamefont
  {Jokinen}, \citenamefont {Kankainen}, \citenamefont {Novikov}, \citenamefont
  {Rinta-Antila}, \citenamefont {de~Roubin},\ and\ \citenamefont
  {Vilen}}]{Nesterenko2019}%
  \BibitemOpen
  \bibfield  {author} {\bibinfo {author} {\bibfnamefont {D.~A.}\ \bibnamefont
  {Nesterenko}}, \bibinfo {author} {\bibfnamefont {L.}~\bibnamefont {Canete}},
  \bibinfo {author} {\bibfnamefont {T.}~\bibnamefont {Eronen}}, \bibinfo
  {author} {\bibfnamefont {A.}~\bibnamefont {Jokinen}}, \bibinfo {author}
  {\bibfnamefont {A.}~\bibnamefont {Kankainen}}, \bibinfo {author}
  {\bibfnamefont {Y.~N.}\ \bibnamefont {Novikov}}, \bibinfo {author}
  {\bibfnamefont {S.}~\bibnamefont {Rinta-Antila}}, \bibinfo {author}
  {\bibfnamefont {A.}~\bibnamefont {de~Roubin}}, \ and\ \bibinfo {author}
  {\bibfnamefont {M.}~\bibnamefont {Vilen}},\ }\href {\doibase
  10.1016/j.ijms.2018.10.038} {\bibfield  {journal} {\bibinfo  {journal}
  {International Journal of Mass Spectrometry}\ }\textbf {\bibinfo {volume}
  {435}},\ \bibinfo {pages} {204} (\bibinfo {year} {2019})}\BibitemShut
  {NoStop}%
\bibitem [{\citenamefont {Moore}\ \emph {et~al.}(2013)\citenamefont {Moore},
  \citenamefont {Eronen}, \citenamefont {Gorelov}, \citenamefont {Hakala},
  \citenamefont {Jokinen}, \citenamefont {Kankainen}, \citenamefont {Kolhinen},
  \citenamefont {Koponen}, \citenamefont {Penttil{\"{a}}}, \citenamefont
  {Pohjalainen}, \citenamefont {Reponen}, \citenamefont {Rissanen},
  \citenamefont {Saastamoinen}, \citenamefont {Rinta-Antila}, \citenamefont
  {Sonnenschein},\ and\ \citenamefont {{\"{A}}yst{\"{o}}}}]{Moore2013}%
  \BibitemOpen
  \bibfield  {author} {\bibinfo {author} {\bibfnamefont {I.~D.}\ \bibnamefont
  {Moore}}, \bibinfo {author} {\bibfnamefont {T.}~\bibnamefont {Eronen}},
  \bibinfo {author} {\bibfnamefont {D.}~\bibnamefont {Gorelov}}, \bibinfo
  {author} {\bibfnamefont {J.}~\bibnamefont {Hakala}}, \bibinfo {author}
  {\bibfnamefont {A.}~\bibnamefont {Jokinen}}, \bibinfo {author} {\bibfnamefont
  {A.}~\bibnamefont {Kankainen}}, \bibinfo {author} {\bibfnamefont {V.~S.}\
  \bibnamefont {Kolhinen}}, \bibinfo {author} {\bibfnamefont {J.}~\bibnamefont
  {Koponen}}, \bibinfo {author} {\bibfnamefont {H.}~\bibnamefont
  {Penttil{\"{a}}}}, \bibinfo {author} {\bibfnamefont {I.}~\bibnamefont
  {Pohjalainen}}, \bibinfo {author} {\bibfnamefont {M.}~\bibnamefont
  {Reponen}}, \bibinfo {author} {\bibfnamefont {J.}~\bibnamefont {Rissanen}},
  \bibinfo {author} {\bibfnamefont {A.}~\bibnamefont {Saastamoinen}}, \bibinfo
  {author} {\bibfnamefont {S.}~\bibnamefont {Rinta-Antila}}, \bibinfo {author}
  {\bibfnamefont {V.}~\bibnamefont {Sonnenschein}}, \ and\ \bibinfo {author}
  {\bibfnamefont {J.}~\bibnamefont {{\"{A}}yst{\"{o}}}},\ }\href {\doibase
  10.1016/j.nimb.2013.06.036} {\bibfield  {journal} {\bibinfo  {journal}
  {Nuclear Instruments and Methods in Physics Research, Section B: Beam
  Interactions with Materials and Atoms}\ }\textbf {\bibinfo {volume} {317}},\
  \bibinfo {pages} {208} (\bibinfo {year} {2013})}\BibitemShut {NoStop}%
\bibitem [{\citenamefont {Eronen}\ and\ \citenamefont
  {Hardy}(2012)}]{Eronen2012}%
  \BibitemOpen
  \bibfield  {author} {\bibinfo {author} {\bibfnamefont {T.}~\bibnamefont
  {Eronen}}\ and\ \bibinfo {author} {\bibfnamefont {J.~C.}\ \bibnamefont
  {Hardy}},\ }\href {\doibase 10.1140/epja/i2012-12048-y} {\bibfield  {journal}
  {\bibinfo  {journal} {European Physical Journal A}\ }\textbf {\bibinfo
  {volume} {48}},\ \bibinfo {pages} {1} (\bibinfo {year} {2012})}\BibitemShut
  {NoStop}%
\bibitem [{\citenamefont {Kolhinen}\ \emph {et~al.}(2013)\citenamefont
  {Kolhinen}, \citenamefont {Eronen}, \citenamefont {Gorelov}, \citenamefont
  {Hakala}, \citenamefont {Jokinen}, \citenamefont {Jokiranta}, \citenamefont
  {Kankainen}, \citenamefont {Koikkalainen}, \citenamefont {Koponen},
  \citenamefont {Kulmala}, \citenamefont {Lantz}, \citenamefont {Mattera},
  \citenamefont {Moore}, \citenamefont {Penttil{\"{a}}}, \citenamefont
  {Pikkarainen}, \citenamefont {Pohjlainen}, \citenamefont {Reponen},
  \citenamefont {Rinta-Antila}, \citenamefont {Rissanen}, \citenamefont
  {{Rodr{\'{i}}guez Triguero}}, \citenamefont {Rytk{\"{o}}nen}, \citenamefont
  {Saastamoinen}, \citenamefont {Solders}, \citenamefont {Sonnenschein},\ and\
  \citenamefont {{\"{A}}yst{\"{o}}}}]{Kolhinen2013}%
  \BibitemOpen
  \bibfield  {author} {\bibinfo {author} {\bibfnamefont {V.~S.}\ \bibnamefont
  {Kolhinen}}, \bibinfo {author} {\bibfnamefont {T.}~\bibnamefont {Eronen}},
  \bibinfo {author} {\bibfnamefont {D.}~\bibnamefont {Gorelov}}, \bibinfo
  {author} {\bibfnamefont {J.}~\bibnamefont {Hakala}}, \bibinfo {author}
  {\bibfnamefont {A.}~\bibnamefont {Jokinen}}, \bibinfo {author} {\bibfnamefont
  {K.}~\bibnamefont {Jokiranta}}, \bibinfo {author} {\bibfnamefont
  {A.}~\bibnamefont {Kankainen}}, \bibinfo {author} {\bibfnamefont
  {M.}~\bibnamefont {Koikkalainen}}, \bibinfo {author} {\bibfnamefont
  {J.}~\bibnamefont {Koponen}}, \bibinfo {author} {\bibfnamefont
  {H.}~\bibnamefont {Kulmala}}, \bibinfo {author} {\bibfnamefont
  {M.}~\bibnamefont {Lantz}}, \bibinfo {author} {\bibfnamefont
  {A.}~\bibnamefont {Mattera}}, \bibinfo {author} {\bibfnamefont {I.~D.}\
  \bibnamefont {Moore}}, \bibinfo {author} {\bibfnamefont {H.}~\bibnamefont
  {Penttil{\"{a}}}}, \bibinfo {author} {\bibfnamefont {T.}~\bibnamefont
  {Pikkarainen}}, \bibinfo {author} {\bibfnamefont {I.}~\bibnamefont
  {Pohjlainen}}, \bibinfo {author} {\bibfnamefont {M.}~\bibnamefont {Reponen}},
  \bibinfo {author} {\bibfnamefont {S.}~\bibnamefont {Rinta-Antila}}, \bibinfo
  {author} {\bibfnamefont {J.}~\bibnamefont {Rissanen}}, \bibinfo {author}
  {\bibfnamefont {C.}~\bibnamefont {{Rodr{\'{i}}guez Triguero}}}, \bibinfo
  {author} {\bibfnamefont {K.}~\bibnamefont {Rytk{\"{o}}nen}}, \bibinfo
  {author} {\bibfnamefont {A.}~\bibnamefont {Saastamoinen}}, \bibinfo {author}
  {\bibfnamefont {A.}~\bibnamefont {Solders}}, \bibinfo {author} {\bibfnamefont
  {V.}~\bibnamefont {Sonnenschein}}, \ and\ \bibinfo {author} {\bibfnamefont
  {J.}~\bibnamefont {{\"{A}}yst{\"{o}}}},\ }\href {\doibase
  10.1016/j.nimb.2013.07.050} {\bibfield  {journal} {\bibinfo  {journal}
  {Nuclear Instruments and Methods in Physics Research, Section B: Beam
  Interactions with Materials and Atoms}\ }\textbf {\bibinfo {volume} {317}},\
  \bibinfo {pages} {506} (\bibinfo {year} {2013})}\BibitemShut {NoStop}%
\bibitem [{\citenamefont {Karvonen}\ \emph {et~al.}(2008)\citenamefont
  {Karvonen}, \citenamefont {Moore}, \citenamefont {Sonoda}, \citenamefont
  {Kessler}, \citenamefont {Penttil{\"{a}}}, \citenamefont
  {Per{\"{a}}j{\"{a}}rvi}, \citenamefont {Ronkanen},\ and\ \citenamefont
  {{\"{A}}yst{\"{o}}}}]{Karvonen2008}%
  \BibitemOpen
  \bibfield  {author} {\bibinfo {author} {\bibfnamefont {P.}~\bibnamefont
  {Karvonen}}, \bibinfo {author} {\bibfnamefont {I.~D.}\ \bibnamefont {Moore}},
  \bibinfo {author} {\bibfnamefont {T.}~\bibnamefont {Sonoda}}, \bibinfo
  {author} {\bibfnamefont {T.}~\bibnamefont {Kessler}}, \bibinfo {author}
  {\bibfnamefont {H.}~\bibnamefont {Penttil{\"{a}}}}, \bibinfo {author}
  {\bibfnamefont {K.}~\bibnamefont {Per{\"{a}}j{\"{a}}rvi}}, \bibinfo {author}
  {\bibfnamefont {P.}~\bibnamefont {Ronkanen}}, \ and\ \bibinfo {author}
  {\bibfnamefont {J.}~\bibnamefont {{\"{A}}yst{\"{o}}}},\ }\href {\doibase
  10.1016/j.nimb.2008.07.022} {\bibfield  {journal} {\bibinfo  {journal}
  {Nuclear Instruments and Methods in Physics Research, Section B: Beam
  Interactions with Materials and Atoms}\ }\textbf {\bibinfo {volume} {266}},\
  \bibinfo {pages} {4794} (\bibinfo {year} {2008})}\BibitemShut {NoStop}%
\bibitem [{\citenamefont {Nieminen}\ \emph {et~al.}(2001)\citenamefont
  {Nieminen}, \citenamefont {Huikari}, \citenamefont {Jokinen}, \citenamefont
  {{\"{A}}yst{\"{o}}}, \citenamefont {Campbell},\ and\ \citenamefont
  {Cochrane}}]{Nieminen2001}%
  \BibitemOpen
  \bibfield  {author} {\bibinfo {author} {\bibfnamefont {A.}~\bibnamefont
  {Nieminen}}, \bibinfo {author} {\bibfnamefont {J.}~\bibnamefont {Huikari}},
  \bibinfo {author} {\bibfnamefont {A.}~\bibnamefont {Jokinen}}, \bibinfo
  {author} {\bibfnamefont {J.}~\bibnamefont {{\"{A}}yst{\"{o}}}}, \bibinfo
  {author} {\bibfnamefont {P.}~\bibnamefont {Campbell}}, \ and\ \bibinfo
  {author} {\bibfnamefont {E.~C.}\ \bibnamefont {Cochrane}},\ }\href {\doibase
  10.1016/S0168-9002(00)00750-6} {\bibfield  {journal} {\bibinfo  {journal}
  {Nuclear Instruments and Methods in Physics Research, Section A:
  Accelerators, Spectrometers, Detectors and Associated Equipment}\ }\textbf
  {\bibinfo {volume} {469}},\ \bibinfo {pages} {244} (\bibinfo {year}
  {2001})}\BibitemShut {NoStop}%
\bibitem [{\citenamefont {Eronen}\ \emph {et~al.}(2008)\citenamefont {Eronen},
  \citenamefont {Elomaa}, \citenamefont {Hager}, \citenamefont {Hakala},
  \citenamefont {Jokinen}, \citenamefont {Kankainen}, \citenamefont {Rahaman},
  \citenamefont {Rissanen}, \citenamefont {Weber},\ and\ \citenamefont
  {{\"{A}}yst{\"{o}}}}]{Eronen2008a}%
  \BibitemOpen
  \bibfield  {author} {\bibinfo {author} {\bibfnamefont {T.}~\bibnamefont
  {Eronen}}, \bibinfo {author} {\bibfnamefont {V.~V.}\ \bibnamefont {Elomaa}},
  \bibinfo {author} {\bibfnamefont {U.}~\bibnamefont {Hager}}, \bibinfo
  {author} {\bibfnamefont {J.}~\bibnamefont {Hakala}}, \bibinfo {author}
  {\bibfnamefont {A.}~\bibnamefont {Jokinen}}, \bibinfo {author} {\bibfnamefont
  {A.}~\bibnamefont {Kankainen}}, \bibinfo {author} {\bibfnamefont
  {S.}~\bibnamefont {Rahaman}}, \bibinfo {author} {\bibfnamefont
  {J.}~\bibnamefont {Rissanen}}, \bibinfo {author} {\bibfnamefont
  {C.}~\bibnamefont {Weber}}, \ and\ \bibinfo {author} {\bibfnamefont
  {J.}~\bibnamefont {{\"{A}}yst{\"{o}}}},\ }\href
  {https://www.actaphys.uj.edu.pl/R/39/2/445/pdf} {\bibfield  {journal}
  {\bibinfo  {journal} {Acta Physica Polonica B}\ }\textbf {\bibinfo {volume}
  {39}},\ \bibinfo {pages} {445} (\bibinfo {year} {2008})}\BibitemShut
  {NoStop}%
\bibitem [{\citenamefont {Nesterenko}\ \emph {et~al.}(2018)\citenamefont
  {Nesterenko}, \citenamefont {Eronen}, \citenamefont {Kankainen},
  \citenamefont {Canete}, \citenamefont {Jokinen}, \citenamefont {Moore},
  \citenamefont {Penttil{\"{a}}}, \citenamefont {Rinta-Antila}, \citenamefont
  {de~Roubin},\ and\ \citenamefont {Vilen}}]{Nesterenko2018}%
  \BibitemOpen
  \bibfield  {author} {\bibinfo {author} {\bibfnamefont {D.~A.}\ \bibnamefont
  {Nesterenko}}, \bibinfo {author} {\bibfnamefont {T.}~\bibnamefont {Eronen}},
  \bibinfo {author} {\bibfnamefont {A.}~\bibnamefont {Kankainen}}, \bibinfo
  {author} {\bibfnamefont {L.}~\bibnamefont {Canete}}, \bibinfo {author}
  {\bibfnamefont {A.}~\bibnamefont {Jokinen}}, \bibinfo {author} {\bibfnamefont
  {I.~D.}\ \bibnamefont {Moore}}, \bibinfo {author} {\bibfnamefont
  {H.}~\bibnamefont {Penttil{\"{a}}}}, \bibinfo {author} {\bibfnamefont
  {S.}~\bibnamefont {Rinta-Antila}}, \bibinfo {author} {\bibfnamefont
  {A.}~\bibnamefont {de~Roubin}}, \ and\ \bibinfo {author} {\bibfnamefont
  {M.}~\bibnamefont {Vilen}},\ }\href {\doibase 10.1140/epja/i2018-12589-y}
  {\bibfield  {journal} {\bibinfo  {journal} {European Physical Journal A}\
  }\textbf {\bibinfo {volume} {54}},\ \bibinfo {pages} {0} (\bibinfo {year}
  {2018})}\BibitemShut {NoStop}%
\bibitem [{\citenamefont {Nesterenko}\ \emph {et~al.}(2021)\citenamefont
  {Nesterenko}, \citenamefont {Eronen}, \citenamefont {Ge}, \citenamefont
  {Kankainen},\ and\ \citenamefont {Vilen}}]{nesterenko2021}%
  \BibitemOpen
  \bibfield  {author} {\bibinfo {author} {\bibfnamefont {D.~A.}\ \bibnamefont
  {Nesterenko}}, \bibinfo {author} {\bibfnamefont {T.}~\bibnamefont {Eronen}},
  \bibinfo {author} {\bibfnamefont {Z.}~\bibnamefont {Ge}}, \bibinfo {author}
  {\bibfnamefont {A.}~\bibnamefont {Kankainen}}, \ and\ \bibinfo {author}
  {\bibfnamefont {M.}~\bibnamefont {Vilen}},\ }\href
  {https://doi.org/10.1140/epja/s10050-021-00608-3} {\bibfield  {journal}
  {\bibinfo  {journal} {Eur. Phys. J. A}\ }\textbf {\bibinfo {volume} {57}},\
  \bibinfo {pages} {302} (\bibinfo {year} {2021})}\BibitemShut {NoStop}%
\bibitem [{\citenamefont {Savard}\ \emph {et~al.}(1991)\citenamefont {Savard},
  \citenamefont {Becker}, \citenamefont {Bollen}, \citenamefont {Kluge},
  \citenamefont {Moore}, \citenamefont {Otto}, \citenamefont {Schweikhard},
  \citenamefont {Stolzenberg},\ and\ \citenamefont {Wiess}}]{Savard1991}%
  \BibitemOpen
  \bibfield  {author} {\bibinfo {author} {\bibfnamefont {G.}~\bibnamefont
  {Savard}}, \bibinfo {author} {\bibfnamefont {S.}~\bibnamefont {Becker}},
  \bibinfo {author} {\bibfnamefont {G.}~\bibnamefont {Bollen}}, \bibinfo
  {author} {\bibfnamefont {H.~J.}\ \bibnamefont {Kluge}}, \bibinfo {author}
  {\bibfnamefont {R.~B.}\ \bibnamefont {Moore}}, \bibinfo {author}
  {\bibfnamefont {T.}~\bibnamefont {Otto}}, \bibinfo {author} {\bibfnamefont
  {L.}~\bibnamefont {Schweikhard}}, \bibinfo {author} {\bibfnamefont
  {H.}~\bibnamefont {Stolzenberg}}, \ and\ \bibinfo {author} {\bibfnamefont
  {U.}~\bibnamefont {Wiess}},\ }\href {\doibase 10.1016/0375-9601(91)91008-2}
  {\bibfield  {journal} {\bibinfo  {journal} {Physics Letters A}\ }\textbf
  {\bibinfo {volume} {158}},\ \bibinfo {pages} {247} (\bibinfo {year}
  {1991})}\BibitemShut {NoStop}%
\bibitem [{\citenamefont {Eliseev}\ \emph {et~al.}(2014)\citenamefont
  {Eliseev}, \citenamefont {Blaum}, \citenamefont {Block}, \citenamefont
  {D{\"{o}}rr}, \citenamefont {Droese}, \citenamefont {Eronen}, \citenamefont
  {Goncharov}, \citenamefont {H{\"{o}}cker}, \citenamefont {Ketter},
  \citenamefont {Ramirez}, \citenamefont {Nesterenko}, \citenamefont
  {Novikov},\ and\ \citenamefont {Schweikhard}}]{Eliseev2014}%
  \BibitemOpen
  \bibfield  {author} {\bibinfo {author} {\bibfnamefont {S.}~\bibnamefont
  {Eliseev}}, \bibinfo {author} {\bibfnamefont {K.}~\bibnamefont {Blaum}},
  \bibinfo {author} {\bibfnamefont {M.}~\bibnamefont {Block}}, \bibinfo
  {author} {\bibfnamefont {A.}~\bibnamefont {D{\"{o}}rr}}, \bibinfo {author}
  {\bibfnamefont {C.}~\bibnamefont {Droese}}, \bibinfo {author} {\bibfnamefont
  {T.}~\bibnamefont {Eronen}}, \bibinfo {author} {\bibfnamefont
  {M.}~\bibnamefont {Goncharov}}, \bibinfo {author} {\bibfnamefont
  {M.}~\bibnamefont {H{\"{o}}cker}}, \bibinfo {author} {\bibfnamefont
  {J.}~\bibnamefont {Ketter}}, \bibinfo {author} {\bibfnamefont {E.~M.}\
  \bibnamefont {Ramirez}}, \bibinfo {author} {\bibfnamefont {D.~A.}\
  \bibnamefont {Nesterenko}}, \bibinfo {author} {\bibfnamefont {Y.~N.}\
  \bibnamefont {Novikov}}, \ and\ \bibinfo {author} {\bibfnamefont
  {L.}~\bibnamefont {Schweikhard}},\ }\href {\doibase
  10.1007/s00340-013-5621-0} {\bibfield  {journal} {\bibinfo  {journal}
  {Applied Physics B: Lasers and Optics}\ }\textbf {\bibinfo {volume} {114}},\
  \bibinfo {pages} {107} (\bibinfo {year} {2014})}\BibitemShut {NoStop}%
\bibitem [{\citenamefont {Eliseev}\ \emph {et~al.}(2013)\citenamefont
  {Eliseev}, \citenamefont {Blaum}, \citenamefont {Block}, \citenamefont
  {Droese}, \citenamefont {Goncharov}, \citenamefont {{Minaya Ramirez}},
  \citenamefont {Nesterenko}, \citenamefont {Novikov},\ and\ \citenamefont
  {Schweikhard}}]{Eliseev2013}%
  \BibitemOpen
  \bibfield  {author} {\bibinfo {author} {\bibfnamefont {S.}~\bibnamefont
  {Eliseev}}, \bibinfo {author} {\bibfnamefont {K.}~\bibnamefont {Blaum}},
  \bibinfo {author} {\bibfnamefont {M.}~\bibnamefont {Block}}, \bibinfo
  {author} {\bibfnamefont {C.}~\bibnamefont {Droese}}, \bibinfo {author}
  {\bibfnamefont {M.}~\bibnamefont {Goncharov}}, \bibinfo {author}
  {\bibfnamefont {E.}~\bibnamefont {{Minaya Ramirez}}}, \bibinfo {author}
  {\bibfnamefont {D.~A.}\ \bibnamefont {Nesterenko}}, \bibinfo {author}
  {\bibfnamefont {Y.~N.}\ \bibnamefont {Novikov}}, \ and\ \bibinfo {author}
  {\bibfnamefont {L.}~\bibnamefont {Schweikhard}},\ }\href {\doibase
  10.1103/PhysRevLett.110.082501} {\bibfield  {journal} {\bibinfo  {journal}
  {Physical Review Letters}\ }\textbf {\bibinfo {volume} {110}},\ \bibinfo
  {pages} {82501} (\bibinfo {year} {2013})}\BibitemShut {NoStop}%
\bibitem [{\citenamefont {Kramida}\ \emph {et~al.}(2020)\citenamefont
  {Kramida}, \citenamefont {{Yu.~Ralchenko}}, \citenamefont {Reader},\ and\
  \citenamefont {{and NIST ASD Team}}}]{NIST_ASD}%
  \BibitemOpen
  \bibfield  {author} {\bibinfo {author} {\bibfnamefont {A.}~\bibnamefont
  {Kramida}}, \bibinfo {author} {\bibnamefont {{Yu.~Ralchenko}}}, \bibinfo
  {author} {\bibfnamefont {J.}~\bibnamefont {Reader}}, \ and\ \bibinfo {author}
  {\bibnamefont {{and NIST ASD Team}}},\ }\href@noop {} {}\bibinfo
  {howpublished} {{NIST Atomic Spectra Database (ver. 5.8), [Online].
  Available: {\tt{https://physics.nist.gov/asd}} [2021, January 19]. National
  Institute of Standards and Technology, Gaithersburg, MD.}} (\bibinfo {year}
  {2020})\BibitemShut {NoStop}%
\bibitem [{\citenamefont {Kellerbauer}\ \emph {et~al.}(2003)\citenamefont
  {Kellerbauer}, \citenamefont {Blaum}, \citenamefont {Bollen}, \citenamefont
  {Herfurth}, \citenamefont {Kluge}, \citenamefont {Kuckein}, \citenamefont
  {Sauvan}, \citenamefont {Scheidenberger},\ and\ \citenamefont
  {Schweikhard}}]{Kellerbauer2003}%
  \BibitemOpen
  \bibfield  {author} {\bibinfo {author} {\bibfnamefont {A.}~\bibnamefont
  {Kellerbauer}}, \bibinfo {author} {\bibfnamefont {K.}~\bibnamefont {Blaum}},
  \bibinfo {author} {\bibfnamefont {G.}~\bibnamefont {Bollen}}, \bibinfo
  {author} {\bibfnamefont {F.}~\bibnamefont {Herfurth}}, \bibinfo {author}
  {\bibfnamefont {H.~J.}\ \bibnamefont {Kluge}}, \bibinfo {author}
  {\bibfnamefont {M.}~\bibnamefont {Kuckein}}, \bibinfo {author} {\bibfnamefont
  {E.}~\bibnamefont {Sauvan}}, \bibinfo {author} {\bibfnamefont
  {C.}~\bibnamefont {Scheidenberger}}, \ and\ \bibinfo {author} {\bibfnamefont
  {L.}~\bibnamefont {Schweikhard}},\ }\href {\doibase
  10.1140/epjd/e2002-00222-0} {\bibfield  {journal} {\bibinfo  {journal}
  {European Physical Journal D}\ }\textbf {\bibinfo {volume} {22}},\ \bibinfo
  {pages} {53} (\bibinfo {year} {2003})}\BibitemShut {NoStop}%
\bibitem [{\citenamefont {Roux}\ \emph {et~al.}(2013)\citenamefont {Roux},
  \citenamefont {Blaum}, \citenamefont {Block}, \citenamefont {Droese},
  \citenamefont {Eliseev}, \citenamefont {Goncharov}, \citenamefont {Herfurth},
  \citenamefont {Ramirez}, \citenamefont {Nesterenko}, \citenamefont
  {Novikov},\ and\ \citenamefont {Schweikhard}}]{Roux2013}%
  \BibitemOpen
  \bibfield  {author} {\bibinfo {author} {\bibfnamefont {C.}~\bibnamefont
  {Roux}}, \bibinfo {author} {\bibfnamefont {K.}~\bibnamefont {Blaum}},
  \bibinfo {author} {\bibfnamefont {M.}~\bibnamefont {Block}}, \bibinfo
  {author} {\bibfnamefont {C.}~\bibnamefont {Droese}}, \bibinfo {author}
  {\bibfnamefont {S.}~\bibnamefont {Eliseev}}, \bibinfo {author} {\bibfnamefont
  {M.}~\bibnamefont {Goncharov}}, \bibinfo {author} {\bibfnamefont
  {F.}~\bibnamefont {Herfurth}}, \bibinfo {author} {\bibfnamefont {E.~M.}\
  \bibnamefont {Ramirez}}, \bibinfo {author} {\bibfnamefont {D.~A.}\
  \bibnamefont {Nesterenko}}, \bibinfo {author} {\bibfnamefont {Y.~N.}\
  \bibnamefont {Novikov}}, \ and\ \bibinfo {author} {\bibfnamefont
  {L.}~\bibnamefont {Schweikhard}},\ }\href {\doibase
  10.1140/epjd/e2013-40110-x} {\bibfield  {journal} {\bibinfo  {journal} {The
  European Physical Journal D}\ }\textbf {\bibinfo {volume} {67}},\ \bibinfo
  {pages} {1} (\bibinfo {year} {2013})}\BibitemShut {NoStop}%
\bibitem [{\citenamefont {Birge}(1932)}]{Birge1932}%
  \BibitemOpen
  \bibfield  {author} {\bibinfo {author} {\bibfnamefont {R.~T.}\ \bibnamefont
  {Birge}},\ }\href {\doibase 10.1103/PhysRev.40.207} {\bibfield  {journal}
  {\bibinfo  {journal} {Physical Review}\ }\textbf {\bibinfo {volume} {40}},\
  \bibinfo {pages} {207} (\bibinfo {year} {1932})}\BibitemShut {NoStop}%
\bibitem [{\citenamefont {Huang}\ \emph {et~al.}(2021)\citenamefont {Huang},
  \citenamefont {Wang}, \citenamefont {Kondev}, \citenamefont {Audi},\ and\
  \citenamefont {Naimi}}]{Huang2021}%
  \BibitemOpen
  \bibfield  {author} {\bibinfo {author} {\bibfnamefont {W.}~\bibnamefont
  {Huang}}, \bibinfo {author} {\bibfnamefont {M.}~\bibnamefont {Wang}},
  \bibinfo {author} {\bibfnamefont {F.}~\bibnamefont {Kondev}}, \bibinfo
  {author} {\bibfnamefont {G.}~\bibnamefont {Audi}}, \ and\ \bibinfo {author}
  {\bibfnamefont {S.}~\bibnamefont {Naimi}},\ }\href {\doibase
  10.1088/1674-1137/abddb0} {\bibfield  {journal} {\bibinfo  {journal} {Chinese
  Physics C}\ }\textbf {\bibinfo {volume} {45}},\ \bibinfo {pages} {030002}
  (\bibinfo {year} {2021})}\BibitemShut {NoStop}%
\bibitem [{\citenamefont {Brown}\ \emph {et~al.}(2005)\citenamefont {Brown},
  \citenamefont {Stone}, \citenamefont {Stone}, \citenamefont {Towner},\ and\
  \citenamefont {Hjorth-Jensen}}]{SN100PN}%
  \BibitemOpen
  \bibfield  {author} {\bibinfo {author} {\bibfnamefont {B.~A.}\ \bibnamefont
  {Brown}}, \bibinfo {author} {\bibfnamefont {N.~J.}\ \bibnamefont {Stone}},
  \bibinfo {author} {\bibfnamefont {J.~R.}\ \bibnamefont {Stone}}, \bibinfo
  {author} {\bibfnamefont {I.~S.}\ \bibnamefont {Towner}}, \ and\ \bibinfo
  {author} {\bibfnamefont {M.}~\bibnamefont {Hjorth-Jensen}},\ }\href {\doibase
  10.1103/PhysRevC.71.044317} {\bibfield  {journal} {\bibinfo  {journal} {Phys.
  Rev. C}\ }\textbf {\bibinfo {volume} {71}},\ \bibinfo {pages} {044317}
  (\bibinfo {year} {2005})}\BibitemShut {NoStop}%
\bibitem [{\citenamefont {Brown}\ and\ \citenamefont {Rae}(2014)}]{Brown2014}%
  \BibitemOpen
  \bibfield  {author} {\bibinfo {author} {\bibfnamefont {B.}~\bibnamefont
  {Brown}}\ and\ \bibinfo {author} {\bibfnamefont {W.}~\bibnamefont {Rae}},\
  }\href {\doibase 10.1016/j.nds.2014.07.022} {\bibfield  {journal} {\bibinfo
  {journal} {Nuclear Data Sheets}\ }\textbf {\bibinfo {volume} {120}},\
  \bibinfo {pages} {115} (\bibinfo {year} {2014})}\BibitemShut {NoStop}%
\bibitem [{\citenamefont {Suhonen}(2017)}]{Suhonen2017}%
  \BibitemOpen
  \bibfield  {author} {\bibinfo {author} {\bibfnamefont {J.~T.}\ \bibnamefont
  {Suhonen}},\ }\href {\doibase 10.3389/fphy.2017.00055} {\bibfield  {journal}
  {\bibinfo  {journal} {Frontiers in Physics}\ }\textbf {\bibinfo {volume} {5}}
  (\bibinfo {year} {2017}),\ 10.3389/fphy.2017.00055}\BibitemShut {NoStop}%
\bibitem [{\citenamefont {Kotila}\ and\ \citenamefont
  {Iachello}(2012)}]{Kotila2012}%
  \BibitemOpen
  \bibfield  {author} {\bibinfo {author} {\bibfnamefont {J.}~\bibnamefont
  {Kotila}}\ and\ \bibinfo {author} {\bibfnamefont {F.}~\bibnamefont
  {Iachello}},\ }\href {\doibase 10.1103/PhysRevC.85.034316} {\bibfield
  {journal} {\bibinfo  {journal} {Physical Review C - Nuclear Physics}\
  }\textbf {\bibinfo {volume} {85}},\ \bibinfo {pages} {34316} (\bibinfo {year}
  {2012})}\BibitemShut {NoStop}%
\bibitem [{\citenamefont {Gove}\ and\ \citenamefont
  {Martin}(1971)}]{GOVE1971205}%
  \BibitemOpen
  \bibfield  {author} {\bibinfo {author} {\bibfnamefont {N.}~\bibnamefont
  {Gove}}\ and\ \bibinfo {author} {\bibfnamefont {M.}~\bibnamefont {Martin}},\
  }\href {\doibase https://doi.org/10.1016/S0092-640X(71)80026-8} {\bibfield
  {journal} {\bibinfo  {journal} {Atomic Data and Nuclear Data Tables}\
  }\textbf {\bibinfo {volume} {10}},\ \bibinfo {pages} {205} (\bibinfo {year}
  {1971})}\BibitemShut {NoStop}%
\bibitem [{\citenamefont {Davis}\ \emph {et~al.}(1968)\citenamefont {Davis},
  \citenamefont {Harmer},\ and\ \citenamefont {Hoffman}}]{Davis68}%
  \BibitemOpen
  \bibfield  {author} {\bibinfo {author} {\bibfnamefont {R.}~\bibnamefont
  {Davis}}, \bibinfo {author} {\bibfnamefont {D.~S.}\ \bibnamefont {Harmer}}, \
  and\ \bibinfo {author} {\bibfnamefont {K.~C.}\ \bibnamefont {Hoffman}},\
  }\href {\doibase 10.1103/PhysRevLett.20.1205} {\bibfield  {journal} {\bibinfo
   {journal} {Phys. Rev. Lett.}\ }\textbf {\bibinfo {volume} {20}},\ \bibinfo
  {pages} {1205} (\bibinfo {year} {1968})}\BibitemShut {NoStop}%
\bibitem [{\citenamefont {Raghavan}(1997)}]{Raghavan97}%
  \BibitemOpen
  \bibfield  {author} {\bibinfo {author} {\bibfnamefont {R.~S.}\ \bibnamefont
  {Raghavan}},\ }\href {\doibase 10.1103/PhysRevLett.78.3618} {\bibfield
  {journal} {\bibinfo  {journal} {Phys. Rev. Lett.}\ }\textbf {\bibinfo
  {volume} {78}},\ \bibinfo {pages} {3618} (\bibinfo {year}
  {1997})}\BibitemShut {NoStop}%
\bibitem [{\citenamefont {Haselschwardt}\ \emph {et~al.}(2023)\citenamefont
  {Haselschwardt}, \citenamefont {Lenardo}, \citenamefont {Daniels},
  \citenamefont {Finch}, \citenamefont {Friesen}, \citenamefont {Howell},
  \citenamefont {Malone}, \citenamefont {Mancil},\ and\ \citenamefont
  {Tornow}}]{Haselschwardt2023}%
  \BibitemOpen
  \bibfield  {author} {\bibinfo {author} {\bibfnamefont {S.~J.}\ \bibnamefont
  {Haselschwardt}}, \bibinfo {author} {\bibfnamefont {B.~G.}\ \bibnamefont
  {Lenardo}}, \bibinfo {author} {\bibfnamefont {T.}~\bibnamefont {Daniels}},
  \bibinfo {author} {\bibfnamefont {S.~W.}\ \bibnamefont {Finch}}, \bibinfo
  {author} {\bibfnamefont {F.~Q.~L.}\ \bibnamefont {Friesen}}, \bibinfo
  {author} {\bibfnamefont {C.~R.}\ \bibnamefont {Howell}}, \bibinfo {author}
  {\bibfnamefont {C.~R.}\ \bibnamefont {Malone}}, \bibinfo {author}
  {\bibfnamefont {E.}~\bibnamefont {Mancil}}, \ and\ \bibinfo {author}
  {\bibfnamefont {W.}~\bibnamefont {Tornow}},\ }\href@noop {} {\enquote
  {\bibinfo {title} {Observation of low-lying isomeric states in $^{136}$cs: a
  new avenue for dark matter and solar neutrino detection in xenon
  detectors},}\ } (\bibinfo {year} {2023}),\ \Eprint
  {http://arxiv.org/abs/2301.11893} {arXiv:2301.11893 [nucl-ex]} \BibitemShut
  {NoStop}%
\bibitem [{\citenamefont {Newstead}\ \emph {et~al.}(2019)\citenamefont
  {Newstead}, \citenamefont {Strigari},\ and\ \citenamefont
  {Lang}}]{Newstead19}%
  \BibitemOpen
  \bibfield  {author} {\bibinfo {author} {\bibfnamefont {J.~L.}\ \bibnamefont
  {Newstead}}, \bibinfo {author} {\bibfnamefont {L.~E.}\ \bibnamefont
  {Strigari}}, \ and\ \bibinfo {author} {\bibfnamefont {R.~F.}\ \bibnamefont
  {Lang}},\ }\href {\doibase 10.1103/PhysRevD.99.043006} {\bibfield  {journal}
  {\bibinfo  {journal} {Phys. Rev. D}\ }\textbf {\bibinfo {volume} {99}},\
  \bibinfo {pages} {043006} (\bibinfo {year} {2019})}\BibitemShut {NoStop}%
\bibitem [{\citenamefont {Appel}\ \emph {et~al.}(2022)\citenamefont {Appel},
  \citenamefont {Bagdasarian}, \citenamefont {Basilico}, \citenamefont
  {Bellini}, \citenamefont {Benziger}, \citenamefont {Biondi}, \citenamefont
  {Caccianiga}, \citenamefont {Calaprice}, \citenamefont {Caminata},
  \citenamefont {Cavalcante}, \citenamefont {Chepurnov}, \citenamefont
  {D'Angelo}, \citenamefont {Derbin}, \citenamefont {Di~Giacinto},
  \citenamefont {Di~Marcello}, \citenamefont {Ding}, \citenamefont
  {Di~Ludovico}, \citenamefont {Di~Noto}, \citenamefont {Drachnev},
  \citenamefont {Franco}, \citenamefont {Galbiati}, \citenamefont {Ghiano},
  \citenamefont {Giammarchi}, \citenamefont {Goretti}, \citenamefont
  {G\"ottel}, \citenamefont {Gromov}, \citenamefont {Guffanti}, \citenamefont
  {Ianni}, \citenamefont {Ianni}, \citenamefont {Jany}, \citenamefont
  {Kobychev}, \citenamefont {Korga}, \citenamefont {Kumaran}, \citenamefont
  {Laubenstein}, \citenamefont {Litvinovich}, \citenamefont {Lombardi},
  \citenamefont {Lomskaya}, \citenamefont {Ludhova}, \citenamefont
  {Lukyanchenko}, \citenamefont {Machulin}, \citenamefont {Martyn},
  \citenamefont {Meroni}, \citenamefont {Miramonti}, \citenamefont {Misiaszek},
  \citenamefont {Muratova}, \citenamefont {Nugmanov}, \citenamefont {Oberauer},
  \citenamefont {Orekhov}, \citenamefont {Ortica}, \citenamefont {Pallavicini},
  \citenamefont {Papp}, \citenamefont {Pelicci}, \citenamefont {Penek},
  \citenamefont {Pietrofaccia}, \citenamefont {Pilipenko}, \citenamefont
  {Pocar}, \citenamefont {Raikov}, \citenamefont {Ranalli}, \citenamefont
  {Ranucci}, \citenamefont {Razeto}, \citenamefont {Re}, \citenamefont
  {Redchuk}, \citenamefont {Rossi}, \citenamefont {Sch\"onert}, \citenamefont
  {Semenov}, \citenamefont {Settanta}, \citenamefont {Skorokhvatov},
  \citenamefont {Singhal}, \citenamefont {Smirnov}, \citenamefont {Sotnikov},
  \citenamefont {Tartaglia}, \citenamefont {Testera}, \citenamefont {Unzhakov},
  \citenamefont {Villante}, \citenamefont {Vishneva}, \citenamefont {Vogelaar},
  \citenamefont {von Feilitzsch}, \citenamefont {Wojcik}, \citenamefont {Wurm},
  \citenamefont {Zavatarelli}, \citenamefont {Zuber},\ and\ \citenamefont
  {Zuzel}}]{Appel22}%
  \BibitemOpen
  \bibfield  {author} {\bibinfo {author} {\bibfnamefont {S.}~\bibnamefont
  {Appel}}, \bibinfo {author} {\bibfnamefont {Z.}~\bibnamefont {Bagdasarian}},
  \bibinfo {author} {\bibfnamefont {D.}~\bibnamefont {Basilico}}, \bibinfo
  {author} {\bibfnamefont {G.}~\bibnamefont {Bellini}}, \bibinfo {author}
  {\bibfnamefont {J.}~\bibnamefont {Benziger}}, \bibinfo {author}
  {\bibfnamefont {R.}~\bibnamefont {Biondi}}, \bibinfo {author} {\bibfnamefont
  {B.}~\bibnamefont {Caccianiga}}, \bibinfo {author} {\bibfnamefont
  {F.}~\bibnamefont {Calaprice}}, \bibinfo {author} {\bibfnamefont
  {A.}~\bibnamefont {Caminata}}, \bibinfo {author} {\bibfnamefont
  {P.}~\bibnamefont {Cavalcante}}, \bibinfo {author} {\bibfnamefont
  {A.}~\bibnamefont {Chepurnov}}, \bibinfo {author} {\bibfnamefont
  {D.}~\bibnamefont {D'Angelo}}, \bibinfo {author} {\bibfnamefont
  {A.}~\bibnamefont {Derbin}}, \bibinfo {author} {\bibfnamefont
  {A.}~\bibnamefont {Di~Giacinto}}, \bibinfo {author} {\bibfnamefont
  {V.}~\bibnamefont {Di~Marcello}}, \bibinfo {author} {\bibfnamefont {X.~F.}\
  \bibnamefont {Ding}}, \bibinfo {author} {\bibfnamefont {A.}~\bibnamefont
  {Di~Ludovico}}, \bibinfo {author} {\bibfnamefont {L.}~\bibnamefont
  {Di~Noto}}, \bibinfo {author} {\bibfnamefont {I.}~\bibnamefont {Drachnev}},
  \bibinfo {author} {\bibfnamefont {D.}~\bibnamefont {Franco}}, \bibinfo
  {author} {\bibfnamefont {C.}~\bibnamefont {Galbiati}}, \bibinfo {author}
  {\bibfnamefont {C.}~\bibnamefont {Ghiano}}, \bibinfo {author} {\bibfnamefont
  {M.}~\bibnamefont {Giammarchi}}, \bibinfo {author} {\bibfnamefont
  {A.}~\bibnamefont {Goretti}}, \bibinfo {author} {\bibfnamefont {A.~S.}\
  \bibnamefont {G\"ottel}}, \bibinfo {author} {\bibfnamefont {M.}~\bibnamefont
  {Gromov}}, \bibinfo {author} {\bibfnamefont {D.}~\bibnamefont {Guffanti}},
  \bibinfo {author} {\bibfnamefont {A.}~\bibnamefont {Ianni}}, \bibinfo
  {author} {\bibfnamefont {A.}~\bibnamefont {Ianni}}, \bibinfo {author}
  {\bibfnamefont {A.}~\bibnamefont {Jany}}, \bibinfo {author} {\bibfnamefont
  {V.}~\bibnamefont {Kobychev}}, \bibinfo {author} {\bibfnamefont
  {G.}~\bibnamefont {Korga}}, \bibinfo {author} {\bibfnamefont
  {S.}~\bibnamefont {Kumaran}}, \bibinfo {author} {\bibfnamefont
  {M.}~\bibnamefont {Laubenstein}}, \bibinfo {author} {\bibfnamefont
  {E.}~\bibnamefont {Litvinovich}}, \bibinfo {author} {\bibfnamefont
  {P.}~\bibnamefont {Lombardi}}, \bibinfo {author} {\bibfnamefont
  {I.}~\bibnamefont {Lomskaya}}, \bibinfo {author} {\bibfnamefont
  {L.}~\bibnamefont {Ludhova}}, \bibinfo {author} {\bibfnamefont
  {G.}~\bibnamefont {Lukyanchenko}}, \bibinfo {author} {\bibfnamefont
  {I.}~\bibnamefont {Machulin}}, \bibinfo {author} {\bibfnamefont
  {J.}~\bibnamefont {Martyn}}, \bibinfo {author} {\bibfnamefont
  {E.}~\bibnamefont {Meroni}}, \bibinfo {author} {\bibfnamefont
  {L.}~\bibnamefont {Miramonti}}, \bibinfo {author} {\bibfnamefont
  {M.}~\bibnamefont {Misiaszek}}, \bibinfo {author} {\bibfnamefont
  {V.}~\bibnamefont {Muratova}}, \bibinfo {author} {\bibfnamefont
  {R.}~\bibnamefont {Nugmanov}}, \bibinfo {author} {\bibfnamefont
  {L.}~\bibnamefont {Oberauer}}, \bibinfo {author} {\bibfnamefont
  {V.}~\bibnamefont {Orekhov}}, \bibinfo {author} {\bibfnamefont
  {F.}~\bibnamefont {Ortica}}, \bibinfo {author} {\bibfnamefont
  {M.}~\bibnamefont {Pallavicini}}, \bibinfo {author} {\bibfnamefont
  {L.}~\bibnamefont {Papp}}, \bibinfo {author} {\bibfnamefont {L.}~\bibnamefont
  {Pelicci}}, \bibinfo {author} {\bibfnamefont {O.}~\bibnamefont {Penek}},
  \bibinfo {author} {\bibfnamefont {L.}~\bibnamefont {Pietrofaccia}}, \bibinfo
  {author} {\bibfnamefont {N.}~\bibnamefont {Pilipenko}}, \bibinfo {author}
  {\bibfnamefont {A.}~\bibnamefont {Pocar}}, \bibinfo {author} {\bibfnamefont
  {G.}~\bibnamefont {Raikov}}, \bibinfo {author} {\bibfnamefont {M.~T.}\
  \bibnamefont {Ranalli}}, \bibinfo {author} {\bibfnamefont {G.}~\bibnamefont
  {Ranucci}}, \bibinfo {author} {\bibfnamefont {A.}~\bibnamefont {Razeto}},
  \bibinfo {author} {\bibfnamefont {A.}~\bibnamefont {Re}}, \bibinfo {author}
  {\bibfnamefont {M.}~\bibnamefont {Redchuk}}, \bibinfo {author} {\bibfnamefont
  {N.}~\bibnamefont {Rossi}}, \bibinfo {author} {\bibfnamefont
  {S.}~\bibnamefont {Sch\"onert}}, \bibinfo {author} {\bibfnamefont
  {D.}~\bibnamefont {Semenov}}, \bibinfo {author} {\bibfnamefont
  {G.}~\bibnamefont {Settanta}}, \bibinfo {author} {\bibfnamefont
  {M.}~\bibnamefont {Skorokhvatov}}, \bibinfo {author} {\bibfnamefont
  {A.}~\bibnamefont {Singhal}}, \bibinfo {author} {\bibfnamefont
  {O.}~\bibnamefont {Smirnov}}, \bibinfo {author} {\bibfnamefont
  {A.}~\bibnamefont {Sotnikov}}, \bibinfo {author} {\bibfnamefont
  {R.}~\bibnamefont {Tartaglia}}, \bibinfo {author} {\bibfnamefont
  {G.}~\bibnamefont {Testera}}, \bibinfo {author} {\bibfnamefont
  {E.}~\bibnamefont {Unzhakov}}, \bibinfo {author} {\bibfnamefont {F.~L.}\
  \bibnamefont {Villante}}, \bibinfo {author} {\bibfnamefont {A.}~\bibnamefont
  {Vishneva}}, \bibinfo {author} {\bibfnamefont {R.~B.}\ \bibnamefont
  {Vogelaar}}, \bibinfo {author} {\bibfnamefont {F.}~\bibnamefont {von
  Feilitzsch}}, \bibinfo {author} {\bibfnamefont {M.}~\bibnamefont {Wojcik}},
  \bibinfo {author} {\bibfnamefont {M.}~\bibnamefont {Wurm}}, \bibinfo {author}
  {\bibfnamefont {S.}~\bibnamefont {Zavatarelli}}, \bibinfo {author}
  {\bibfnamefont {K.}~\bibnamefont {Zuber}}, \ and\ \bibinfo {author}
  {\bibfnamefont {G.}~\bibnamefont {Zuzel}} (\bibinfo {collaboration} {Borexino
  Collaboration}),\ }\href {\doibase 10.1103/PhysRevLett.129.252701} {\bibfield
   {journal} {\bibinfo  {journal} {Phys. Rev. Lett.}\ }\textbf {\bibinfo
  {volume} {129}},\ \bibinfo {pages} {252701} (\bibinfo {year}
  {2022})}\BibitemShut {NoStop}%
\bibitem [{\citenamefont {Bahcall}(1994)}]{Bahcall94}%
  \BibitemOpen
  \bibfield  {author} {\bibinfo {author} {\bibfnamefont {J.~N.}\ \bibnamefont
  {Bahcall}},\ }\href {\doibase 10.1103/PhysRevD.49.3923} {\bibfield  {journal}
  {\bibinfo  {journal} {Phys. Rev. D}\ }\textbf {\bibinfo {volume} {49}},\
  \bibinfo {pages} {3923} (\bibinfo {year} {1994})}\BibitemShut {NoStop}%
\bibitem [{\citenamefont {Redshaw}\ \emph {et~al.}(2007)\citenamefont
  {Redshaw}, \citenamefont {Wingfield}, \citenamefont {McDaniel},\ and\
  \citenamefont {Myers}}]{Matthew07}%
  \BibitemOpen
  \bibfield  {author} {\bibinfo {author} {\bibfnamefont {M.}~\bibnamefont
  {Redshaw}}, \bibinfo {author} {\bibfnamefont {E.}~\bibnamefont {Wingfield}},
  \bibinfo {author} {\bibfnamefont {J.}~\bibnamefont {McDaniel}}, \ and\
  \bibinfo {author} {\bibfnamefont {E.~G.}\ \bibnamefont {Myers}},\ }\href
  {\doibase 10.1103/PhysRevLett.98.053003} {\bibfield  {journal} {\bibinfo
  {journal} {Phys. Rev. Lett.}\ }\textbf {\bibinfo {volume} {98}},\ \bibinfo
  {pages} {053003} (\bibinfo {year} {2007})}\BibitemShut {NoStop}%
\bibitem [{\citenamefont {Dror}\ \emph {et~al.}(2020)\citenamefont {Dror},
  \citenamefont {Elor},\ and\ \citenamefont {McGehee}}]{Dror2020}%
  \BibitemOpen
  \bibfield  {author} {\bibinfo {author} {\bibfnamefont {J.~A.}\ \bibnamefont
  {Dror}}, \bibinfo {author} {\bibfnamefont {G.}~\bibnamefont {Elor}}, \ and\
  \bibinfo {author} {\bibfnamefont {R.}~\bibnamefont {McGehee}},\ }\href
  {\doibase 10.1103/PhysRevLett.124.181301} {\bibfield  {journal} {\bibinfo
  {journal} {Phys. Rev. Lett.}\ }\textbf {\bibinfo {volume} {124}},\ \bibinfo
  {pages} {181301} (\bibinfo {year} {2020})}\BibitemShut {NoStop}%
\end{thebibliography}%

\end{document}